\documentclass[fleqn,10pt,usenames,dvipsnames]{wlscirep}

\usepackage[utf8]{inputenc} 
\usepackage{booktabs}
\usepackage{float}
\usepackage{pdfpages}
\usepackage{soul}
\usepackage{fancybox}
\usepackage{xspace}
\usepackage{textcomp}
\usepackage{graphicx}
\usepackage{amsmath,amsfonts,amssymb}
\usepackage{multirow}
\usepackage[normalem]{ulem} 
\usepackage{pdflscape}
\usepackage{bbold} 

\usepackage{changepage}

\usepackage{textcomp,marvosym}

\usepackage{cite}

\usepackage{nameref}

\usepackage{microtype}
\DisableLigatures[f]{encoding = *, family = * }

\usepackage{array}

\newcolumntype{+}{!{\vrule width 2pt}}

\newlength\savedwidth

\usepackage{multirow}

\PassOptionsToPackage{hyphens}{url}\usepackage{hyperref}

\usepackage{xspace}
\newcommand{\ie}{i.e.\xspace}
\newcommand{\eg}{e.g.\xspace}
\newcommand{\et}{et al.\xspace}

\newcommand{\mde}{\texttt{MDE}\xspace}
\newcommand{\bgt}{\texttt{BGT}\xspace}
\newcommand{\sao}{\texttt{SAO}\xspace}

\newcommand{\all}{\texttt{all}\xspace}
\newcommand{\work}{\texttt{work}\xspace}
\newcommand{\nonwork}{\texttt{nonwork}\xspace}

\newcommand{\men}{\texttt{men}\xspace}
\newcommand{\women}{\texttt{women}\xspace}

\newcommand{\lowerc}{\texttt{lower}\xspace}
\newcommand{\middlec}{\texttt{middle}\xspace}
\newcommand{\upperc}{\texttt{upper}\xspace}

\newcommand{\avg}[1]{\langle #1 \rangle}
\newcommand{\kde}{{\rm KDE}}

\usepackage{xr}
\usepackage{xr-hyper}
\usepackage{hyperref}
\usepackage{booktabs,tabularx}

\graphicspath{{figures/}}

\title{Differences in the spatial landscape of urban mobility: gender and socioeconomic perspectives}

\author[1,*]{Mariana Macedo}
\author[2]{Laura Lotero}
\author[3,4]{Alessio Cardillo}
\author[1]{Ronaldo Menezes}
\author[1]{Hugo Barbosa}

\affil[1]{BioComplex Lab, University of Exeter -- Exeter, United Kingdom}
\affil[2]{Faculty of Industrial Engineering, Universidad Pontificia Bolivariana -- Medell\'in, Colombia}
\affil[3]{Department of Computer Science and Mathematics, University Rovira i Virgili, E-43007 Tarragona, Spain}
\affil[4]{GOTHAM Lab -- Institute for Biocomputation and Physics of Complex Systems (BIFI), University of Zaragoza, E-50018 Zaragoza, Spain}

\affil[*]{mmacedo@biocomplexlab.org}

\keywords{human mobility, gender, socioeconomic status, commuting patterns}

\begin{abstract}
Many of our routines and activities are linked to our ability to move; be it commuting to work, shopping for groceries, or meeting friends. Yet, factors that limit the individuals' ability to fully realise their mobility needs will ultimately affect the opportunities they can have access to (\eg  cultural activities, professional interactions). One important aspect frequently overlooked in human mobility studies is how gender-centred issues can amplify other sources of mobility disadvantages (\eg  socioeconomic inequalities), unevenly affecting the pool of opportunities men and women have access to. In this work, we leverage on a combination of computational, statistical, and information-theoretical approaches to investigate the existence of systematic discrepancies in the mobility diversity (\ie the diversity of travel destinations) of {\it (1)} men and women from different socioeconomic backgrounds, and {\it (2)} work and non-work travels. Our analysis is based on datasets containing multiple instances of large-scale, official, travel surveys carried out in three major metropolitan areas in South America: Medell\'in and Bogot\'a in Colombia, and S\~ao Paulo in Brazil.  Our results indicate the presence of general discrepancies in the urban mobility diversities related to the gender and socioeconomic characteristics of the individuals. Lastly, this paper sheds new light on the possible origins of gender-level human mobility inequalities, contributing to the general understanding of disaggregated patterns in human mobility.
\end{abstract}

\begin{document}

\flushbottom
\maketitle

\thispagestyle{empty}

	\section*{Introduction}
	
	Human travelling behaviours are linked to a myriad of problems in cities such as traffic congestion, disease spreading, and criminality. Conversely, many of our social and economic activities, such as working, shopping, and socialising hinge on our ability to move. Not surprisingly, human mobility plays a key role in the social and economic development of cities \cite{xu2018human,bettencourt2013origins,barbosa2020uncovering}. One example is the economic and social impacts of the mobility restrictions imposed by governments worldwide in 2020 due to the COVID-19 pandemic~\cite{Warren2020,bonaccorsi2020economic,Ho2020,kraemer2020effect}. In fact, the economic and social hardships arising from the mobility restrictions unevenly affected different segments of the populations, exacerbating inequalities of economic, social ~\cite{bonaccorsi2020economic,coven2020disparities,Blundell2020} and gender~\cite{collins2020covid,Myers2020} roots. On the other hand, the scientific literature in human mobility has vast pieces of evidence indicating the existence of persistent mobility differences across socioeconomic and gender groups~\cite{salon2010mobility, adeel2017gender}. In many instances, urban mobility differences are rooted in the economic landscape of a city and the spatial distribution of opportunities (\eg employment) in urban areas. 
	
	Thus, understanding the mobility necessities and characteristics of the different segments of the society -- especially the less advantaged populations -- is crucial to reduce social, economic and gender inequalities, objectives contemplated by the United Nations in their Sustainable Development Goals \cite{un_goals}.
	
	With regards to the socioeconomic facets of urban mobility, in previous works, we have shown that, in Colombia, middle-income populations tend to distribute their visits to most of the areas of a city while upper and lower-income groups are more likely to concentrate their trips towards a smaller fraction of the zones~\cite{Lotero2016,lotero2016rich}. Furthermore, it has been shown that in Brazil, populations from different socioeconomic strata tend to use different transportation modes~\cite{vasconcellos2018urban}.
	
	Moreira \et suggest that in Brazil, public safety can play an important role in how people move~ \cite{Moreira2020}. Even though safety is a problem faced by all genders, empirical evidences indicate that, when possible, women are more likely to opt for longer (or more costly) journeys in favour of a trip perceived as safer~\cite{Moreira2020,MejiaDorantes2020,ng2018understanding, Singh2019}. Nevertheless, in general, women are more likely to make shorter trips than men~\cite{macedo2020gender}. Furthermore, women with care duties are more likely to work at locations having shorter commuting travel time, leading them to display different patterns of mobility when compared with men \cite{Stoet2019, Kaufman2020,Petrongolo2020}. Hence, the spatial distribution of job opportunities in a city, combined with the gender division of labour and imbalances in the workloads with care responsibilities may all contribute to gender-centred differences in mobility.

	Thus, of all the sociodemographic dimensions known to influence human travelling behaviours, in this work, we concentrate on the interaction between gender and the socioeconomic characteristics of travellers and their mobility patterns. We argue that previously-observed socioeconomic differences in urban mobility~\cite{Lotero2016,lotero2016rich,Moreira2020,macedo2020gender}  could be connected with how different groups concentrate/distribute their travels throughout the urban area. Our goal, therefore, is to quantify how concentrated/dispersed the travelling behaviours of different segments of a population are. 
	
	Indeed, certain characteristics of the urban areas, combined with the opportunity landscape of the cities, will attract people in different ways -- and with different magnitudes, -- in connection with their sociodemographic characteristics. However, it is noteworthy that the travelling behaviours of a society are not static in time but, rather, evolve alongside the population, in response to underlying cultural, social, and economic changes. Data-driven, longitudinal studies related to sociodemographic processes in human mobility are frequently hindered by data limitations, with few exceptions. Using credit card record data, Lenormand \et showed that women in Spain tend to travel shorter distances, frequently closer to their trajectories' centre of mass, while men tend to display longer journeys~\cite{lenormand2015influence}. We hypothesise that these discrepancies in the mobility patterns of women and men can be exacerbated when combined with other dimensions, such as the socioeconomic status.  
	
	With this objective in mind, we analyse urban mobility through the lenses of mobility diversity~\cite{lenormand2020entropy,Pappalardo2015,Pappalardo2016}. In our formulation, the mobility diversity is measured as the Shannon entropy of the empirical probability distribution of travels made towards the set of zones or sub-areas (\eg census tracts) of a city. Our analyses are based on multiple waves of household travel surveys from three metropolitan areas in South America carried out in different points in time: Medell\'in (2005, 2017) and Bogot\'a (2012, 2019) in Colombia, and S\~ao Paulo (1997, 2007, 2017) in Brazil. Each survey is composed of three parts: the travel questionnaire, focusing on the trips  themselves (\eg destination, modal, and purpose), the household  (\eg number of residents and family arrangement), and the sociodemographic characteristics of the respondents (\eg gender, age, and socioeconomic stratum). 
	
	Our results indicate that, in the areas we analysed, the travel distribution of men and women are marked -- consistently -- different. Moreover, such differences are not uniform across socioeconomic groups. In fact, the socioeconomic mechanisms operating on the mobility landscape seem to emphasise and amplify the gender-centred differences in mobility. Our findings shed new light on the potential mechanisms contributing to gender and socioeconomic disadvantages in urban areas. In a broader perspective, our results suggest that a  possible combination of gender biases in employment opportunities with the specialisation and spatial organisation of the areas of the urban fabric, spurs imbalances in the mobility travel costs sustained by men and women.

	\section*{Data and methods}
	\label{sec:datamethods}

	
	In this work, we analyse household travel surveys from three large urban areas in South America, being two in Colombia and one in Brazil. The Colombian datasets correspond to the metropolitan areas  of Medell\'in (henceforth stylised as \mde) and Bogot\'a (\bgt) while the Brazilian dataset covers the metropolitan area of S\~ao Paulo (\sao). For each area, we analysed the data collected in different years: \{2005, 2017\} for \mde, \{2012, 2019\} for \bgt, and \{1997, 2007, 2017\} for \sao, respectively. Table~\ref{tab:dataset} summarises the main characteristics of each dataset, providing information such as the number of zones covered by the surveys ($N_Z$), their total area ($\mathcal{A}$), the number of travellers ($N_P$), the number of travels ($N_T$), the fractions of travellers ($f^X$),  and travels ($f_T^X$) per gender $X \in \{M, W\}$. The data of different gender and socioeconomic groups are detailed in S1~Section of the Supplementary Material.
	
	\begin{table}[htbp]
		\begin{adjustwidth}{-0in}{0in}
			\centering
			\caption{\textbf{Main properties of the raw datasets analysed in our study. For each region, we have the total area covered $\mathcal{A}$, and the number of zones into which it is divided, $N_Z$.} Then, for each year we have the number of travellers $N_P$, the fraction of men (women) travellers $f^M$ ($f^W$), the number of travels $N_T$, and the fraction of travels made by men (women) $f_T^M$ ($f_T^W$).}
			
			\tabcolsep=0.15cm
			\begin{tabular}{c r c c c c c r c c}
				\toprule
				\textbf{Region} & $\mathcal{A}$ ($km^2$) & $N_Z$ & Year & $N_P$ & $f^M$ & $f^W$ & \multicolumn{1}{c}{$N_T$} & $f_{T}^M$ & $f_{T}^W$ \\
				\midrule
				\multirow{2}{*}\mde & \multirow{2}{*}{1,167}  & \multirow{2}{*}{215} & 2005 & 22,840 & 0.48 & 0.52 & 70,773 & 0.48 & 0.52 \\
				&     &     & 2017 & 38,048 & 0.49 & 0.51 & 87,614 & 0.49 & 0.51 \\ \midrule
				\multirow{2}{*}\bgt & \multirow{2}{*}{24,477} & \multirow{2}{*}{400} & 2012 & 11,677 & 0.46 & 0.54 & 41,440 & 0.45 & 0.55 \\ 
				&     &     & 2019 & 47,149 & 0.48 & 0.52 & 164,931 & 0.48 & 0.52 \\ \midrule
				\multirow{3}{*}\sao & \multirow{3}{*}{9,486}  & \multirow{3}{*}{248} & 1997 & 37,316 & 0.48 & 0.52 & 93,376 & 0.48 & 0.52 \\
				&     &     & 2007 & 51,103 & 0.49 & 0.51 & 137,411 & 0.49 & 0.51 \\
				&     &     & 2017 & 48,085 & 0.50 & 0.50 & 125,544 & 0.50 & 0.50 \\
				\bottomrule
			\end{tabular}
			\label{tab:dataset}
		\end{adjustwidth}
	\end{table}

	The goal of the mobility surveys used in this work is to capture the set of recent and regular trips performed by people. Each travel entry comes with information on its origin and destination zones, departure and arrival times, the purpose of the travel (also known as \emph{demand}), and the transportation mode(s) used. Additionally, the surveys include also questions related to the sociodemographic characteristics of the respondents, such as their gender, occupation, and socioeconomic status. Each person lists its travels done to fulfil the purposes of work, going home, study, have fun, shop, and go to a health-related appointment. Moreover, each travel can be broken down into at most five pieces, each made using a different transportation mode. Finally, most of the surveys' questionnaires are made of multiple-choice questions to limit inconsistencies.
	
	The populations' samples are based on the sociodemographic composition of the population resident within the metropolitan area. To account for the representativeness of a respondent's answers -- according to their socioeconomic and demographic characteristics relative to the general population -- each response in the survey is associated with an \emph{expansion factor}. Such a factor scales up the sample estimating to the population from which the sample was drawn. We carry out our analysis on the ``expanded datasets,'' with the sole exception of the \mde survey of 2017, for which the expansion factors are unavailable. Notwithstanding, it is worth mentioning that the expanded datasets are not too different from the original ones (see S3.2~Section of the Supplementary Material). Among the plethora of attributes available to discriminate travellers, we argue that travels related to work purposes are of special interest because work activities represent better differences in social strata and gender.
	
	To ensure the consistency of our comparative analyses, we harmonised the spatial partitioning of the cities as well as the socioeconomic categorisation of the respondents. A few zones in each city were split into smaller areas to accommodate changes in their underlying population numbers occurred between consecutive waves of the survey.  Therefore, we decided to use the partitioning corresponding of the first year available in each metropolitan area, and merge together those zones that split in the following years. We ensured that our aggregation methodology does not alter the overall distributions of travel time, travel distance, and the fraction of travels. Summing up, throughout our manuscript, the spatial division of the data corresponds to the area divisions of 2005 (\mde), 2012 (\bgt), and 1997 (\sao), respectively. A spatial visualisation of the final divisions can be seen in Figs~\ref{fig:mapgender} and \ref{fig:mapsocio}, as well as in S2~Section of the Supplementary Material. We also ensured that the socioeconomic classification of the populations was consistent across years, regions and -- to a lesser extent -- countries.

	Ensuring the consistency of socioeconomic status is less straightforward than spatial partitioning. The reason is that not only the classification might change across time, but also, different countries adopt different criteria/schemes. To interpret the results for Colombia and Brazil from a common framework, we rearranged the socioeconomic classifications for both countries into three socioeconomic strata: \lowerc, \middlec, and \upperc. The population distributions obtained from this rearrangement (\ref{S-tab:alltravelsexp} and \ref{S-tab:worktravelsexp}~Tables in the Supplementary Material) were similar to what is frequently observed in modern societies~\cite{leo2016socioeconomic,Gao2019}. More information on the methodological details on the socioeconomic classification of the populations is available in S1.1~Section of the Supplementary Material.


	After ensuring that the data are aggregated consistently both in terms of spatial partitioning and socioeconomic classification of the travellers, we can proceed to analyse the mobility patterns across population groups. We decided to study the evolution in time of the mobility patterns and its similarities/differences between cities using an approach based on information theory.
	
	Specifically, we compute a modified version of the \emph{mobility diversity} indicator proposed by Pappalardo \et in~\cite{Pappalardo2015}, and use it similarly to what is done in the work of Lenormand \et \cite{lenormand2020entropy}. More precisely, Pappalardo \et \cite{Pappalardo2015} compute the mobility diversity based on each individual's mobility, whereas Lenormand \et \cite{lenormand2020entropy} consider the attractiveness of a location taking into account where people live and visit. Contrarily to what done by Lenormand \et \cite{lenormand2020entropy}, we are not interested in the zone where people live, but in capturing the trip-chaining characteristics of overall travels.
	
	Given a set of travels made by a group of travellers $X$ to satisfy/fulfil purpose $d$, the \emph{mobility diversity} of such a group, $H^{X}_{d}$, is -- up to a multiplicative factor, -- the Shannon entropy of its \emph{spatial coverage}. The latter corresponds to the probability that travellers from a group $X$ visit a given zone $i$ to satisfy/fulfil purpose $d$, $p^{X}_{d}(i)$, yielding:
	\begin{equation}
		\label{eq:mobdiv}
		H_{d}^{X} = - \frac{1}{\log_2 N_Z} \sum^{N_Z}_{i=1} p^{X}_{d}(i) \, \log_2 p^{X}_{d}(i) \,,
	\end{equation}
	where
	\begin{equation}
		\label{eq:probability}
		p^{X}_{d}(i) = \frac{N^{X}_{d}(i)}{N^{X}_{d}} \,.
	\end{equation}

	Here, $N^{X}_{d}(i)$ denotes the number of travels made by a group $X$ to fulfil purpose $d$ whose destination is zone $i$; whereas $N^{X}_{d}$ denotes the total number of travels made by a group $X$ to fulfil purpose $d$. According to Eq~\eqref{eq:mobdiv}, $H_{d}^{X} \in [0,1]$ with the boundary values corresponding to two distinct mobility scenarios. The case $H_{d}^{X} = 0$ corresponds to the scenario where all travels have the same destination zone. The case $H_{d}^{X} = 1$, instead, corresponds to the scenario where travels cover uniformly all the available zones (\ie Eq~\eqref{eq:probability} is independent on the zone). The detailed calculations of the boundary values of $H$ are available in S3.1~Section of the Supplementary Material. Finally, as mentioned previously, the group $X$ can be chosen according to several criteria based on gender, socioeconomic status, or a combination of them.

	To account for the effect of variations in sample and population sizes and estimate the variations in mobility diversity in the populations, we employed a bootstrapping strategy and estimated the $H$ values from random samples of the data. More precisely, given the set of all the travels made by a certain group of travellers, $X$, fulfilling a given purpose, $d$, we sample 60\% of such travels and then compute the quantity we are interested in (\eg  the value of $H_{d}^{X}$ using Eq~\eqref{eq:mobdiv}); we repeat the sampling 1000 times. Considering in the bootstrapping percentages of travels ranging from 60\% to 90\% does not affect qualitatively the results (see \ref{S-fig:percentage_effect} Fig of the Supplementary Material). The analyses presented in the next section were performed on the distributions of the mobility diversities obtained from the bootstrapping. From these distributions,  we used different statistical methods to verify the differences in the diversity distributions across groups. Details on the statistical verification methods and results are provided in the S4~Section of the Supplementary Material.

	There are several caveats associated with bootstrapping. The first is that the results might depend on the size of the bootstrap sample. To account for such a possibility, we analysed the evolution of the values of $H$ with respect to the size of the bootstrap sample. We found that the value of $H$ saturates as the sample's size increases (see S5~Section of the Supplementary Material). The second factor that might affect the values of $H$ is the fact that the areas of the zones in the tessellation are not equal. For this reason, we compared the empirical values of $H$ with those obtained using five null models accounting for both the non-homogeneous size of the areas, and the non-uniform density of inhabitants (see S6~Section of the Supplementary Material). Finally, the third potential issue is related to the effects of the sample's composition (\eg{} the presence of more poor travellers than rich ones) on the value of $H$. We argue, however, that such an imbalance is part of the intrinsic nature of the Colombian and Brazilian societies and, therefore, does not constitute a bias in our results.
	
	In summary, our results are valid even when accounting for the sample size effect and when comparing them with null models. Indeed, we find that the following factors are essential to mobility: travel distances following a truncated power-law distribution and non-homogeneous residential distribution \cite{Barbosa2018}. However, these factors are insufficient to explain the differences in mobility diversity across gender and socioeconomic groups.

	
	\section*{Results}
	\label{sec:results}
	
	In this section, we explore the existence of systematic differences in the mobility patterns of men and women that could represent potential sources of additional disadvantages and inequalities. In line with our hypothesis on the existence of structural, gender-centred mobility disadvantages, reflected from the labour market, we focus our attention on work-related trips. The rationale is that potential gender inequalities permeating the socioeconomic fabric (\eg{} employment landscape) would manifest themselves as differences in the commuting behaviours of men and women,  even more so across socioeconomic groups.
	
	To explore the role played by gender and socioeconomic factors on urban mobility, we focus our analysis on measuring the mobility diversity of the overall travels (\all) performed by each segment, including, for instance,  travels related to \textit{shopping}, \textit{health}, and \textit{leisure}, and their differences with work (\work) and non-work (\nonwork) travels. Next, we analyse the mobility diversity of the populations across gender or socioeconomic strata. Finally, we investigate the mobility diversity distributions obtained from the combined effects of both gender and socioeconomic strata. Given that our focus is on the disadvantages endured by different segments of the society, we conducted our last set of experiments specifically for the work travels, without further partitioning the data into other travel categories. 
	
	The visual exploration of the data (Figs~\ref{fig:mapgender} and \ref{fig:mapsocio}, and S2~Section of the Supplementary Material) confirms our hypotheses on the role of gender and socioeconomic status on mobility. We observe, for example, that the majority of the areas in \bgt are covered by a high density of \work travels performed by men and by the middle class. As expansions factors are unavailable for the 2017 survey of \mde, we are unable to make any claim on the temporal evolution in \mde. However, in the discussion section, we will comment about longitudinal changes of $H$ for \mde.

	\subsection*{Analysis of the travel's purpose}
	\label{sub:travel_purpose}
	
\begin{figure}[htbp]
		\centering
		\includegraphics[width=0.975\textwidth]{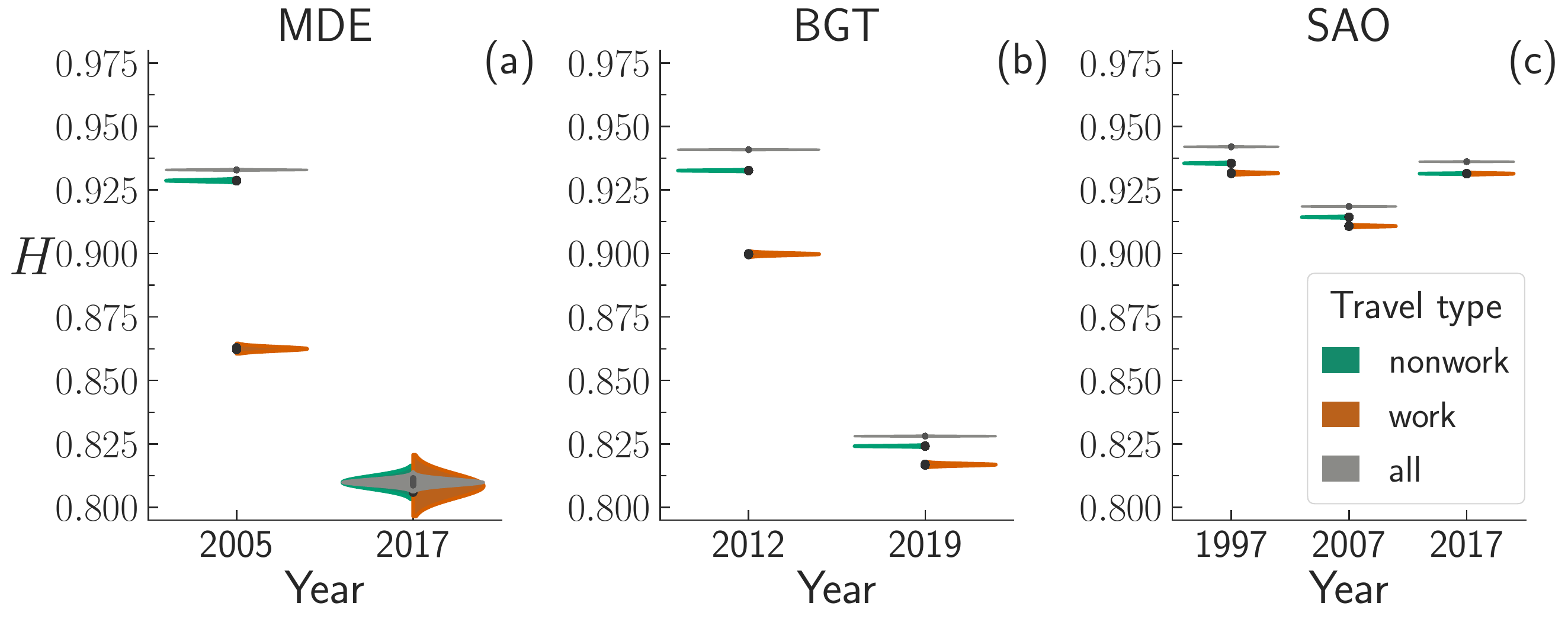}
		\caption{\textbf{Violin plots of the bootstrapped mobility diversity, $H$, for \all, \work and \nonwork travels made in each region and year.} To better visualise the overlap (or not) between the distributions of \all, \work, and \nonwork travels, we show the distributions for \all travels duplicated (entire grey violins instead of half-violins).}
		\label{fig:aggregatedpatterns}
	\end{figure}

	To assess the extent to which different purpose of travel shape mobility, we divide the travels into three groups: those related to work (\work), those related to any purpose except for work (\nonwork) and, finally, all the travels regardless of their purpose (\all). Then, we compute the mobility diversity, $H$, of travels belonging to each of the aforementioned groups. Fig~\ref{fig:aggregatedpatterns} displays the evolution across time of the distribution of the values of $H$ for each travels' group and each city. By computing the Welch's $t$-test between each pair of distributions, we ensure that they are statistically distinct ($p$-value $<$ 0.001). The visual inspection of Fig~\ref{fig:aggregatedpatterns} reveals that the peaks of the distributions of $H$ are all located above 0.80, meaning that the travels are -- more or less -- evenly distributed across all the zones available regardless of the purpose, city, or year considered. In general, we observe that travels of the \work group display smaller values of $H$ than those belonging to the other groups. This suggests that job opportunities are more spatially concentrated throughout urban areas than other sources of mobility demands together, like education or leisure. Looking at the evolution in time of the diversity of \work travels, we observe that both Colombian cities display a monotone decrease of $H$ overtime, whereas \sao{} $H$ displays a decrease (1997--2007) followed by an increase (2007--2017). Such a decrease may denote that job opportunities might have declined in some zones, making the travels to work more spatially concentrated.

	\subsection*{Effects of gender on mobility's diversity}\label{sub:gender}

	
	Men and women display different patterns in mobility such as average travel time, preferences on the mode of transportation, and commuting travel distance~\cite{Moreira2020, MejiaDorantes2020, Singh2019,macedo2020gender, gustafson2006work, Rodriguez2020, Psylla2017,Farre2020}. Here, we analyse whether mobility diversity is a suitable candidate to grasp differences in mobility in a gender-centred manner.
	
\begin{figure}[htbp]
		\centering
                \includegraphics[width=0.975\textwidth]{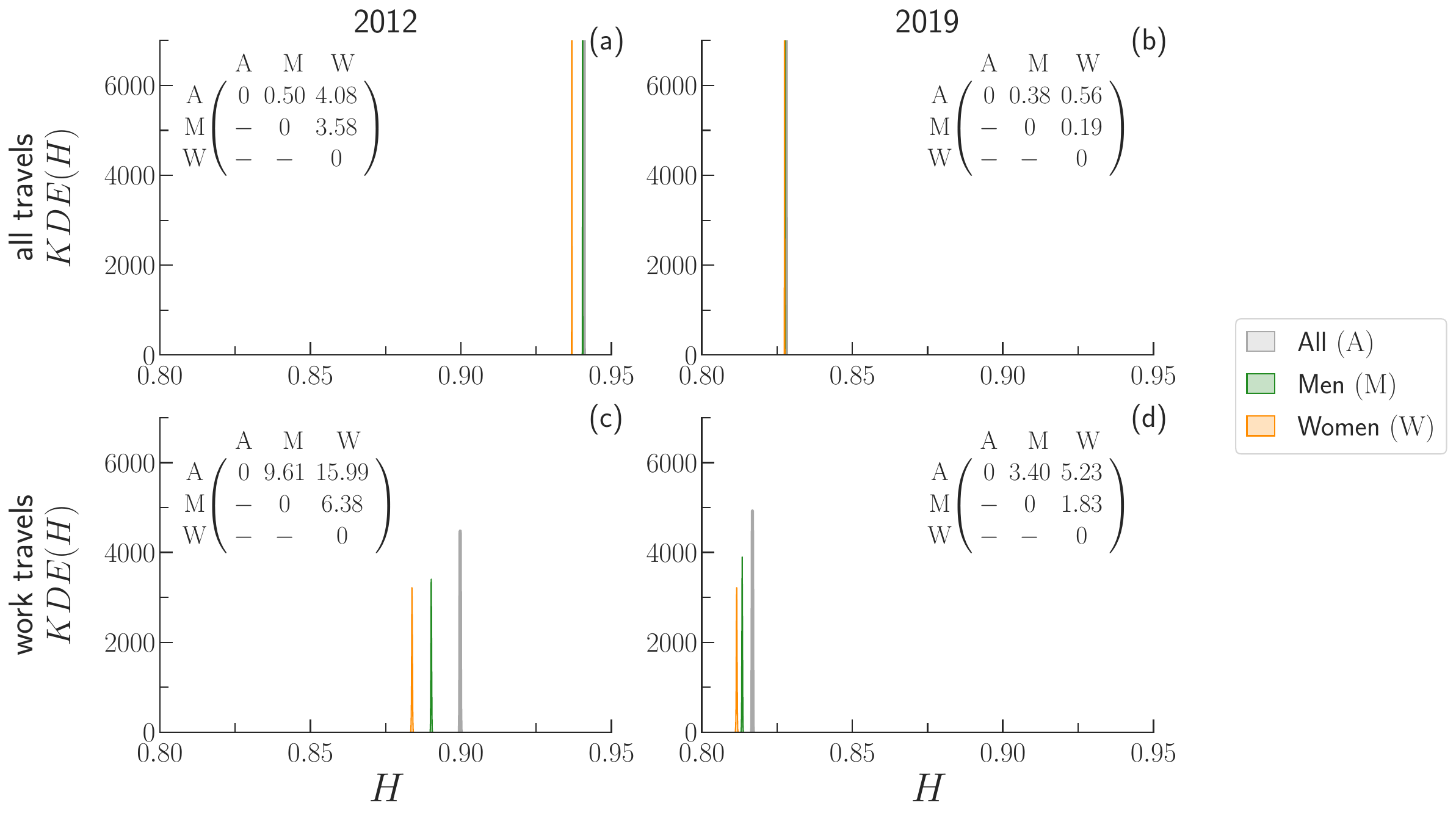}
		\caption{\textbf{Kernel Density Estimation plots of the mobility diversity, $H$, for \all travels (panels \textbf{a} and \textbf{b}) and \work travels (panels \textbf{c} and \textbf{d}) in the urban area of \bgt.} For each travel purpose, we plot $\kde(H)$ for travels made by men ($M$), women ($W$), and all travellers ($A$). The matrix appearing within each graphic summarises the distances between the medians of the distributions (peak-to-peak distances multiplied by $10^{-3}$).}
		\label{fig:genderbogota}
	\end{figure}

	As an example, we consider the case of \bgt. In Fig~\ref{fig:genderbogota}, we display the Kernel Density Estimation (KDE) of the mobility diversity, $H$, of travels made by men ($M$), women ($W$), and all ($A$) travellers either regardless of the purpose of travel (\all), and for work travels only (\work). A quick inspection of Fig~\ref{fig:genderbogota} reveals that the envelopes of the KDEs tend to get closer (smaller distance between them). The fact that travellers, regardless of their gender, display similar values of $H$ means that they tend to cover the metropolitan area in a similar way. Such a phenomenon is also corroborated by the values of the peak-to-peak distance between the KDEs of $M$, $W$, and $A$ travellers shown in the matrices appearing within each panel. Moreover, the fact that the value of $H$ of a given group of travellers decreases across time confirms that travel destinations are less uniformly scattered over the metropolitan area in 2019 than in 2012.

	Another feature is that the average values of $H$ obtained for women travellers are always smaller than the one of men. The comparison between the KDEs confirms that women tend to distribute their mobility over the metropolitan area less than men. We ensure that the differences between the KDEs are statistically significant by computing the Welch's $t$-test between all the possible pairs of distributions ($p$-value $<$ 0.001). In light of the results found for \bgt, we repeat the same analysis also for the data available for \mde and \sao (see S3.3~Section of the Supplementary Material).

	Fig~\ref{fig:genderregions} provides an overview of the effects of gender on $H$ for all the urban areas together over all the available years. Even for the complete set of areas and time snapshots, the Welch's $t$-test confirmed that the distributions are statistically different ($p$-value $<$ 0.001). The sole exception is the case of \men and \all travellers of \mde in 2017.
	
	In general, the value of $H$ associated with men's mobility is higher than that of women regardless of the purpose of travel in agreement with the results observed in Fig~\ref{fig:genderbogota}. The sole exception, however, is the case of \sao in 1997 and 2007 for which $H^W > H^M$. The violin plots show also that, in general, $\Delta H^{MA} < \Delta H^{WA}$, where $\Delta H^{X Y} = \left\vert \avg{H^X} - \avg{H^Y} \right\vert$ and $X,Y \in \{ A, M, W \}$.
	
	We stress that the values of $H$ computed for travels performed by both genders can be greater than those computed considering travels made only by women or men. Such a difference is due to the fact that the distribution of the probabilities of visiting a location is more uniform for the non gendered case than the gendered one. However, it is worth noting that the opposite can also happen. This is true because when we consider only women/men travels, the probabilities may display skewed values which, instead, become more homogeneous when we consider travels made by both genders. Further details on the effects of sample size across groups can be found in S5~Section of the Supplementary Material.

	Focusing on \bgt{} and \sao{}, we observe that $H$ decreases over the years regardless of the travel's purpose or the traveller's gender (in Fig~\ref{fig:genderregions}). The \sao{} urban area displays the same V-shaped pattern (\ie{} the value of $H$ decreases between 1997 and 2007, and increases between 2007 and 2017) observed in Fig~\ref{fig:aggregatedpatterns}. In particular, we do not notice any qualitative differences between the distributions of $H_{\text{\texttt{all}}}$ and $H_{\text{\texttt{work}}}$.

	It is worth mentioning that although in our dataset women are more likely to perform more short travels than men (see \ref{S-fig:traveldistance} Fig and \ref{S-tab:traveldistance} and \ref{S-tab:traveldistancetests} Tables of the Supplementary Material)~\cite{macedo2020gender,gauvin2020gender}, the lower values of $H^W$ do not stem from the preference of women to remain within the same zone (see \ref{S-tab:probabilitysamezones}~Table of the Supplementary Material). Moreover, our sample does not show a high difference in the percentages of men and women living and working in the same zone (see \ref{S-tab:probabilitysamezones}~Table of the Supplementary Material). We also investigated whether the number of travel destinations chosen by men is higher than what women choose, but we have not found any statistically significant difference (see \ref{S-fig:workmorezones}~Fig of the Supplementary Material). Therefore, individually, women and men have similar likelihoods of performing travel to work in the same number of destinations (zones).

	The data analysis confirms that the fraction of \work travels made by \men, $P^{M}_{work}$, is higher than its women counterpart. On the other hand, we have found that non-work related travels are proportionally higher for women than men (see \ref{S-tab:higherfractiontravelsperarea}~Table of the Supplementary Material). Moreover, except for \mde in 2017, the travel's destination for women and men follows different distributions regardless of the purpose of travel (tested by Student $t$-test and Kolmogorov–Smirnov test with $p$-value $< 0.01$). In S7~Section of the Supplementary Material, we also present that the gender differences in mobility diversity are not a sole by-product from endogenous mobility and residential segregation.

	The small differences between the values of $H$ displayed in Fig~\ref{fig:genderregions}, and the balance between genders in the composition of travellers' groups, push us to ask whether such differences are concealed by other factors related, for instance, with the socioeconomic status of travellers. For this reason, we study the effects of socioeconomic status in mobility.

	\subsection*{Effects of gender \& socioeconomic status}
	\label{sub:socio}

	
	Finally, we explore the effect of socioeconomic status and gender in mobility diversity. However, before studying the effects of these two aspects combined, we must gauge the role of socioeconomic status alone. For this reason, we grouped travellers according to the three socioeconomic classes defined in the Data and Methods section (\ie \lowerc $Low$, \middlec $Mid$, and \upperc $Up$). We computed the values of $H$ of travels made by travellers belonging to each socioeconomic class, as well as for travels made by all travellers combined ($A$), and for travels made for either \all or \work purposes.

\begin{figure}[htbp]
		\centering
		\includegraphics[width=0.95\textwidth]{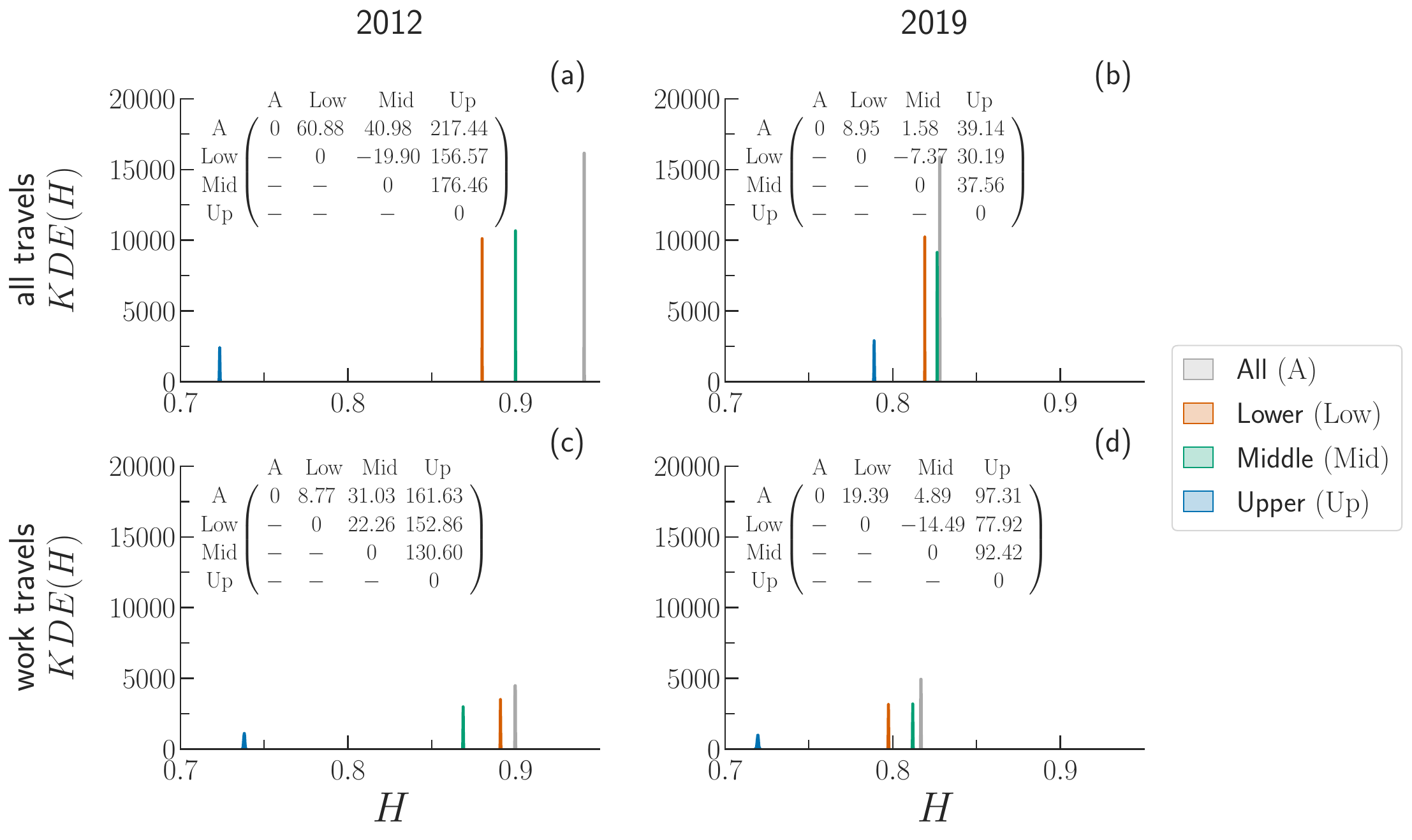}
		\caption{\textbf{Kernel Density Estimation ($\kde$) plots of the mobility diversity, $H$, for travels made by travellers belonging to different socioeconomic status in the \bgt area of 2012 (panels a and c) and 2019 (panels b and d).} The top row (panels a and b) displays the values obtained considering \all travels, whereas the bottom row (panels c and d) displays the values obtained considering only travels associated with the \work purpose. We show the $\kde(H)$ for travels made by all travellers ($A$), as well as for those belonging to the \lowerc ($Low$), \middlec ($Mid$), or \upperc ($Up$) socioeconomic class. The matrix appearing within each plot encapsulates the distance between the median of the distributions (peak-to-peak distances multiplied by $10^{-3}$).}
		\label{fig:socio_kde_bgt}
	\end{figure}

	In Fig~\ref{fig:socio_kde_bgt}, we display the $\kde(H)$ for the \bgt area for the years 2012 and 2019, respectively. In agreement with the trend observed thus far, we observe a decrease over time of the mobility diversity regardless of the travel's purpose. Travellers belonging to the \upperc class attain the lowest value of $H$, suggesting that they cover less uniformly the space. In other words, upper-income individuals might be more ``selective'' in their mobility than people belonging to the other classes. On the other hand, in general, \middlec class travellers display the highest values of $H$. The higher values of $H^{Mid}$ over the other classes suggest that individuals in the \middlec class cover the space more uniformly than other classes.

	People belonging to upper-income class might move to fewer zones because they concentrate their travels in areas that has a high volume of opportunities. Conversely, people belonging to lower-income class are obliged to move since they cannot afford to live in zones next to those with more opportunities. In addition, people belonging to \lowerc class may not have access to as many areas as \middlec class due to the lack of affordable (public) transportation and opportunities. For instance, \lowerc{} class people might not be able to reach areas in which the public transportation system is insufficient or inadequate. Finally, we observe that the peak-to-peak distances between the $\kde$ become smaller over the years, indicating a possible decrease of socioeconomic inequalities in \bgt.

	Regarding the other urban areas, we observe that the KDE plots (see \ref{S-fig:kdesociomde} and \ref{S-fig:kdesociosao}~Figs of the Supplementary Material) confirm that travellers belonging to the \upperc class attain the lowest values of $H$, whereas those belonging to the \middlec class cover more uniformly the available space. Such socioeconomic magnitudes of $H$ do not depend on the travel's purpose, albeit each urban area displays its own peculiarities.
	
\begin{figure}[htbp]
		\centering
                \includegraphics[width=0.95\textwidth]{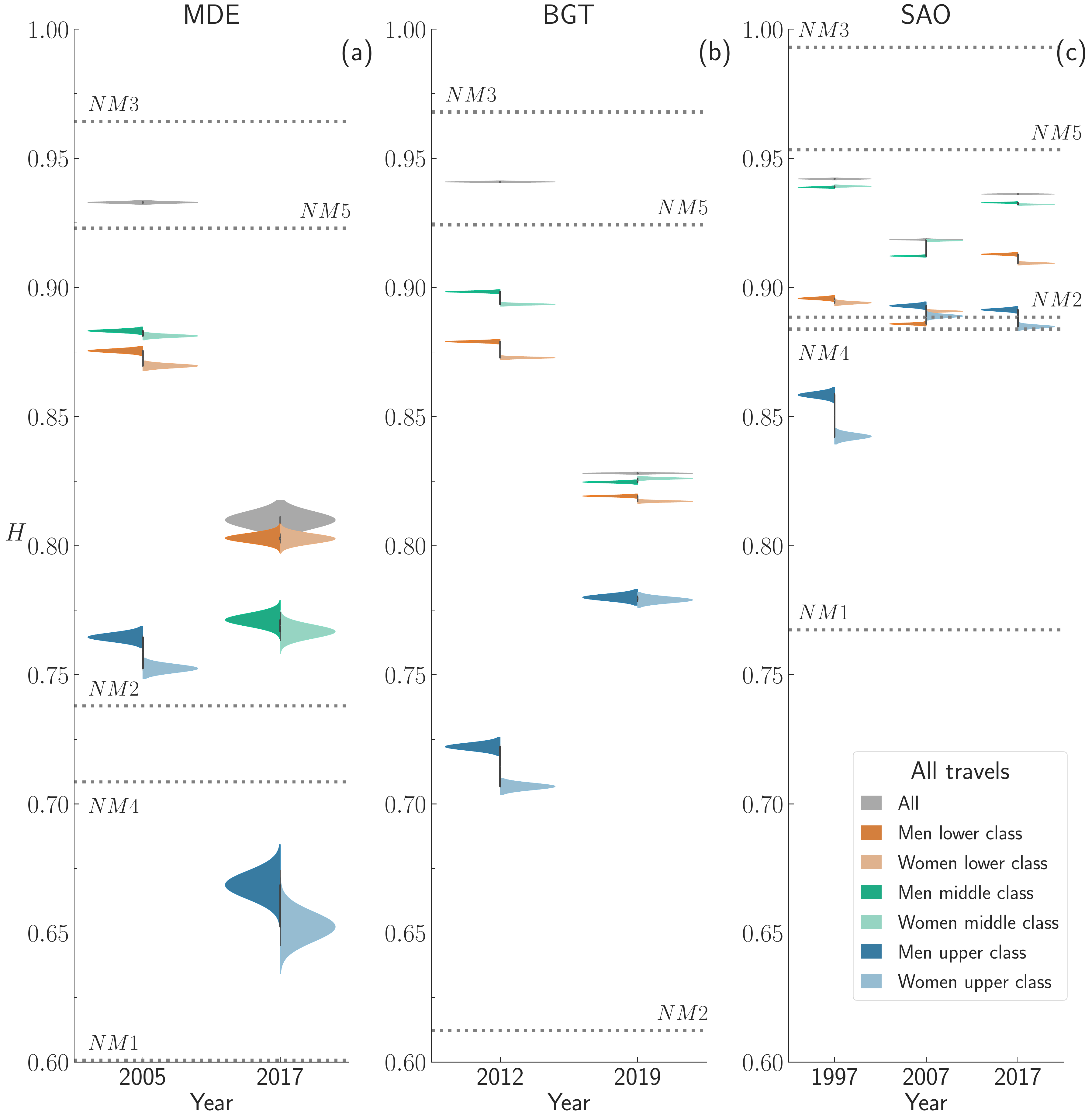}
		\caption{\textbf{Violin plots of the mobility diversity, $H$, of travels made for \all purposes by travellers grouped according to their socioeconomic status and gender.} Each column refers to a different region, and for each region, we consider all the available years. For each socioeconomic status (\upperc, \middlec, and \lowerc) a darker hue denotes men travellers, whereas lighter hue denotes women ones. Dotted lines in grey denote the values of $H$ computed from travels generated using null models (See details in S6~Section of the Supplementary Material).}
		\label{fig:allgendersocioregions}
	\end{figure}

	After analysing the role of socioeconomic status alone, we are ready to look at the combined effect of gender and socioeconomic status. To this aim, we compute the mobility diversity of travels made by travellers having a certain social status (\eg  \middlec) and gender (\eg  $W$). In Fig~\ref{fig:allgendersocioregions}, we display the violin plots of $H$ computed for travels made for \all purposes by all combinations of gender and socioeconomic status.\newline
	First, we noticed that socioeconomic status shapes the mobility of people considerably, whereas gender exerts a smaller effect. Yet, we can observe a gender distinction, with men tending to display higher values of $H$ than women within the same socioeconomic class. We noticed that the starker differences between genders occur for travellers belonging to the \upperc class (see \ref{S-tab:genderdifferences}~Table of the Supplementary Material). Similar conclusions can be drawn from the mobility diversity computed for travels made for \work purposes by all combinations of gender and socioeconomic status (see \ref{S-fig:workgendersocioregions}~Fig of the Supplementary Material). We want to stress that we can reject the hypothesis that all the distributions are statistically similar (confirmed by a Welch's $t$-test with $p$-value $< 0.01$).

	To assess the contribution of the gender and socioeconomic attributes (alone and combined) in the mobility diversity $H$, we apply three statistical tests. First, we apply the ANOVA one-way test to investigate further if the averages of the mobility diversity distributions computed separately by either the gender or socioeconomic status groups are statistically different. Then, we apply the ANOVA two-way test to investigate if the averages of the mobility diversity distributions computed by gender and socioeconomic status together are statistically different. Finally, we apply the Tukey's HSD post hoc test to identify within attributes what are the groups with statistically different average values of $H$. All the detailed explanation and specific values of $F$ and $p$-values from ANOVA and Tukey's HSD post hoc tests are detailed in S4~Section of the Supplementary Material.
	
	Based on the above-mentioned tests, we can reject the hypothesis that the mean values of the mobility diversity, $H$, from the travels performed by gender (men, women and all travellers) and socioeconomic groups (\lowerc, \middlec, \upperc and all travellers) are similar. When considering only the gender or the socioeconomic status, there is no exception in the statistical tests. 
	
	Lastly, we compare the mobility diversity distributions of gender and socioeconomic status taken together. We reject the null hypothesis that the mean values of the mobility diversity, $H$, distributions are similar from the majority of pairwise comparisons, except for the \work travels taking place in \mde during 2017 and performed by (i) \all and \men travellers, and (ii) \men of the \upperc{} class. Moreover, the value of $H$ computed for the entire population is higher than that of the travels made by travellers of a given gender and socioeconomic group. Such a difference is due to the fact that gender and socioeconomic status play a role in the spatial concentration of travels in fewer areas. Further analyses on the effect of sample size, residential distribution, and spatial tessellation are collected in S5 and S6~Sections of the Supplementary Material.
	
	Summing up, in general, the patterns in mobility diversity for different groups of gender and socioeconomic classes taken separately or together are statistically different. 
	Without exception, we can claim that the socioeconomic group consistently accounts for the highest gap in mobility diversity.

	\section*{Discussion}
	\label{sec:discussion}
	
	In search to understand general patterns -- universalities -- in human mobility, a consistent body of literature assumes that travellers are indistinguishable from one another~\cite{gonzalez2008understanding,simini2012universal,Asgari2013,barbosa2015effect,Barbosa2018,alessandretti2020scales}. However, travellers are different, and they can be differentiated according to several features. To this aim, we analyse the travel records collected by surveys conducted across several years in two Colombian (Medell\'in and Bogot\'a) and one Brazilian (S\~ao Paulo) metropolitan areas. We demonstrate that features like gender and socioeconomic status exert a strong influence on the individuals' mobility patterns at the urban level/scale.
	
	Using information theory, we measured the spatial diversity of the travels performed by different ensembles of people belonging to different gender and socioeconomic groups through a modified version of Shannon's entropy. Such a quantity, named mobility diversity, depends on where the travels take place, \ie the probability that a zone is the travel's destination. Thus, mobility diversity can be thought as a proxy of the ``predictability'' of a group's mobility~\cite{Pappalardo2015,song2010limits}.
	
	The travel records collected by the surveys come with meta-data such as age, gender, socioeconomic status, and family relationships. Such attributes may be used to group travels (and travellers) according to several criteria. We decided to focus on three class of attributes: the purpose of the travel, gender of the traveller, and his/her socioeconomic status. To decipher how each attribute shapes the mobility, we analysed the role of each attribute alone first, and then the role of the attributes taken altogether.
	
	The literature shows that the purpose of travels (\eg  going home and to work) shapes mobility differently in several spatio-temporal characteristics such as the probability of returning to the last visited location, the fraction of travels over time, the most frequent visited locations, and the amount of money spent related to the distance travelled~\cite{barbosa2015effect, Barbosa2018, ballis2020revealing,Lenormand2020}. To unveil the role played by the \emph{purpose} of the travel, we have divided travels into three groups/classes: one made of travels due to work activities (\work), one made of travels due to any purpose except work (\nonwork), and another made by all travels together regardless of their purpose (\all).  We have found that work-related travels -- in general -- are less homogeneously distributed than the other types of travels. This could be due to the fact that areas offering a higher amount of job opportunities concentrate a higher number of travels related to work. However, we have observed that each urban area evolves differently over time, suggesting that mobility is strongly intertwined with the economic context where it takes place. Thus, the patterns observed in the present work cannot be considered universal.
	
	The analysis of the role played by gender has revealed the existence of a distinction between the mobility of men and women, with the former being more entropic/diverse than the latter. In our analysis, such a phenomenon is independent of the region, time, and purpose of the travel. Such a difference between genders has also been highlighted by other studies on mobility~\cite{vasconcellos2018urban,Moreira2020,MejiaDorantes2020,ng2018understanding,Singh2019,macedo2020gender,lenormand2015influence,gustafson2006work,Rodriguez2020, Psylla2017, gauvin2020gender} which have found, among other things, that women tend to make shorter travels than men, and avoid to travel to certain destinations particularly during late hours. Moreover, we have observed that the gender differences in mobility diversity get smaller over time. Although each area/country has undergone different financial and social changes, such a reduction might be due to the effects of policies aimed either at reducing the gender gap directly or mitigating factors hindering women's mobility (\eg   insecurity)~\cite{Moreira2020,MejiaDorantes2020,Petrongolo2020,Farre2020,montoya2020gender,Chant2013}.
	
	When it comes to the role of the socioeconomic status, travellers can be analysed as a group. Our analysis highlights that the diversity gap between socioeconomic classes (SESs) is starker than the gender case. Both the differences between the values of $H$ attained by each SES, and the overall range of values of $H$, point to wealth playing a crucial role in shaping the urban mobility. Moreover, we observe a distinction among SESs, with \upperc and \lowerc being the classes exploring the space in the least diverse way, and \middlec travellers displaying the most diverse mobility patterns~\cite{carlsson1999travel, carra2016modelling, alonso1964location, frias2012relation}. However, the reasons behind the lower values of $H$ attained by \upperc and \lowerc are not the same. The \upperc class, in fact, appears to be more selective in their destinations (and also move less) possibly because they can afford a broad range of options (\ie buying a car or living in expensive areas closer to where they work). The \lowerc class, instead, limit their exploration of the available space because they lack affordable ways to move across the metropolitan areas~\cite{vasconcellos2018urban,lenormand2015influence,Gao2019,de2005urban}. 
	
	Finally, we examined the effect of combining gender and socioeconomic status in one analysis. Although SES plays a major role in discriminating travellers, we noticed that each SES displays a further separation based on gender, with men attaining higher values of $H$ than women. Such separation is independent of the purpose of travel, region, and year. Hence, we can conclude that the gender gap in mobility is a widespread phenomenon affecting women \emph{tout-court}. In fact, regardless the region, the upper-class displays the highest gender difference, and this may be true because there is a higher gender inequality in highly qualified jobs that women are less likely to pursue~\cite{stier2014occupational,clarke2020gender}.
	
	Comparing our results with the null models, we observe that the spatial organisation of the residential landscape, and the travel distance distribution both play major roles on the magnitude of $H$. The size of the sample used to compute $H$ seems to not play a significant role on the results, regardless of the combination of features used to discriminate the travellers. We observe, in fact, that using just 40\% of the travels is enough to capture the overall mobility patterns. Finally, we indicate that removing from the samples endogenous mobility and mobility made to the residential areas for each individual do not remove the gender effect observed in mobility diversity.
	
	Past studies concluded that features like gender and age of the travellers do not play a remarkable role in our ability to predict their mobility behaviour~\cite{song2010limits,Alessandretti2018}. More recent studies, instead, have highlighted that gender plays a role in the discrimination of travellers~\cite{Singh2019,Petrongolo2020,Martin2019}. Although some studies have addressed the role of the gender and the socioeconomic status of travellers separately~\cite{lotero2016rich,macedo2020gender,lenormand2015influence,gauvin2020gender}; to the best of our knowledge, their combined effect has been studied seldom. We presented here a systematic study of how gender and socioeconomic status are intertwined and shape urban mobility. In particular, we observed that there is a gap between gender regardless of the socioeconomic status of the traveller. In fact, women report disadvantages in several aspects of their life such as income, free time, and career's progression~\cite{Chant2013}, and this work shows that mobility (and access to transport) is another aspect in which women suffer~\cite{MejiaDorantes2020,Singh2019,gauvin2020gender,levin20199}. 
	
	Nevertheless, our analysis has some limitations. The first is that results are intimately tied to the type of data used. Despite being very detailed and rich in meta-data, surveys are very expensive to carry out (both economically and time-wise). Moreover, the information collected could be exposed to subjective biases on both the interviewer and the interviewed sides. Other limitations include the composition of the sample, and the tendency to capture only routine behaviours. Expansion factors surely mitigate the aforementioned issues, but they are not always available. Finally, the spatio-temporal resolution/accuracy of surveys is not comparable with other types of data, such as mobile phones or credit card records.
	
	Future extensions of our work could involve the study of other travel purposes (\eg  related to study or leisure), or the effects of the traveller's age/profession on its mobility pattern. Finally, expanding the pool of countries/cities, including cases from countries located outside Latin America, could help to unveil more generalised trends on the gender gap in urban mobility.

	
	\section*{List of abbreviations}
	
	\begin{description}
		\item[\mde{}:] The urban area of Medell\'in.
		\item[\bgt{}:] The urban area of Bogot\'a.
		\item[\sao{}:] The urban area of S\~ao Paulo.
		\item[\all{}:] The whole travels available.
		\item[\work{}:] Travels related with work activities.
		\item[\nonwork{}:] Travels not related with work activities.
		\item[\women{}:] Travels performed by women.
		\item[\men{}:] Travels performed by men.
		\item[SES:] Socioeconomic status.
		\item[\lowerc{}:] Travels performed by travellers belonging to the lower socioeconomic class.
		\item[\middlec{}:] Travels performed by travellers belonging to the middle socioeconomic class.
		\item[\upperc{}:] Travels performed by travellers belonging to the upper socioeconomic class.
	\end{description}

	\section*{Competing interests}
	The authors declare that they have no competing interests.
	
	\section*{Author contributions}
	MM and HB designed the study; LL and MM contributed with the data; MM and HB performed the analysis; MM and HB analysed the results; MM and AC wrote the paper. MM and AC prepared the graphics. All authors read, reviewed, and approved the final manuscript.
	
	\section*{Availability of data and materials}
	All the data analysed in this paper are publicly available online: Medell\'in and Bogot\'a \cite{datosabiertos, secretariadistritalmovilidad} and S\~ao Paulo \cite{datasaopaulo}. Moreover, the code is available on \url{github.com/marianagmmacedo/mobility_diversity}.
	
	\section*{Funding}
	RM, HB and MM acknowledge support in part by the U. S. Army Research Office (ARO) under grant number W911NF-18-1-0421. LL acknowledges partial support from the Leicester Institute for Advanced Studies under the Rutherford Fellowship Scheme funded by the Department for Business, Energy and Industrial Strategy, and administered by Universities UK International. AC acknowledges the support of the Spanish Ministerio de Ciencia e Innovaci\'on (MICINN) through Grant IJCI-2017-34300. 
	
	\section*{Acknowledgements}
	We thank Area Metropolitana del Valle de Aburr\'a, in Medell\'in, and Secretarıa Distrital de Movilidad, in Bogot\'a, for the Origin-Destination Surveys Datasets. This work was made possible by a Visiting Fellowship of LL in the Leicester Institute for Advanced Studies at the University of Leicester.
	
	We acknowledge that the surveys released under the Creative Commons licence the shape files of the urban areas with the specific spatial tessellation: Medell\'in and Bogot\'a \cite{datosabiertos, secretariadistritalmovilidad}, and S\~ao Paulo \cite{datasaopaulo}. The open data rules for every urban area considered in our study allow to share, modify and publish the shape files as CC BY 4.0 licence without any restriction: Medell\'in  \cite{aburra_copyright}, Bogot\'a \cite{bogota_copyright} and S\~ao Paulo \cite{saopaulo_copyright}. Finally, we acknowledge that the maps plotted in this paper uses the basemap by CartoDB, under CC BY 3.0 and data by OpenStreetMap, under ODbL  \cite{basemap_copyright}.

 	\bibliography{references}

\begin{thebibliography}{10}
\urlstyle{rm}
\expandafter\ifx\csname url\endcsname\relax
  \def\url#1{\texttt{#1}}\fi
\expandafter\ifx\csname urlprefix\endcsname\relax\def\urlprefix{URL }\fi
\expandafter\ifx\csname doiprefix\endcsname\relax\def\doiprefix{DOI: }\fi
\providecommand{\bibinfo}[2]{#2}
\providecommand{\eprint}[2][]{\url{#2}}

\bibitem{xu2018human}
\bibinfo{author}{Xu, Y.}, \bibinfo{author}{Belyi, A.}, \bibinfo{author}{Bojic,
  I.} \& \bibinfo{author}{Ratti, C.}
\newblock \bibinfo{journal}{\bibinfo{title}{Human mobility and socioeconomic
  status: Analysis of singapore and boston}}.
\newblock {\emph{\JournalTitle{Computers, Environment and Urban Systems}}}
  \textbf{\bibinfo{volume}{72}}, \bibinfo{pages}{51--67}
  (\bibinfo{year}{2018}).

\bibitem{bettencourt2013origins}
\bibinfo{author}{Bettencourt, L.~M.}
\newblock \bibinfo{journal}{\bibinfo{title}{The origins of scaling in cities}}.
\newblock {\emph{\JournalTitle{Science}}} \textbf{\bibinfo{volume}{340}},
  \bibinfo{pages}{1438--1441}, \doiprefix\url{10.1126/science.1235823}
  (\bibinfo{year}{2013}).

\bibitem{barbosa2020uncovering}
\bibinfo{author}{Barbosa, H.} \emph{et~al.}
\newblock \bibinfo{journal}{\bibinfo{title}{Uncovering the socioeconomic facets
  of human mobility}}.
\newblock {\emph{\JournalTitle{Scientific Reports}}}
  \textbf{\bibinfo{volume}{11}}, \bibinfo{pages}{8616},
  \doiprefix\url{10.1038/s41598-021-87407-4} (\bibinfo{year}{2021}).

\bibitem{Warren2020}
\bibinfo{author}{Warren, M.~S.} \& \bibinfo{author}{Skillman, S.~W.}
\newblock \bibinfo{journal}{\bibinfo{title}{{Mobility Changes in Response to
  COVID-19}}}.
\newblock {\emph{\JournalTitle{arXiv}}} \bibinfo{pages}{2003.14228}
  (\bibinfo{year}{2020}).
\newblock \eprint{2003.14228}.

\bibitem{bonaccorsi2020economic}
\bibinfo{author}{Bonaccorsi, G.} \emph{et~al.}
\newblock \bibinfo{journal}{\bibinfo{title}{Economic and social consequences of
  human mobility restrictions under {COVID}-19}}.
\newblock {\emph{\JournalTitle{Proceedings of the National Academy of
  Sciences}}} \textbf{\bibinfo{volume}{117}}, \bibinfo{pages}{15530--15535},
  \doiprefix\url{10.1073/pnas.2007658117} (\bibinfo{year}{2020}).

\bibitem{Ho2020}
\bibinfo{author}{Ho, K.~W.}, \bibinfo{author}{Kheng, M.} \&
  \bibinfo{author}{Tan, T.}
\newblock \bibinfo{journal}{\bibinfo{title}{Challenges to social mobility in
  singapore}}.
\newblock {\emph{\JournalTitle{SSRN}}} \doiprefix\url{10.2139/ssrn.3643266}
  (\bibinfo{year}{2020}).

\bibitem{kraemer2020effect}
\bibinfo{author}{Kraemer, M.~U.} \emph{et~al.}
\newblock \bibinfo{journal}{\bibinfo{title}{The effect of human mobility and
  control measures on the {COVID}-19 epidemic in {C}hina}}.
\newblock {\emph{\JournalTitle{Science}}} \textbf{\bibinfo{volume}{368}},
  \bibinfo{pages}{493--497}, \doiprefix\url{10.1126/science.abb4218}
  (\bibinfo{year}{2020}).

\bibitem{coven2020disparities}
\bibinfo{author}{Coven, J.} \& \bibinfo{author}{Gupta, A.}
\newblock \bibinfo{title}{Disparities in mobility responses to {COVID}-19}.
\newblock \bibinfo{type}{Tech. Rep.}, \bibinfo{institution}{NYU Stern Working
  Paper} (\bibinfo{year}{2020}).

\bibitem{Blundell2020}
\bibinfo{author}{Blundell, R.}, \bibinfo{author}{{Costa Dias}, M.},
  \bibinfo{author}{Joyce, R.} \& \bibinfo{author}{Xu, X.}
\newblock \bibinfo{journal}{\bibinfo{title}{{COVID-19 and Inequalities}}}.
\newblock {\emph{\JournalTitle{Fiscal Studies}}} \textbf{\bibinfo{volume}{41}},
  \bibinfo{pages}{291--319}, \doiprefix\url{10.1111/1475-5890.12232}
  (\bibinfo{year}{2020}).

\bibitem{collins2020covid}
\bibinfo{author}{Collins, C.}, \bibinfo{author}{Landivar, L.~C.},
  \bibinfo{author}{Ruppanner, L.} \& \bibinfo{author}{Scarborough, W.~J.}
\newblock \bibinfo{journal}{\bibinfo{title}{{COVID}-19 and the gender gap in
  work hours}}.
\newblock {\emph{\JournalTitle{Gender, Work \& Organization}}}
  \bibinfo{pages}{1--12}, \doiprefix\url{10.1111/gwao.12506}
  (\bibinfo{year}{2020}).

\bibitem{Myers2020}
\bibinfo{author}{Myers, K.~R.} \emph{et~al.}
\newblock \bibinfo{journal}{\bibinfo{title}{{Unequal effects of the COVID-19
  pandemic on scientists}}}.
\newblock {\emph{\JournalTitle{Nature Human Behaviour}}}
  \textbf{\bibinfo{volume}{4}}, \bibinfo{pages}{880--883},
  \doiprefix\url{10.1038/s41562-020-0921-y} (\bibinfo{year}{2020}).

\bibitem{salon2010mobility}
\bibinfo{author}{Salon, D.} \& \bibinfo{author}{Gulyani, S.}
\newblock \bibinfo{journal}{\bibinfo{title}{Mobility, poverty, and gender:
  travel ‘choices’ of slum residents in nairobi, kenya}}.
\newblock {\emph{\JournalTitle{Transport Reviews}}}
  \textbf{\bibinfo{volume}{30}}, \bibinfo{pages}{641--657},
  \doiprefix\url{10.1080/01441640903298998} (\bibinfo{year}{2010}).

\bibitem{adeel2017gender}
\bibinfo{author}{Adeel, M.}, \bibinfo{author}{Yeh, A.~G.} \&
  \bibinfo{author}{Zhang, F.}
\newblock \bibinfo{journal}{\bibinfo{title}{Gender inequality in mobility and
  mode choice in pakistan}}.
\newblock {\emph{\JournalTitle{Transportation}}} \textbf{\bibinfo{volume}{44}},
  \bibinfo{pages}{1519--1534}, \doiprefix\url{10.1007/s11116-016-9712-8}
  (\bibinfo{year}{2017}).

\bibitem{un_goals}
\bibinfo{title}{United nations | sustainable development goals (sdgs)}.
\newblock \bibinfo{howpublished}{Available at:
  \url{https://sdgs.un.org/goals}}.

\bibitem{Lotero2016}
\bibinfo{author}{Lotero, L.}, \bibinfo{author}{Cardillo, A.},
  \bibinfo{author}{Hurtado, R.} \& \bibinfo{author}{G{\'{o}}mez-Garde{\~{n}}es,
  J.}
\newblock \bibinfo{title}{{Several Multiplexes in the Same City: The Role of
  Socioeconomic Differences in Urban Mobility}}.
\newblock In \bibinfo{editor}{Garas, A.} (ed.)
  \emph{\bibinfo{booktitle}{Interconnected Networks. Understanding Complex
  Systems}}, Understanding Complex Systems, \bibinfo{pages}{149--164},
  \doiprefix\url{10.1007/978-3-319-23947-7\_9} (\bibinfo{publisher}{Springer
  International Publishing}, \bibinfo{address}{Cham}, \bibinfo{year}{2016}).

\bibitem{lotero2016rich}
\bibinfo{author}{Lotero, L.}, \bibinfo{author}{Hurtado, R.~G.},
  \bibinfo{author}{Flor{\'\i}a, L.~M.} \&
  \bibinfo{author}{G{\'o}mez-Garde{\~n}es, J.}
\newblock \bibinfo{journal}{\bibinfo{title}{Rich do not rise early:
  spatio-temporal patterns in the mobility networks of different socio-economic
  classes}}.
\newblock {\emph{\JournalTitle{Royal Society open science}}}
  \textbf{\bibinfo{volume}{3}}, \bibinfo{pages}{150654},
  \doiprefix\url{10.1098/rsos.150654} (\bibinfo{year}{2016}).

\bibitem{vasconcellos2018urban}
\bibinfo{author}{Vasconcellos, E.~A.}
\newblock \bibinfo{journal}{\bibinfo{title}{Urban transport policies in
  {B}razil: The creation of a discriminatory mobility system}}.
\newblock {\emph{\JournalTitle{Journal of transport geography}}}
  \textbf{\bibinfo{volume}{67}}, \bibinfo{pages}{85--91},
  \doiprefix\url{10.1016/j.jtrangeo.2017.08.014} (\bibinfo{year}{2018}).

\bibitem{Moreira2020}
\bibinfo{author}{Moreira, G.~C.} \& \bibinfo{author}{Ceccato, V.~A.}
\newblock \bibinfo{journal}{\bibinfo{title}{{Gendered mobility and violence in
  the S{\~{a}}o Paulo metro, Brazil}}}.
\newblock {\emph{\JournalTitle{Urban Studies}}} \textbf{\bibinfo{volume}{58}},
  \bibinfo{pages}{1--20}, \doiprefix\url{10.1177/0042098019885552}
  (\bibinfo{year}{2020}).

\bibitem{MejiaDorantes2020}
\bibinfo{author}{Mej{\'{i}}a-Dorantes, L.} \& \bibinfo{author}{{Soto
  Villagr{\'{a}}n}, P.}
\newblock \bibinfo{journal}{\bibinfo{title}{{A review on the influence of
  barriers on gender equality to access the city: A synthesis approach of
  Mexico City and its Metropolitan Areaa}}}.
\newblock {\emph{\JournalTitle{Cities}}} \textbf{\bibinfo{volume}{96}},
  \bibinfo{pages}{102439}, \doiprefix\url{10.1016/j.cities.2019.102439}
  (\bibinfo{year}{2020}).

\bibitem{ng2018understanding}
\bibinfo{author}{Ng, W.-S.} \& \bibinfo{author}{Acker, A.}
\newblock \bibinfo{title}{Understanding urban travel behaviour by gender for
  efficient and equitable transport policies}.
\newblock \bibinfo{howpublished}{International Transport Forum Discussion
  Paper. Available at:
  \url{https://www.itf-oecd.org/understanding-urban-travel-behaviour-gender-efficient-and-equitable-transport-policies}}
  (\bibinfo{year}{2018}).

\bibitem{Singh2019}
\bibinfo{author}{Singh, Y.~J.}
\newblock \bibinfo{journal}{\bibinfo{title}{Is smart mobility also
  gender-smart?}}
\newblock {\emph{\JournalTitle{Journal of Gender Studies}}}
  \textbf{\bibinfo{volume}{29}}, \bibinfo{pages}{832--846},
  \doiprefix\url{10.1080/09589236.2019.1650728} (\bibinfo{year}{2020}).

\bibitem{macedo2020gender}
\bibinfo{author}{Macedo, M.}, \bibinfo{author}{Lotero, L.},
  \bibinfo{author}{Cardillo, A.}, \bibinfo{author}{Barbosa, H.} \&
  \bibinfo{author}{Menezes, R.}
\newblock \bibinfo{title}{Gender patterns of human mobility in {C}olombia:
  Reexamining {R}avenstein’s laws of migration}.
\newblock In \emph{\bibinfo{booktitle}{Complex Networks XI: Proceedings of the
  11th Conference on Complex Networks CompleNet 2020}},
  \bibinfo{pages}{269--281}, \doiprefix\url{10.1007/978-3-030-40943-2\_23}
  (\bibinfo{organization}{Springer Nature}, \bibinfo{year}{2020}).

\bibitem{Stoet2019}
\bibinfo{author}{Stoet, G.} \& \bibinfo{author}{Geary, D.~C.}
\newblock \bibinfo{journal}{\bibinfo{title}{{A simplified approach to measuring
  national gender inequality}}}.
\newblock {\emph{\JournalTitle{PLoS ONE}}} \textbf{\bibinfo{volume}{14}},
  \bibinfo{pages}{1--18}, \doiprefix\url{10.1371/journal.pone.0205349}
  (\bibinfo{year}{2019}).

\bibitem{Kaufman2020}
\bibinfo{author}{Kaufman, G.} \& \bibinfo{author}{Taniguchi, H.}
\newblock \bibinfo{journal}{\bibinfo{title}{{Gender equality and work–family
  conflict from a cross-national perspective}}}.
\newblock {\emph{\JournalTitle{International Journal of Comparative
  Sociology}}} \textbf{\bibinfo{volume}{60}}, \bibinfo{pages}{385--408},
  \doiprefix\url{10.1177/0020715219893750} (\bibinfo{year}{2020}).

\bibitem{Petrongolo2020}
\bibinfo{author}{Petrongolo, B.} \& \bibinfo{author}{Ronchi, M.}
\newblock \bibinfo{journal}{\bibinfo{title}{{Gender gaps and the structure of
  local labor markets}}}.
\newblock {\emph{\JournalTitle{Labour Economics}}}
  \textbf{\bibinfo{volume}{64}}, \bibinfo{pages}{101819},
  \doiprefix\url{10.1016/j.labeco.2020.101819} (\bibinfo{year}{2020}).

\bibitem{lenormand2015influence}
\bibinfo{author}{Lenormand, M.} \emph{et~al.}
\newblock \bibinfo{journal}{\bibinfo{title}{Influence of sociodemographic
  characteristics on human mobility}}.
\newblock {\emph{\JournalTitle{Scientific {R}eports}}}
  \textbf{\bibinfo{volume}{5}}, \bibinfo{pages}{10075},
  \doiprefix\url{10.1038/srep10075} (\bibinfo{year}{2015}).

\bibitem{lenormand2020entropy}
\bibinfo{author}{Lenormand, M.} \emph{et~al.}
\newblock \bibinfo{journal}{\bibinfo{title}{Entropy as a measure of
  attractiveness and socioeconomic complexity in rio de janeiro metropolitan
  area}}.
\newblock {\emph{\JournalTitle{Entropy}}} \textbf{\bibinfo{volume}{22}},
  \bibinfo{pages}{368}, \doiprefix\url{10.3390/e22030368}
  (\bibinfo{year}{2020}).

\bibitem{Pappalardo2015}
\bibinfo{author}{Pappalardo, L.}, \bibinfo{author}{Pedreschi, D.},
  \bibinfo{author}{Smoreda, Z.} \& \bibinfo{author}{Giannotti, F.}
\newblock \bibinfo{journal}{\bibinfo{title}{{Using big data to study the link
  between human mobility and socio-economic development}}}.
\newblock {\emph{\JournalTitle{Proceedings - 2015 IEEE International Conference
  on Big Data, IEEE Big Data 2015}}} \bibinfo{pages}{871--878},
  \doiprefix\url{10.1109/BigData.2015.7363835} (\bibinfo{year}{2015}).

\bibitem{Pappalardo2016}
\bibinfo{author}{Pappalardo, L.} \emph{et~al.}
\newblock \bibinfo{journal}{\bibinfo{title}{{An analytical framework to nowcast
  well-being using mobile phone data}}}.
\newblock {\emph{\JournalTitle{International Journal of Data Science and
  Analytics}}} \textbf{\bibinfo{volume}{2}}, \bibinfo{pages}{75--92},
  \doiprefix\url{10.1007/s41060-016-0013-2} (\bibinfo{year}{2016}).
\newblock \eprint{1606.06279}.

\bibitem{leo2016socioeconomic}
\bibinfo{author}{Leo, Y.}, \bibinfo{author}{Fleury, E.},
  \bibinfo{author}{Alvarez-Hamelin, J.~I.}, \bibinfo{author}{Sarraute, C.} \&
  \bibinfo{author}{Karsai, M.}
\newblock \bibinfo{journal}{\bibinfo{title}{Socioeconomic correlations and
  stratification in social-communication networks}}.
\newblock {\emph{\JournalTitle{Journal of The Royal Society Interface}}}
  \textbf{\bibinfo{volume}{13}}, \bibinfo{pages}{20160598},
  \doiprefix\url{10.1098/rsif.2016.0598} (\bibinfo{year}{2016}).

\bibitem{Gao2019}
\bibinfo{author}{Gao, J.}, \bibinfo{author}{Zhang, Y.~C.} \&
  \bibinfo{author}{Zhou, T.}
\newblock \bibinfo{journal}{\bibinfo{title}{{Computational socioeconomics}}}.
\newblock {\emph{\JournalTitle{Physics Reports}}}
  \textbf{\bibinfo{volume}{817}}, \bibinfo{pages}{1--104},
  \doiprefix\url{10.1016/j.physrep.2019.05.002} (\bibinfo{year}{2019}).

\bibitem{Barbosa2018}
\bibinfo{author}{Barbosa, H.} \emph{et~al.}
\newblock \bibinfo{journal}{\bibinfo{title}{{Human mobility: Models and
  applications}}}.
\newblock {\emph{\JournalTitle{Physics Reports}}}
  \textbf{\bibinfo{volume}{734}}, \bibinfo{pages}{1--74},
  \doiprefix\url{10.1016/j.physrep.2018.01.001} (\bibinfo{year}{2018}).

\bibitem{gustafson2006work}
\bibinfo{author}{Gustafson, P.}
\newblock \bibinfo{journal}{\bibinfo{title}{Work-related travel, gender and
  family obligations}}.
\newblock {\emph{\JournalTitle{Work, employment and society}}}
  \textbf{\bibinfo{volume}{20}}, \bibinfo{pages}{513--530},
  \doiprefix\url{10.1177/0950017006066999} (\bibinfo{year}{2006}).

\bibitem{Rodriguez2020}
\bibinfo{author}{Rodr{\'{i}}guez, L.}, \bibinfo{author}{Palanca, J.},
  \bibinfo{author}{del Val, E.} \& \bibinfo{author}{Rebollo, M.}
\newblock \bibinfo{journal}{\bibinfo{title}{{Analyzing urban mobility paths
  based on users' activity in social networks}}}.
\newblock {\emph{\JournalTitle{Future Generation Computer Systems}}}
  \textbf{\bibinfo{volume}{102}}, \bibinfo{pages}{333--346},
  \doiprefix\url{10.1016/j.future.2019.07.072} (\bibinfo{year}{2020}).

\bibitem{Psylla2017}
\bibinfo{author}{Psylla, I.}, \bibinfo{author}{Sapiezynski, P.},
  \bibinfo{author}{Mones, E.} \& \bibinfo{author}{Lehmann, S.}
\newblock \bibinfo{journal}{\bibinfo{title}{{The role of gender in social
  network organization}}}.
\newblock {\emph{\JournalTitle{PLoS ONE}}} \textbf{\bibinfo{volume}{12}},
  \bibinfo{pages}{1--21}, \doiprefix\url{10.1371/journal.pone.0189873}
  (\bibinfo{year}{2017}).

\bibitem{Farre2020}
\bibinfo{author}{Farr\'e, L.}, \bibinfo{author}{Jofre-Monseny, J.} \&
  \bibinfo{author}{Torrecillas, J.}
\newblock \bibinfo{title}{{Commuting Time and the Gender Gap in Labor Market
  Participation}}.
\newblock \bibinfo{howpublished}{IZA Discussion Papers, 13213. Available at:
  \url{https://www.iza.org/publications/dp/13213/commuting-time-and-the-gender-gap-in-labor-market-participation}}
  (\bibinfo{year}{2020}).

\bibitem{gauvin2020gender}
\bibinfo{author}{Gauvin, L.} \emph{et~al.}
\newblock \bibinfo{journal}{\bibinfo{title}{Gender gaps in urban mobility}}.
\newblock {\emph{\JournalTitle{Humanities and Social Sciences Communications}}}
  \textbf{\bibinfo{volume}{7}}, \bibinfo{pages}{1--13},
  \doiprefix\url{10.1057/s41599-020-0500-x} (\bibinfo{year}{2020}).

\bibitem{gonzalez2008understanding}
\bibinfo{author}{Gonzalez, M.~C.}, \bibinfo{author}{Hidalgo, C.~A.} \&
  \bibinfo{author}{Barab\'asi, A.-L.}
\newblock \bibinfo{journal}{\bibinfo{title}{Understanding individual human
  mobility patterns}}.
\newblock {\emph{\JournalTitle{Nature}}} \textbf{\bibinfo{volume}{453}},
  \bibinfo{pages}{779}, \doiprefix\url{10.1038/nature06958}
  (\bibinfo{year}{2008}).

\bibitem{simini2012universal}
\bibinfo{author}{Simini, F.}, \bibinfo{author}{Gonz{\'a}lez, M.~C.},
  \bibinfo{author}{Maritan, A.} \& \bibinfo{author}{Barab{\'a}si, A.-L.}
\newblock \bibinfo{journal}{\bibinfo{title}{A universal model for mobility and
  migration patterns}}.
\newblock {\emph{\JournalTitle{Nature}}} \textbf{\bibinfo{volume}{484}},
  \bibinfo{pages}{96}, \doiprefix\url{10.1038/nature10856}
  (\bibinfo{year}{2012}).

\bibitem{Asgari2013}
\bibinfo{author}{Asgari, F.}, \bibinfo{author}{Gauthier, V.} \&
  \bibinfo{author}{Becker, M.}
\newblock \bibinfo{journal}{\bibinfo{title}{{A survey on Human Mobility and its
  applications}}}.
\newblock {\emph{\JournalTitle{arXiv}}} \bibinfo{pages}{1307.0814}
  (\bibinfo{year}{2013}).
\newblock \eprint{1307.0814}.

\bibitem{barbosa2015effect}
\bibinfo{author}{Barbosa, H.}, \bibinfo{author}{de~Lima-Neto, F.~B.},
  \bibinfo{author}{Evsukoff, A.} \& \bibinfo{author}{Menezes, R.}
\newblock \bibinfo{journal}{\bibinfo{title}{The effect of recency to human
  mobility}}.
\newblock {\emph{\JournalTitle{EPJ Data Science}}}
  \textbf{\bibinfo{volume}{4}}, \bibinfo{pages}{21},
  \doiprefix\url{epjds/s13688-015-0059-8} (\bibinfo{year}{2015}).

\bibitem{alessandretti2020scales}
\bibinfo{author}{Alessandretti, L.}, \bibinfo{author}{Aslak, U.} \&
  \bibinfo{author}{Lehmann, S.}
\newblock \bibinfo{journal}{\bibinfo{title}{The scales of human mobility}}.
\newblock {\emph{\JournalTitle{Nature}}} \textbf{\bibinfo{volume}{587}},
  \bibinfo{pages}{402--407}, \doiprefix\url{10.1038/s41586-020-2909-1}
  (\bibinfo{year}{2020}).

\bibitem{song2010limits}
\bibinfo{author}{Song, C.}, \bibinfo{author}{Qu, Z.}, \bibinfo{author}{Blumm,
  N.} \& \bibinfo{author}{Barab{\'a}si, A.-L.}
\newblock \bibinfo{journal}{\bibinfo{title}{Limits of predictability in human
  mobility}}.
\newblock {\emph{\JournalTitle{Science}}} \textbf{\bibinfo{volume}{327}},
  \bibinfo{pages}{1018--1021}, \doiprefix\url{10.1126/science.1177170}
  (\bibinfo{year}{2010}).

\bibitem{ballis2020revealing}
\bibinfo{author}{Ballis, H.} \& \bibinfo{author}{Dimitriou, L.}
\newblock \bibinfo{journal}{\bibinfo{title}{Revealing personal activities
  schedules from synthesizing multi-period origin-destination matrices}}.
\newblock {\emph{\JournalTitle{Transportation research part B:
  methodological}}} \textbf{\bibinfo{volume}{139}}, \bibinfo{pages}{224--258},
  \doiprefix\url{10.1016/j.trb.2020.06.007} (\bibinfo{year}{2020}).

\bibitem{Lenormand2020}
\bibinfo{author}{Lenormand, M.}, \bibinfo{author}{Arias, J.~M.},
  \bibinfo{author}{San~Miguel, M.} \& \bibinfo{author}{Ramasco, J.~J.}
\newblock \bibinfo{journal}{\bibinfo{title}{On the importance of trip
  destination for modelling individual human mobility patterns}}.
\newblock {\emph{\JournalTitle{Journal of The Royal Society Interface}}}
  \textbf{\bibinfo{volume}{17}}, \bibinfo{pages}{20200673},
  \doiprefix\url{10.1098/rsif.2020.0673} (\bibinfo{year}{2020}).

\bibitem{montoya2020gender}
\bibinfo{author}{Montoya-Robledo, V.} \emph{et~al.}
\newblock \bibinfo{journal}{\bibinfo{title}{Gender stereotypes affecting active
  mobility of care in {B}ogot{\'a}}}.
\newblock {\emph{\JournalTitle{Transportation Research Part D: Transport and
  Environment}}} \textbf{\bibinfo{volume}{86}}, \bibinfo{pages}{102470},
  \doiprefix\url{10.1016/j.trd.2020.102470} (\bibinfo{year}{2020}).

\bibitem{Chant2013}
\bibinfo{author}{Chant, S.}
\newblock \bibinfo{journal}{\bibinfo{title}{{Cities through a ``gender lens'':
  A golden ``urban age'' for women in the global South?}}}
\newblock {\emph{\JournalTitle{Environment and Urbanization}}}
  \textbf{\bibinfo{volume}{25}}, \bibinfo{pages}{9--29},
  \doiprefix\url{10.1177/0956247813477809} (\bibinfo{year}{2013}).

\bibitem{carlsson1999travel}
\bibinfo{author}{Carlsson-Kanyama, A.} \& \bibinfo{author}{Linden, A.-L.}
\newblock \bibinfo{journal}{\bibinfo{title}{Travel patterns and environmental
  effects now and in the future:: implications of differences in energy
  consumption among socio-economic groups}}.
\newblock {\emph{\JournalTitle{Ecological Economics}}}
  \textbf{\bibinfo{volume}{30}}, \bibinfo{pages}{405--417},
  \doiprefix\url{10.1016/S0921-8009(99)00006-3} (\bibinfo{year}{1999}).

\bibitem{carra2016modelling}
\bibinfo{author}{Carra, G.}, \bibinfo{author}{Mulalic, I.},
  \bibinfo{author}{Fosgerau, M.} \& \bibinfo{author}{Barthelemy, M.}
\newblock \bibinfo{journal}{\bibinfo{title}{Modelling the relation between
  income and commuting distance}}.
\newblock {\emph{\JournalTitle{Journal of the Royal Society Interface}}}
  \textbf{\bibinfo{volume}{13}}, \doiprefix\url{10.1098/rsif.2016.0306}
  (\bibinfo{year}{2016}).

\bibitem{alonso1964location}
\bibinfo{author}{Alonso, W.}
\newblock \bibinfo{journal}{\bibinfo{title}{Location and land use. toward a
  general theory of land rent.}}
\newblock {\emph{\JournalTitle{Harvard University Press}}}
  (\bibinfo{year}{1964}).

\bibitem{frias2012relation}
\bibinfo{author}{Frias-Martinez, V.}, \bibinfo{author}{Virseda-Jerez, J.} \&
  \bibinfo{author}{Frias-Martinez, E.}
\newblock \bibinfo{journal}{\bibinfo{title}{On the relation between
  socio-economic status and physical mobility}}.
\newblock {\emph{\JournalTitle{Information Technology for Development}}}
  \textbf{\bibinfo{volume}{18}}, \bibinfo{pages}{91--106},
  \doiprefix\url{10.1080/02681102.2011.630312} (\bibinfo{year}{2012}).

\bibitem{de2005urban}
\bibinfo{author}{De~Vasconcellos, E.~A.}
\newblock \bibinfo{journal}{\bibinfo{title}{Urban change, mobility and
  transport in {S}{\~a}o {P}aulo: three decades, three cities}}.
\newblock {\emph{\JournalTitle{Transport Policy}}}
  \textbf{\bibinfo{volume}{12}}, \bibinfo{pages}{91--104},
  \doiprefix\url{10.1016/j.tranpol.2004.12.001} (\bibinfo{year}{2005}).

\bibitem{stier2014occupational}
\bibinfo{author}{Stier, H.} \& \bibinfo{author}{Yaish, M.}
\newblock \bibinfo{journal}{\bibinfo{title}{Occupational segregation and gender
  inequality in job quality: a multi-level approach}}.
\newblock {\emph{\JournalTitle{Work, employment and society}}}
  \textbf{\bibinfo{volume}{28}}, \bibinfo{pages}{225--246},
  \doiprefix\url{10.1177/0950017013510758} (\bibinfo{year}{2014}).

\bibitem{clarke2020gender}
\bibinfo{author}{Clarke, H.~M.}
\newblock \bibinfo{title}{Gender stereotypes and gender-typed work}.
\newblock In \emph{\bibinfo{booktitle}{Handbook of Labor, Human Resources and
  Population Economics}}, \bibinfo{pages}{1--23},
  \doiprefix\url{10.1007/978-3-319-57365-6\_21-1}
  (\bibinfo{publisher}{Springer}, \bibinfo{year}{2020}).

\bibitem{Alessandretti2018}
\bibinfo{author}{Alessandretti, L.}, \bibinfo{author}{Sapiezynski, P.},
  \bibinfo{author}{Sekara, V.}, \bibinfo{author}{Lehmann, S.} \&
  \bibinfo{author}{Baronchelli, A.}
\newblock \bibinfo{journal}{\bibinfo{title}{{Evidence for a conserved quantity
  in human mobility}}}.
\newblock {\emph{\JournalTitle{Nature Human Behaviour}}}
  \textbf{\bibinfo{volume}{2}}, \bibinfo{pages}{485--491},
  \doiprefix\url{10.1038/s41562-018-0364-x} (\bibinfo{year}{2018}).

\bibitem{Martin2019}
\bibinfo{author}{Martin, F.} \& \bibinfo{author}{Dragojlovic, A.}
\newblock \bibinfo{journal}{\bibinfo{title}{{Gender, Mobility Regimes, and
  Social Transformation in Asia}}}.
\newblock {\emph{\JournalTitle{Journal of Intercultural Studies}}}
  \textbf{\bibinfo{volume}{40}}, \bibinfo{pages}{275--286},
  \doiprefix\url{10.1080/07256868.2019.1599166} (\bibinfo{year}{2019}).

\bibitem{levin20199}
\bibinfo{author}{Levin, L.} \& \bibinfo{author}{Thoresson, K.}
\newblock \bibinfo{journal}{\bibinfo{title}{Gender equality and ``smart''
  mobility}}.
\newblock {\emph{\JournalTitle{Gendering Smart Mobilities}}}
  \bibinfo{pages}{143--161}, \doiprefix\url{10.4324/9780429466601-9}
  (\bibinfo{year}{2019}).

\bibitem{datosabiertos}
\bibinfo{title}{Datos abiertos del area metropolitana de valle de aburr\'a}.
\newblock \bibinfo{howpublished}{Available at:
  \url{https://datosabiertos.metropol.gov.co/search/field_topic/movilidad-y-transporte-2/type/dataset?sort_by=changed}}
  (\bibinfo{year}{Accessed: 2021-02-02}).

\bibitem{secretariadistritalmovilidad}
\bibinfo{title}{Secretaría distrital de movilidad}.
\newblock \bibinfo{howpublished}{Available at: \url{https://www.simur.gov.co/}}
  (\bibinfo{year}{Accessed: 2021-02-02}).

\bibitem{datasaopaulo}
\bibinfo{title}{Metr{\^o} {S}{\~a}o {P}aulo | pesquisa origem e destino: 50
  anos.}
\newblock \bibinfo{howpublished}{Available at:
  \url{http://www.metro.sp.gov.br/pesquisa-od/pesquisa-od-50-anos.aspx}}
  (\bibinfo{year}{Accessed: 2021-02-02}).

\bibitem{aburra_copyright}
\bibinfo{author}{Área Metropolitana del Valle~de Aburrá, D.~A.}
\newblock \bibinfo{title}{Políticas de uso de la información de datos
  abiertos}.
\newblock \bibinfo{howpublished}{Available at:
  \url{https://datosabiertos.metropol.gov.co/pol\%C3\%ADticas-de-uso-de-la-informaci\%C3\%B3n-de-datos-abiertos}}
  (\bibinfo{year}{2021}).

\bibitem{bogota_copyright}
\bibinfo{author}{Bogotá, D.~A.}
\newblock \bibinfo{title}{Acerca de datos abiertos bogotá}.
\newblock \bibinfo{howpublished}{Available at:
  \url{https://datosabiertos.bogota.gov.co/about}} (\bibinfo{year}{2021}).

\bibitem{saopaulo_copyright}
\bibinfo{author}{SP, P. G.~A.}
\newblock \bibinfo{title}{Pesquisa origem e destino}.
\newblock \bibinfo{howpublished}{Available at:
  \url{http://catalogo.governoaberto.sp.gov.br/dataset/869-pesquisa-origem-e-destino}}
  (\bibinfo{year}{2021}).

\bibitem{basemap_copyright}
\bibinfo{author}{QGIS}.
\newblock \bibinfo{title}{Copyright of the dark matter basemap}.
\newblock \bibinfo{howpublished}{Available at:
  \url{https://qms.nextgis.com/geoservices/529/}} (\bibinfo{year}{2021}).

\end{thebibliography}


\begin{thebibliography}{1}

\bibitem{Barbosa2018}
Hugo Barbosa, Marc Barthelemy, Gourab Ghoshal, Charlotte~R. James, Maxime
  Lenormand, Thomas Louail, Ronaldo Menezes, Jos{\'{e}}~J. Ramasco, Filippo
  Simini, and Marcello Tomasini.
\newblock {Human mobility: Models and applications}.
\newblock {\em Physics Reports}, 734:1--74, 2018.

\bibitem{brown2005new}
Angus~M Brown.
\newblock A new software for carrying out one-way {ANOVA} post hoc tests.
\newblock {\em Computer methods and programs in biomedicine}, 79(1):89--95,
  2005.

\bibitem{de2010impact}
Jos{\'e} Roberto~Mendon{\c{c}}a de~Barros.
\newblock The impact of the international financial crisis on {B}razil ({ARI}).
\newblock {\em International Cooperation \& Development}, 2010.

\bibitem{delacre2017psychologists}
Marie Delacre, Daniel Lakens, and Christophe Leys.
\newblock Why psychologists should by default use {W}elch’s $t$-test instead
  of {S}tudent’s $t$-test.
\newblock {\em International Review of Social Psychology}, 30(1):92--101, 2017.

\bibitem{ferrari2011brazil}
Fernando Ferrari~Filho.
\newblock Brazil's response: how did financial regulation and monetary policy
  influence recovery?
\newblock {\em Brazilian Journal of Political Economy}, 31(5):880--888, 2011.

\bibitem{levy2013sampling}
Paul~S Levy and Stanley Lemeshow.
\newblock {\em Sampling of populations: methods and applications}.
\newblock John Wiley \& Sons, 2013.

\bibitem{tabachnick2007experimental}
Barbara~G Tabachnick and Linda~S Fidell.
\newblock {\em Experimental designs using {ANOVA}}.
\newblock Thomson/Brooks/Cole, Belmont (CA), USA, 2007.

\end{thebibliography}


		\begin{figure}[htbp]
			\centering
                        \includegraphics[width=0.85\textwidth]{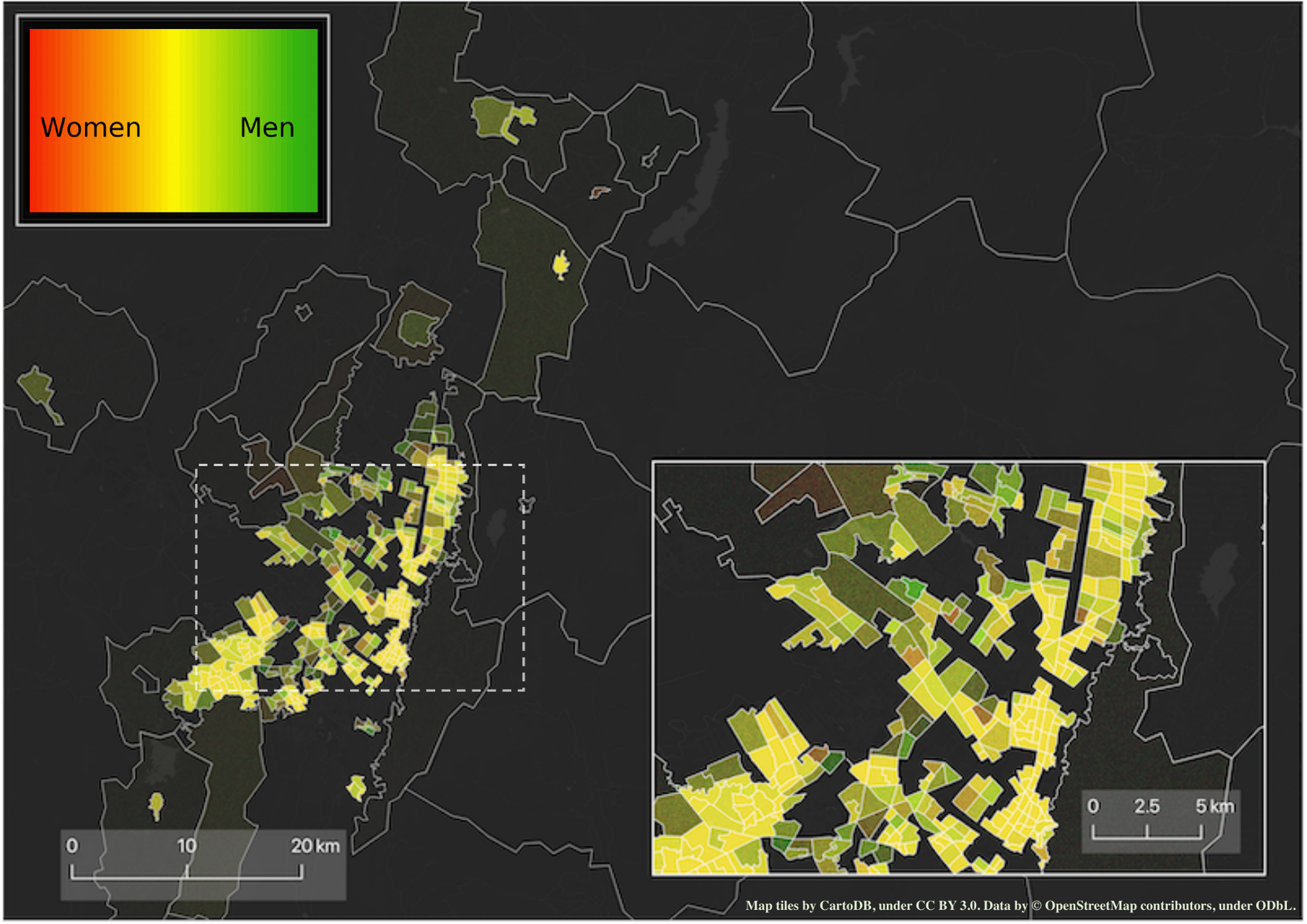}
			\caption{\textbf{Density map of \work travels made in \bgt during the year 2019. Brighter colours represent a higher density of travels to work.} The hue denotes whether for a given zone the majority of travels were made by women (red), men (green), or by both (yellow). The inset portrays a zoom of the city centre. Figure contains information from OpenStreetMap and OpenStreetMap Foundation, which is made available under the Open Database License.
			}
			\label{fig:mapgender}
		\end{figure}

	\begin{figure}[htbp]
		\centering
                \includegraphics[width=0.85\textwidth]{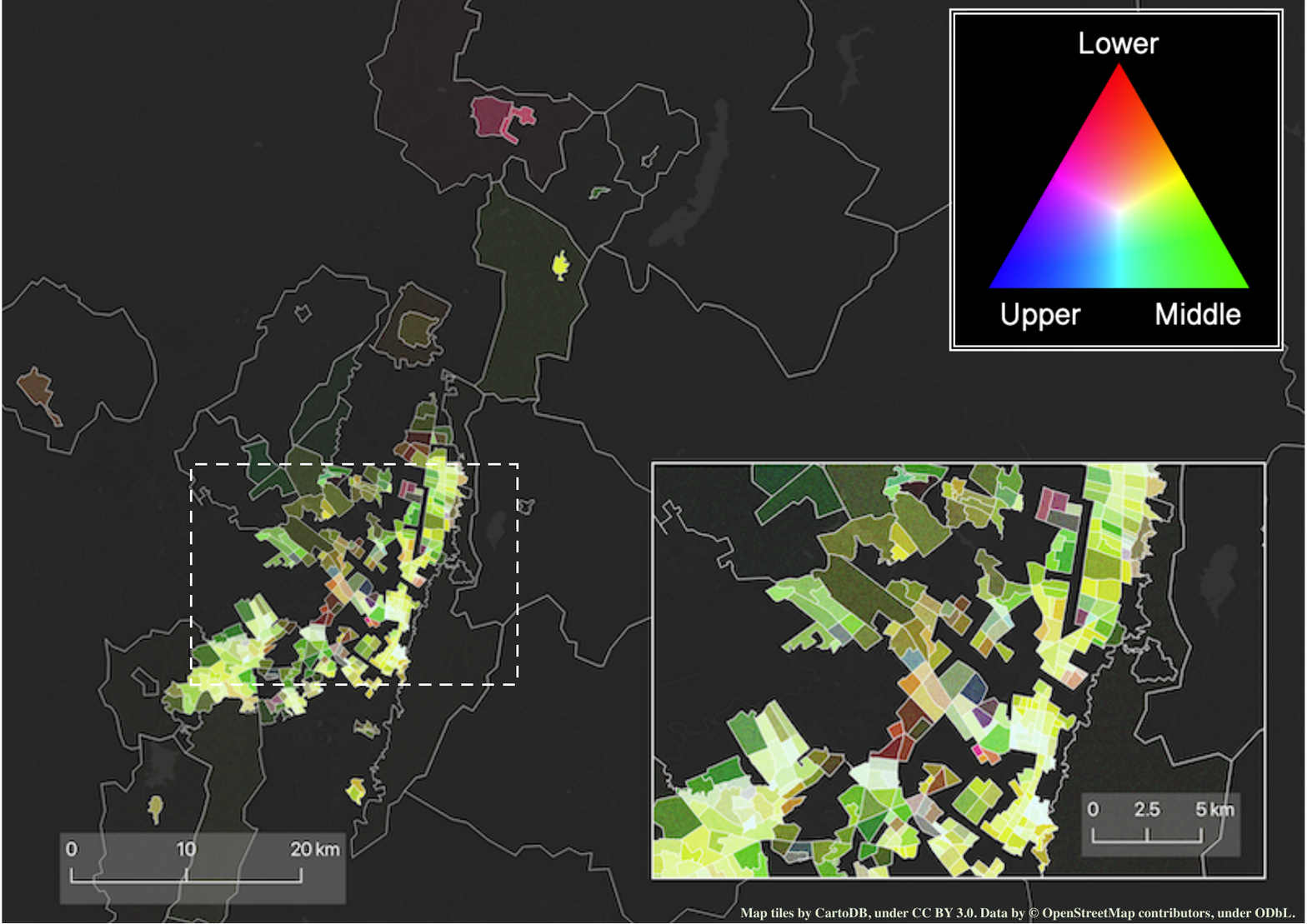}
		\caption{\textbf{Density map of \work travels made in \bgt during the year 2019.} Brighter colours represent a higher density of travels to work. The hue denotes whether for a given zone the majority of travels were made by travellers belonging to the \lowerc (red), \middlec (green), \upperc (blue) or all three socioeconomic status. The inset portrays a zoom of the city centre. Figure contains information from OpenStreetMap and OpenStreetMap Foundation, which is made available under the Open Database License.}
		\label{fig:mapsocio}
	\end{figure}

		\begin{figure}[htbp]
		\centering
		\includegraphics[width=0.95\textwidth]{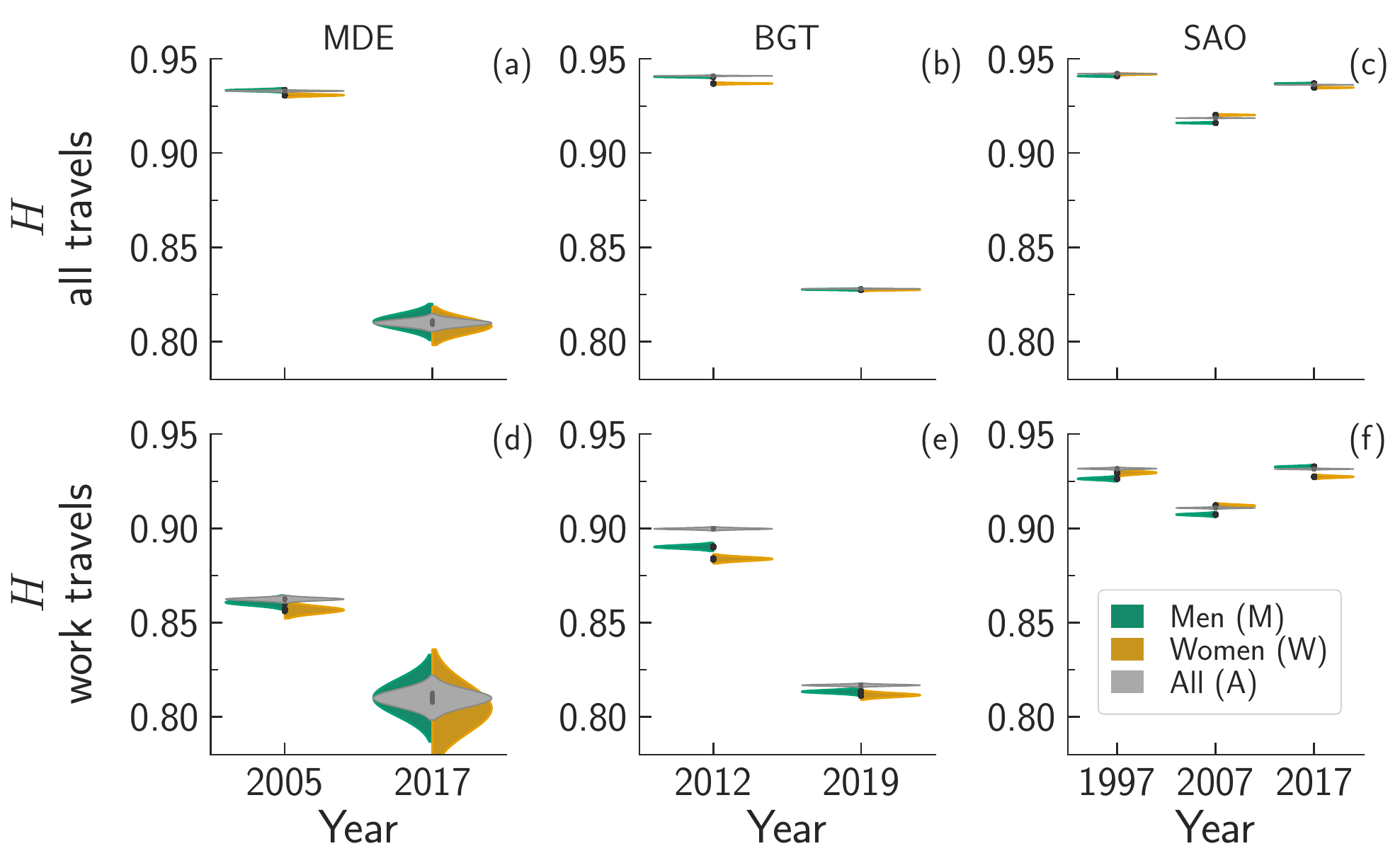}
		\caption{\textbf{Violin plots of the bootstrapped mobility diversity, $H$, for \all travels (top row, panels a-c), and \work travels (bottom row, panels d-f).} Each column refers to a different region: \mde (panels a and d), \bgt (panels b and e), and \sao (panels c and f). For each region, we display the distribution of the values of $H$ in each year. We show the distributions for \all travels duplicated (entire violins), and the distributions for \men and \women travels in half-violins.}
		\label{fig:genderregions}
	\end{figure}

%
%

\makeatletter

\makeatother

\end{document}


	
	\setlength{\abovedisplayskip}{10pt}
	\setlength{\belowdisplayskip}{10pt}
	
	\maketitle

	\section{The mobility surveys}
	\label{sec:datasummary}

	As described in the main manuscript, in this work, we analyse the data collected from travel surveys carried out in three large South American urban areas: two in Colombia and one in Brazil. The Colombian datasets correspond to the metropolitan area surrounding the city of Medell\'in (henceforth indicated as \mde), and the metropolitan area of Bogot\'a (\bgt). The Brazilian dataset corresponds to the mobility taking place in the metropolitan area of S\~ao Paulo (\sao). For each area, we analysed the data collected in different years: \{2005, 2017\} for \mde, \{2012, 2019\} for \bgt, and \{1997, 2007, 2017\} for \sao, respectively.

	The surveys make use of  detailed questionnaires in which respondents were asked to answer objective questions regarding the recurrent mobility habits (\eg  number of daily trips, their purposes, and transport modes) of the different members of the household. For each individual trip, regardless of its purpose (\eg study, work or leisure) the survey captured: the origin and destination zones, departure and arrival times, and transportation modes. Additionally, respondents also answered basic questions regarding the socioeconomic and demographic characteristics of the members of the household such as gender, age, and income. Further details on the socioeconomic classification adopted in our study are available in \ref{sec:sociobrazil} Section.
	
	Furthermore, each entry in the data is associated with an \emph{expansion factor}, a sample weighting factor  that accounts for the representativeness of a respondent's answers relative to the general population~\cite{levy2013sampling}, benchmarked by other  socioeconomic and demographic characteristics. Finally, the spatial partitioning of the regions and the sampling criteria utilised on the surveys were determined by the respective census authorities of each country, in line with the best statistical methods for sampled population surveys. In our data collection, one exception is the \mde{} 2017 survey, whose data were provided without their associated \emph{expansion factors}. This means that our analyses on that specific dataset are based exclusively on the sample data.  In \ref{sec:data_charac} Section, we explore the data sample sizes using the expansion factor for each group, city, and year.
	
	\subsection{Harmonising the socioeconomic classification across years and cities}
	\label{sec:sociobrazil}
	
	Despite the fact that Brazil and Colombia are both developing countries from South America,  the socioeconomic characteristics of the three cities and their populations are different. Furthermore, at the time scale of the travel surveys, there are significant economic changes even at a city level. Thus, one important step in our analyses is harmonising the socioeconomic classification across years, cities, and countries.
	
	For the Colombian datasets, the socioeconomic classification of the respondents is kept consistent across years and cities. Households are split into six strata, and this classification has been widely used as a proxy of the socioeconomic status of individuals, with stratum $1$ corresponding to people with the lowest income, and stratum $6$ corresponding to people with the highest income, instead. The mapping between the aforementioned strata and our partition is: \lowerc (strata $1$ and $2$), \middlec (strata $3$ and $4$), and \upperc (strata $5$ and $6$), respectively.
	
	In our data for the city of S\~ao Paulo, however,  the socioeconomic classification of the population is based on the methodological standards adopted by the Brazilian census authority and their socio-demographic research institute at the time of the survey. Not surprisingly, the classification methodology changes over time to better capture the current picture of the socioeconomic characteristics of the population. 
	More precisely, for the 1997 data, respondents were classified into five socio-economic classes labelled as $\textit{A}$ (upper), $\textit{B}$, (mid-upper), $\textit{C}$ (middle), $\textit{D}$ (mid-lower), and $\textit{E}$ (lower). It is noteworthy that this division takes into account not only overall incomes but other characteristics such as standard of living, purchase power, housing conditions, and access to amenities and transport infrastructure. More recently, Brazilian institutes such as IBGE (Brazilian geography and statistics institute) and ABEP (Brazilian association for population studies) adopted sub-divisions of these major groups to provide a more precise picture of the population's realities in terms of their socioeconomic statuses. The 2007 and 2017 S\~ao Paulo's travel survey data also utilised these subdivisions. The division we adopted in terms of our partition is presented in \ref{tab:ses_mapping}~Table.
	
	%
	%
	%
	\begin{table}[H]
		\caption{\textbf{Mapping of the Brazilian classification scheme into the \lowerc, \middlec{}, and \upperc{} socioeconomic classes (SES) for the three years of the survey.}}
		\label{tab:ses_mapping}
		\centering
		\resizebox{0.45\linewidth}{!}{%
			\begin{tabular}{l c c c}
				\toprule
				& \multicolumn{3}{c}{Year} \\\cline{2-4}
				SES & 1997 & 2007 & 2017 \\\midrule
				\lowerc & $\textit{D}, \textit{E}$ & $\textit{C2}, \textit{D}, \textit{E}$ & $\textit{C2}, \textit{D}, \textit{E}$ \\
				\middlec{} & $\textit{B}, \textit{C}$ & $\textit{B1}, \textit{B2}, \textit{C1}$ & $\textit{B1}, \textit{B2}, \textit{C1}$ \\
				\upperc{} & $\textit{A}$ & $\textit{A1}, \textit{A2}$ & $\textit{A}$ \\
				\bottomrule
			\end{tabular}%
		}
	\end{table}

	\subsection{Data characterisation} 
	\label{sec:data_charac}
	
	In this section, we provide a general overview of our datasets and their compositions in terms of their numbers of underlying populations and their travels. We also provide their partition across the socioeconomic and gender dimensions. \ref{tab:alltravelsexp}~Table displays the composition of the complete datasets, whereas \ref{tab:worktravelsexp}~Table reports the same quantities for the subsets containing the \work travels only. In both tables, quantities denoted with the symbol $N$ represent counts, while quantities denoted with the symbol $f$ represent fractions corresponding to distinct groups. Furthermore, in our notation, the superscript text refers to the group, and a subscript $T$ is used whenever we refer to the travels. Notice that these fractions are computed using the expanded data, meaning that they are not relative to our sample sizes but rather to how many people/travels they represent. 
	
	%
	%
	%
	\begin{table}[H]
		\caption{\textbf{Summary of the composition of all the expanded data sets for travels made for \all purposes.} For a given location and year, we report: the number of travellers $N_P$, the number of travels $N_T$, the fraction of men (women) travellers $f^M$ ($f^W$), and the fraction of travels made by men (women) $f_T^M$ ($f_T^W$). We report also the fraction of travellers belonging to the \lowerc ($f^{\text{\lowerc}}$), \middlec{} ($f^{\text{\middlec{}}}$), and \upperc{} ($f^{\text{\upperc{}}}$) socioeconomic classes, and the same quantities discriminated by gender (\eg{} $f^{\text{\lowerc} W}$). Finally, we report the fraction of travels made by travellers with a given socioeconomic class and gender (\eg{} $f_T^{\text{\lowerc} W}$). The data sets are obtained applying the expansion factors to the raw data from the surveys.}
		\label{tab:alltravelsexp}
		\resizebox{\linewidth}{!}{
			\begin{tabular}{l|rr|rr|rrr}
				\toprule
				Location & \multicolumn{2}{c|}{Medell\'in (\mde)} & \multicolumn{2}{c|}{Bogot\'a (\bgt)} & \multicolumn{3}{c}{S\~ao Paulo (\sao)} \\
				\midrule
				Year & \multicolumn{1}{r}{2005} & \multicolumn{1}{r|}{2017} & \multicolumn{1}{r}{2012} & \multicolumn{1}{r|}{2019} & \multicolumn{1}{r}{1997} & \multicolumn{1}{r}{2007} & \multicolumn{1}{r}{2017} \\
				\midrule
				$N_P$ & 22,702 & 38,048 & 11,672 & 47,149 & 37,316 & 51,103 & 48,085 \\
				$N_T$ & 7,102,052 & 123,449 & 25,628,970 & 88,620,670 & 54,939,650 & 83,313,240 & 95,948,930 \\
				\midrule
				$f^M$ & 0.52 & 0.51 & 0.46 & 0.48 & 0.52 & 0.49 & 0.50 \\
				$f^W$ & 0.48 & 0.49 & 0.54 & 0.52 & 0.48 & 0.51 & 0.50 \\
				\midrule
				$f^M_T$ & 0.52 & 0.51 & 0.42 & 0.47 & 0.51 & 0.49 & 0.50 \\
				$f^W_T$ & 0.48 & 0.49 & 0.58 & 0.53 & 0.49 & 0.51 & 0.50 \\
				\midrule
				$f^{\text{\lowerc}}$ & 0.50 & 0.55 & 0.46 & 0.49 & 0.30 & 0.21 & 0.20 \\
				$f^{\text{\middlec{}}}$ & 0.46 & 0.38 & 0.48 & 0.46 & 0.63 & 0.63 & 0.65 \\
				$f^{\text{\upperc{}}}$ & 0.04 & 0.07 & 0.06 & 0.05 & 0.07 & 0.16 & 0.15 \\
				\midrule
				$f^{\text{\lowerc}}_T$ & 0.41 & 0.54 & 0.52 & 0.50 & 0.26 & 0.24 & 0.22 \\
				$f^{\text{\middlec{}}}_T$ & 0.52 & 0.38 & 0.43 & 0.45 & 0.68 & 0.66 & 0.68 \\
				$f^{\text{\upperc{}}}_T$ & 0.07 & 0.08 & 0.05 & 0.05 & 0.06 & 0.10 & 0.10 \\
				\midrule
				$f^{\text{\lowerc}\, M}$ & 0.26 & 0.28 & 0.21 & 0.24 & 0.16 & 0.10 & 0.09 \\
				$f^{\text{\middlec{}}\, M}$ & 0.23 & 0.19 & 0.22 & 0.22 & 0.32 & 0.31 & 0.32 \\
				$f^{\text{\upperc{}}\, M}$ & 0.02 & 0.04 & 0.03 & 0.02 & 0.04 & 0.08 & 0.08 \\
				$f^{\text{\lowerc}\, W}$ & 0.24 & 0.26 & 0.25 & 0.25 & 0.14 & 0.11 & 0.10 \\
				$f^{\text{\middlec{}}\, W}$ & 0.23 & 0.19 & 0.26 & 0.24 & 0.30 & 0.32 & 0.33 \\
				$f^{\text{\upperc{}}\, W}$ & 0.02 & 0.04 & 0.03 & 0.03 & 0.04 & 0.08 & 0.08 \\
				\midrule
				$f^{\text{\lowerc}\, M}_T$ & 0.22 & 0.28 & 0.21 & 0.23 & 0.13 & 0.11 & 0.10 \\
				$f^{\text{\middlec{}}\, M}_T$ & 0.26 & 0.19 & 0.19 & 0.21 & 0.35 & 0.39 & 0.35 \\
				$f^{\text{\upperc{}}\, M}_T$ & 0.04 & 0.04 & 0.02 & 0.02 & 0.03 & 0.05 & 0.05 \\
				$f^{\text{\lowerc}\, W}_T$ & 0.19 & 0.26 & 0.32 & 0.27 & 0.13 & 0.13 & 0.12 \\
				$f^{\text{\middlec{}}\, W}_T$ & 0.26 & 0.19 & 0.24 & 0.23 & 0.33 & 0.32 & 0.35 \\
				$f^{\text{\upperc{}}\, W}_T$ & 0.03 & 0.04 & 0.03 & 0.03 & 0.03 & 0.05 & 0.05 \\
				\bottomrule
			\end{tabular}
		}
	\end{table}
	%
	
	%
	%
	%
	\begin{table}[H]
		\caption{\textbf{Summary of the composition of all the expanded data sets for travels made only for \work purpose.} See the caption of \ref{tab:alltravelsexp}~Table for the description of each row.}
		\label{tab:worktravelsexp}
		\resizebox{\linewidth}{!}{
			\begin{tabular}{l|rr|rr|rrr}
				\toprule
				Location & \multicolumn{2}{c|}{Medell\'in (\mde)} & \multicolumn{2}{c|}{Bogot\'a (\bgt)} & \multicolumn{3}{c}{S\~ao Paulo (\sao)} \\
				\midrule
				Year & \multicolumn{1}{r}{2005} & \multicolumn{1}{r|}{2017} & \multicolumn{1}{r}{2012} & \multicolumn{1}{r|}{2019} & \multicolumn{1}{r}{1997} & \multicolumn{1}{r}{2007} & \multicolumn{1}{r}{2017} \\
				\midrule
				$N_P$ & 9,081  & 17,466  & 6,844  & 20,208  & 17,806  & 29,640 & 25,333 \\
				$N_T$ & 349,963 & 18,814 & 1,437,599 & 3,916,047 &  5,939,612 &  9,038,745 &  10,363,550 \\
				\midrule
				$f^M$ & 0.61 &    0.62 & 0.55 & 0.56 &  0.62 &  0.55 &  0.55 \\
				$f^W$ & 0.39 &    0.38 & 0.45 & 0.44 &  0.38 &  0.45 &  0.45 \\
				\midrule
				$f^M_T$ & 0.63 &    0.63 & 0.58 & 0.58 &  0.68 &  0.61 &  0.59 \\
				$f^W_T$ & 0.37 &    0.37 & 0.42 & 0.42 &  0.32 &  0.39 &  0.41 \\
				\midrule
				$f^{\text{\lowerc}}$ & 0.50 & 0.54 & 0.47 & 0.48 & 0.30 & 0.19 & 0.17 \\
				$f^{\text{\middlec}}$ & 0.46 & 0.38 & 0.46 & 0.46 & 0.63 & 0.64 & 0.66\\
				$f^{\text{\upperc}}$ & 0.04 & 0.08 & 0.07 & 0.06 & 0.07 & 0.17 & 0.17 \\
				\midrule
				$f^{\text{\lowerc}}_T$ & 0.39 & 0.53 & 0.45 & 0.47 & 0.25 & 0.22 & 0.19 \\
				$f^{\text{\middlec}}_T$ & 0.52 & 0.39 & 0.48 & 0.47 & 0.69 & 0.68 & 0.70 \\
				$f^{\text{\upperc{}}}_T$ & 0.09 & 0.08 & 0.07 & 0.06 & 0.06 & 0.10 & 0.11 \\
				\midrule
				$f^{\text{\lowerc}\, M}$ & 0.32 & 0.35 & 0.27 & 0.28 & 0.19 & 0.10 & 0.09 \\
				$f^{\text{\middlec{}}\, M}$ & 0.27 & 0.23 & 0.25 & 0.25 & 0.39 & 0.35 & 0.36 \\
				$f^{\text{\upperc{}}\, M}$ & 0.02 & 0.05 & 0.04 & 0.03 & 0.04 & 0.10 & 0.10 \\
				$f^{\text{\lowerc}\, W}$ & 0.18 & 0.19 & 0.20 & 0.20 & 0.11 & 0.08 & 0.07 \\
				$f^{\text{\middlec{}}\, W}$ & 0.19 & 0.15 & 0.21 & 0.21 & 0.25 & 0.29 & 0.30 \\
				$f^{\text{\upperc{}}\, W}$ & 0.02 & 0.03 & 0.03 & 0.03 & 0.02 & 0.08 & 0.08 \\
				\midrule
				$f^{\text{\lowerc}\, M}_T$ & 0.26 & 0.34 & 0.26 & 0.28 & 0.17 & 0.13 & 0.11 \\
				$f^{\text{\middlec{}}\, M}_T$ & 0.31 & 0.24 & 0.28 & 0.26 & 0.47 & 0.42 & 0.42 \\
				$f^{\text{\upperc{}}\, M}_T$ & 0.06 & 0.05 & 0.04 & 0.04& 0.04 & 0.06 & 0.06 \\
				$f^{\text{\lowerc}\, W}_T$ & 0.13 & 0.18 & 0.19 & 0.19 & 0.08 & 0.09 & 0.08 \\
				$f^{\text{\middlec{}}\, W}_T$ & 0.21 & 0.15 & 0.20 & 0.21 & 0.22 & 0.26 & 0.29 \\
				$f^{\text{\upperc{}}\, W}_T$ & 0.03 & 0.04 & 0.03 & 0.02 & 0.02 & 0.04 & 0.04 \\
				\bottomrule
			\end{tabular}
		}
	\end{table}

	\section{Spatial distribution of travels and their population compositions}\label{sec:maps}
	
	Here, we provide some additional visual insights to the underlying composition of the travellers by means of density maps. In a density map, thousands of points of different colours are scattered within each area. The number of points is proportional to a measure of interest, whereas their colours encode the groups they belong to. Such an encoding means that denser areas will appear brighter in the map, while the group composition will be reflected on the colour of the area. In our case, the number of points in an area is proportional to the number of \work travels having each area (or zone) as their destination, whereas the colours correspond to either the gender or the socioeconomic groups to which the travellers belong. It is noteworthy that density maps are not intended to provide an accurate, quantitative representation of the population compositions but, rather, to give an overall perspective on the spatial distribution of the trips in terms of their density, mixing, and segregation. For brevity, here we show the visualisations only for the most recent data for each city.

	\subsection{Gender composition}

\begin{figure}[H]
		\centering
		\includegraphics[width=0.75\textwidth]{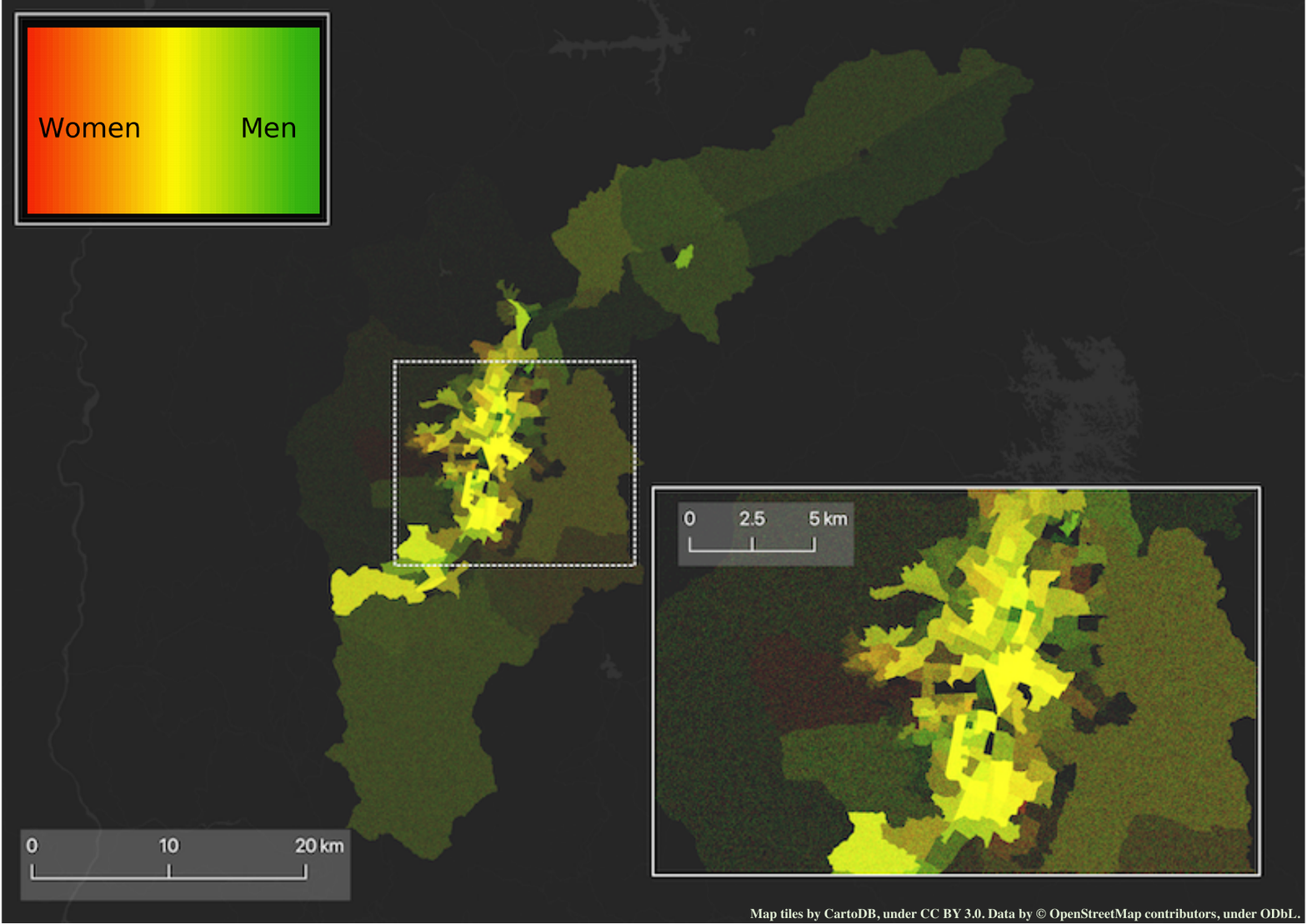}
		\caption{\textbf{Density map of \work travels made in \mde during the year 2017.} Brighter colours represent a higher density of travels to work. The hue denotes whether for a given zone the majority of travels were made by women (red), men (green), or by both (yellow). The inset portrays a zoom of the city centre. Figure contains information from OpenStreetMap and OpenStreetMap Foundation, which is made available under the Open Database License.
		}
		\label{fig:mdegender}
	\end{figure}

\begin{figure}[H]
		\centering
		\includegraphics[width=0.75\textwidth]{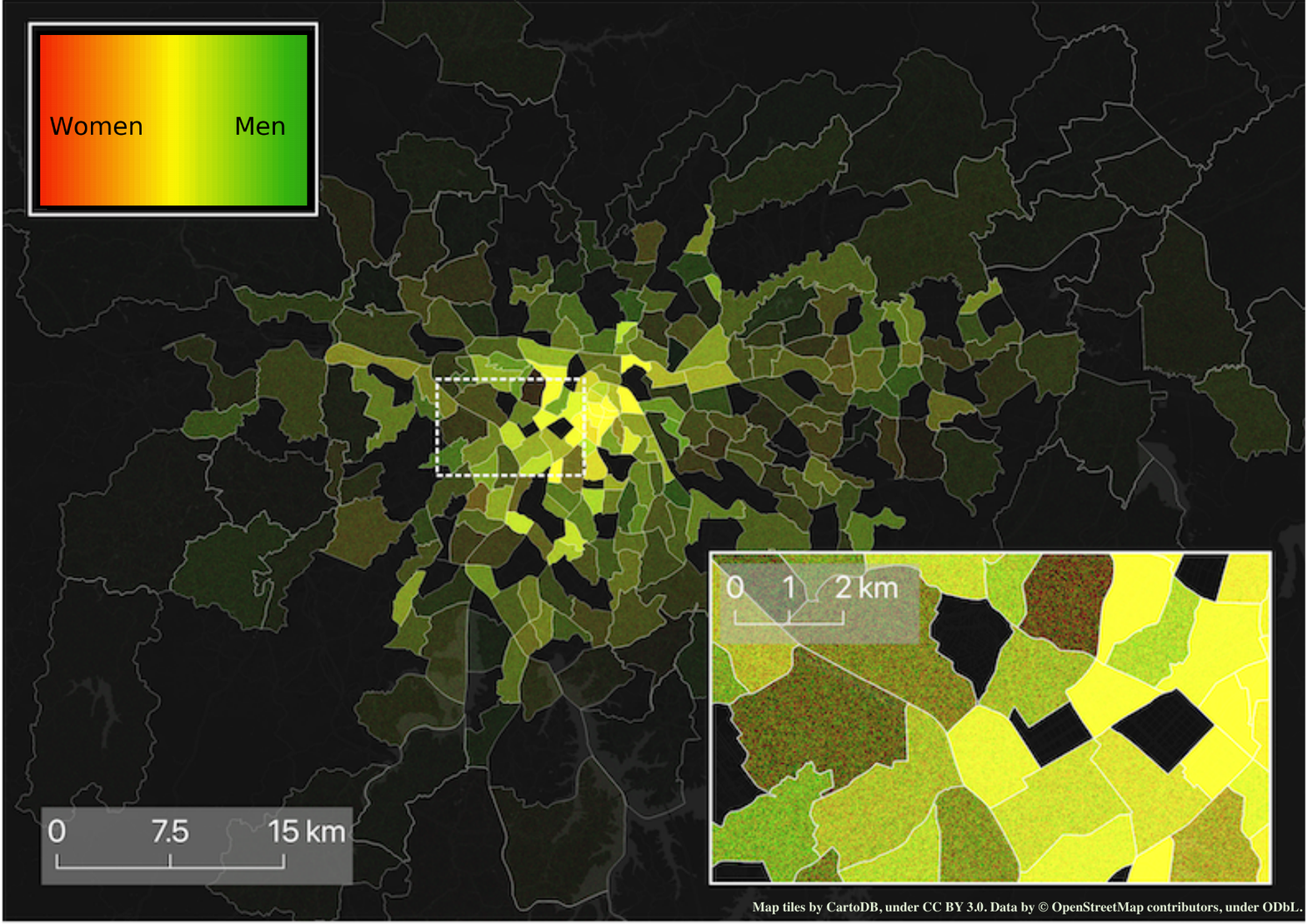}
		\caption{\textbf{Density map of \work travels made in \sao during the year 2017.} Brighter colours represent a higher density of travels to work. The hue denotes whether for a given zone the majority of travels were made by women (red), men (green), or by both (yellow). The inset portrays a zoom of the city centre. Figure contains information from OpenStreetMap and OpenStreetMap Foundation, which is made available under the Open Database License.
		}
		\label{fig:saogender}
	\end{figure}

	We looked at the gender composition of the work-related travels for the cities of \mde and \sao (\ref{fig:mdegender} and \ref{fig:saogender} Figs) respectively. First, in both cities, it is evident the presence of a larger concentration of travels in their central areas.  Furthermore, we can see also that in the centre of the cities, the work travels are more gender-balanced, hence the predominance of brighter white zones. However, some less dense areas exhibit small fluctuations in their gender balances, with a slight prevalence of areas coloured in green. Therefore, it is evident that the origins of the significant differences in the number of work travels made by men and women reported in \ref{tab:worktravelsexp}~Table come from the less dense areas of the cities.


	\subsection{Socioeconomic composition}
	
\begin{figure}[H]
		\centering
		\includegraphics[width=0.8\textwidth]{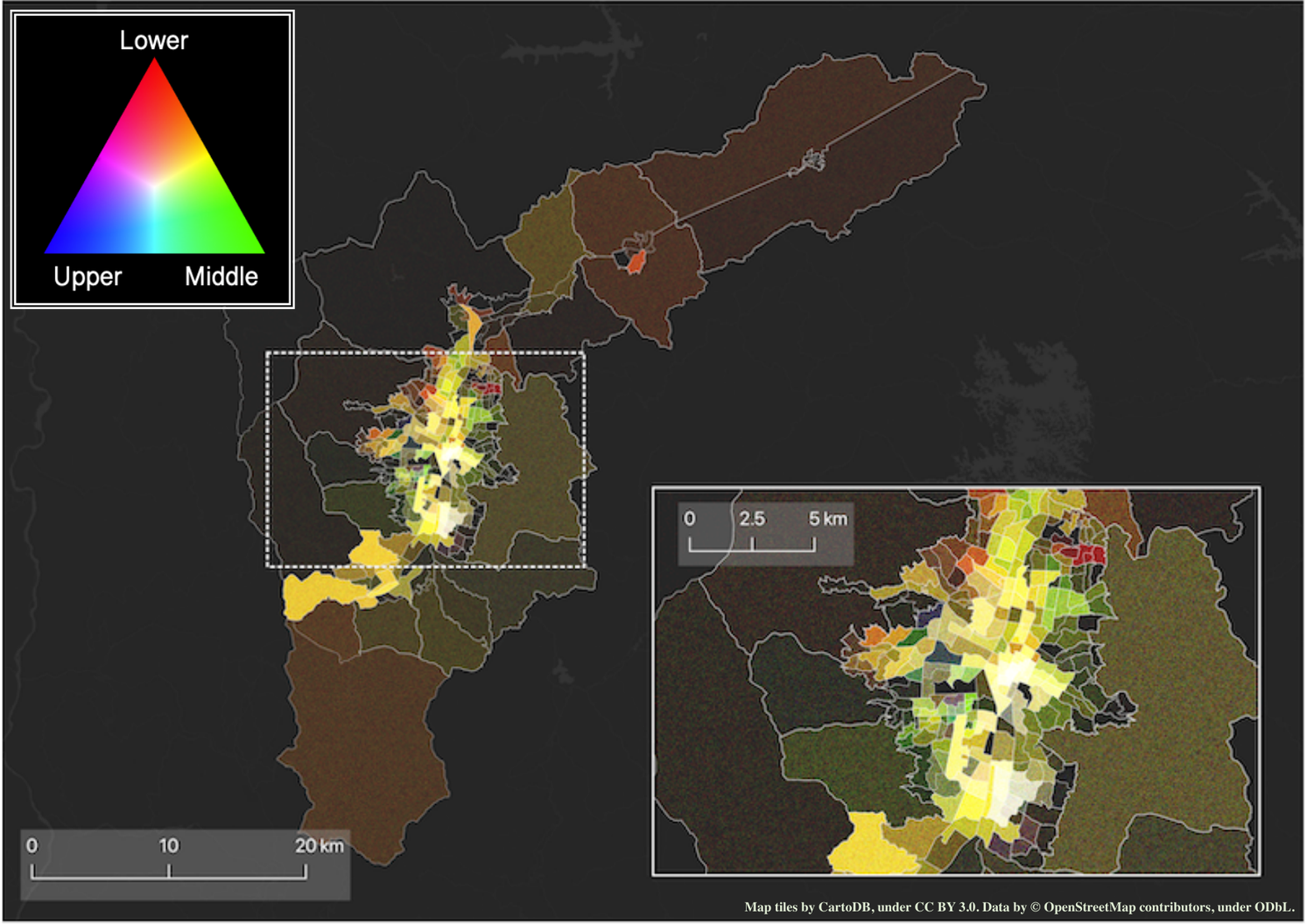}
		\caption{\textbf{Density map of \work travels made in \mde during the year 2017.} Brighter colours represent a higher density of travels to work. The hue denotes whether for a given zone the majority of travels were made by travellers belonging to the \lowerc (red), \middlec{} (green), \upperc{} (blue) or all three socioeconomic status. The inset portrays a zoom of the city centre. Figure contains information from OpenStreetMap and OpenStreetMap Foundation, which is made available under the Open Database License.}
		\label{fig:mdesocio}
	\end{figure}

\begin{figure}[H]
		\centering
		\includegraphics[width=0.8\textwidth]{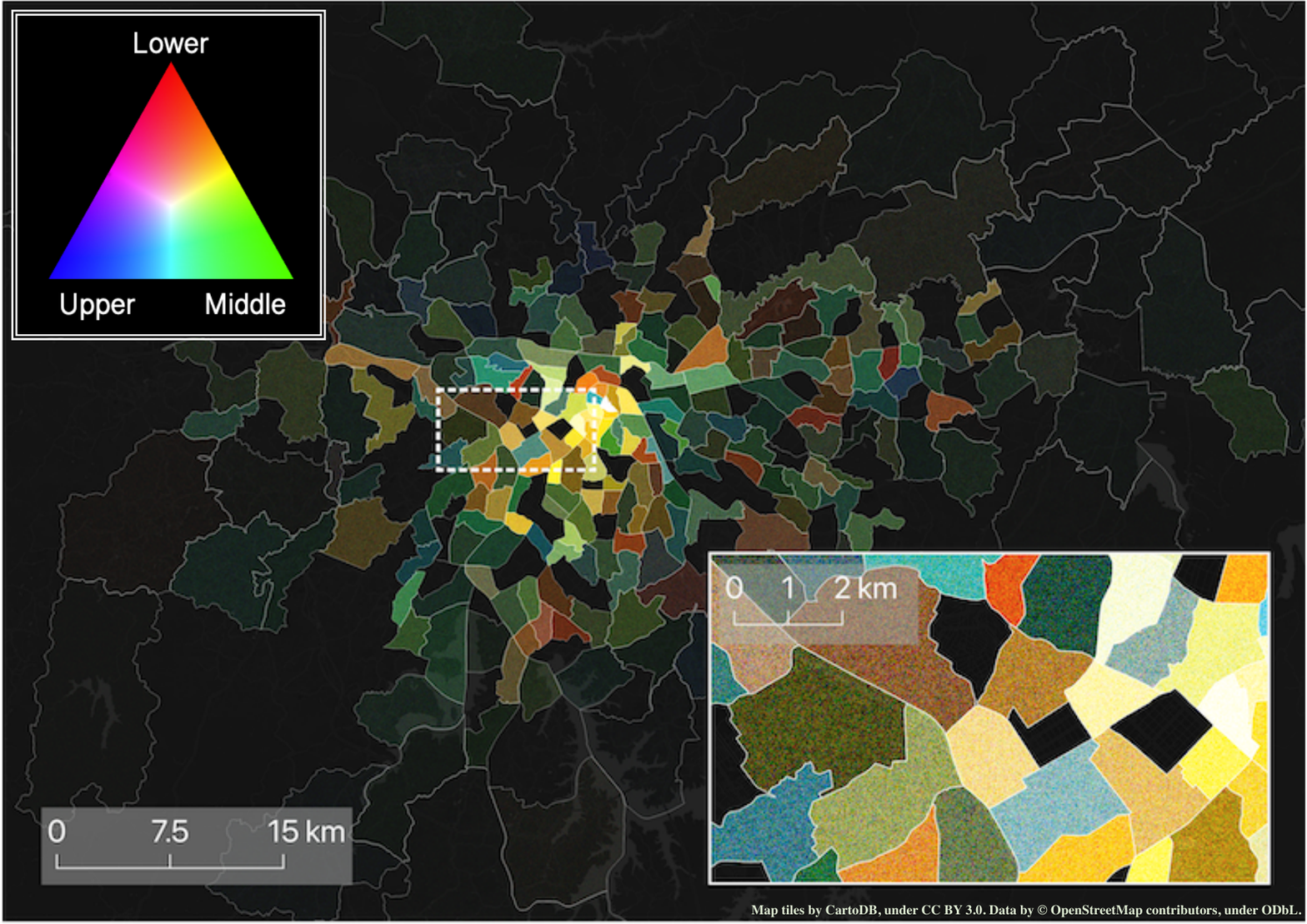}
		\caption{\textbf{Density map of \work travels made in \sao during the year 2017.} Brighter colours represent a higher density of travels to work. The hue denotes whether for a given zone the majority of travels were made by travellers belonging to the \lowerc (red), \middlec{} (green), \upperc{} (blue) or all three socioeconomic status. The inset portrays a zoom of the city centre. Figure contains information from OpenStreetMap and OpenStreetMap Foundation, which is made available under the Open Database License.}
		\label{fig:saosocio}
	\end{figure}

	Another way to look at the travel distribution is through their socioeconomic compositions. One striking feature observed in both Medell\'in (\ref{fig:mdesocio}~Fig) and S\~ao Paulo (\ref{fig:saosocio}~Fig) is that the more visited areas of the cities are also homogeneous with regards to the socioeconomic characteristics of their visiting populations. This is caused by the fact that the central districts of these cities tend to concentrate a large portion of their economic activities and businesses, therefore attracting workers from a broader range of segments, sectors, and backgrounds. 
	
	Despite these marked socioeconomic homogeneities at the centre, we can also observe in \ref{fig:mdesocio} and \ref{fig:saosocio} Figs that there are indeed areas incline to attract more predominantly workers from specific socioeconomic groups. Both in Medell\'in and S\~ao Paulo, it is possible to observe areas coloured in red, indicating a stronger concentration of work travels by lower-income people. Additionally, outside the dense core of the cities, we can observe that both cities tend to have areas that seem to be \textit{less} attractive to specific income groups. For instance,  in Medell\'in, most of the areas are coloured in yellow shades, indicating that those zones attract more workers of lower and middle income and less of upper income. A similar pattern can also be observed, albeit in a lesser extent, in S\~ao Paulo. In fact, S\~ao Paulo tends to have more zones coloured with blue and red hues than Medell\'in, suggesting that S\~ao Paulo is a city in which the economic landscape tend to be more \textit{segregated}.

	\section{Mobility diversity}
	
	\subsection{Boundary values of the mobility diversity}
	\label{sec:boundaries_entropy}
	
	Following Eq~\eqref{M-eq:mobdiv}, one could demonstrate that the mobility diversity of a group of travellers, $X$, travelling to fulfil purpose, $d$, is bounded (\ie{} $H_d^X \in [0,1]$). Such boundary values have a clear, physical, meaning which is related to the characteristics of the probability that travels have as their destination a given zone $i$, $p^{X}_{d}(i)$, presented in Eq~\eqref{M-eq:probability}. In the following, we compute the boundary values. Noteworthy, these boundaries do not depend on either the group of travellers, $X$, or the purpose of travel, $d$, under consideration.
	
	The least diverse mobility pattern corresponds to the case where travellers travel exclusively to one zone (say, $i = \tilde{i}$). Under such an assumption, Eq~\eqref{M-eq:probability} becomes:
	\begin{equation}
		\label{eq:prob_onezone}
		p^{X}_{d}(i) =  %
		\begin{cases} 1& \text{for } i = \tilde{i}\\
			0& \text{otherwise}
		\end{cases}\,.
	\end{equation}
	By replacing $p$ in Eq~\eqref{M-eq:mobdiv}, $H_{d}^{X}$ reads:
	\begin{equation}
		\label{eq:diversity_bound1}
		H_{d}^{X} =  - \frac{1}{\log_2 N_Z}  \Biggl[ \left( 1 \, \log_2 1\right) + \sum_{\substack{i=1\\ i \neq \tilde{i}}}^{N_Z} 0 \, \log_2 0 \Biggr]  = - \frac{1}{\log_2 N_Z}  \, \left( 0 + 0 \right) = 0 \,.
	\end{equation}
	If, instead, we assume that the travellers cover all the available zones uniformly, then each destination is reached by the same number of travels, corresponding to the most diverse mobility pattern. Under such circumstances, Eq~\eqref{M-eq:probability} becomes:
	\begin{equation}
		\label{eq:prob_allzones}
		p^{X}_{d}(i) = \frac{N^{X}_{d}(i)}{N^{X}_{d}} = \frac{N^{X}_{d}/N_Z}{N^{X}_{d}} = \frac{N^{X}_{d}}{N_Z} \frac{1}{N^{X}_{d}} = \frac{1}{N_Z} \; \forall i \,,
	\end{equation}
	where $N_{d}^{X}(i)$ is the total number of trips made by a group $X$ with a purpose $d$ to a destination area $i$ and\newline $N_{d}^{X}$ = $\sum_{i}^{N_{Z}} N_{d}^{X}(i)$. Replacing \eqref{eq:prob_allzones} in Eq~\eqref{M-eq:mobdiv}, gives:
	\begin{equation}
		\label{eq:diversity_bound2_1}
		H_{d}^{X} = - \frac{1}{\log_2 N_Z} \sum^{N_Z}_{i=1} \frac{1}{N_Z} \, \log_2 \frac{1}{N_Z} \,,
	\end{equation}
	as the argument of the sum does not depend on $i$, we can write:
	\begin{equation}
		\label{eq:diversity_bound2_2}
		H_{d}^{X} = - \frac{1}{\log_2 N_Z} N_Z \left[ \frac{1}{N_Z} \, \log_2 \frac{1}{N_Z} \right] = - \frac{1}{\log_2 N_Z} \Bigl( - \log_2 N_Z \Bigr) = 1 \,.
	\end{equation}

	\subsection{Mobility diversity from sampled data}
	\label{sec:data_sampling}
	
	As detailed in \ref{sec:datasummary}~Section, in line with methodological standards in sociodemographic surveys~\cite{levy2013sampling}, each entry in our data is associated to an expansion factor, a weight that accounts for the representativeness of that entry (\eg trip or individual) relative to the universe (\ie{} the entirety of a population).
	
	In traditional sociodemographic surveys, the expansion factors are calculated based on sampling probabilities backed by other factors (\eg area of residence and sociodemographic characteristics).
	In addition to the population-level expansion factor, the data in our household travel surveys also contain the \emph{travel-level} expansion factors accounting for the representativeness of the \emph{trips}. This is particularly important given that the focus of our work is on the mobility diversity, an information-theoretic metric computed from the travelling behaviours of a population.  Therefore, to ensure the validity of our cross-years comparisons, it is crucial that we assess whether the mobility diversity, when computed from the non-expanded data, can still support similar qualitative conclusions. This step is also important to assess the usefulness of the \mde{} 2017 data, whose expansion factors are not available. Thus, for the datasets that contained the expansion factors, we show the mobility diversity distributions obtained from the \textit{unweighted} samples with the ones produced by the expanded samples.

	\begin{figure}[h]
		\centering
		\includegraphics[width=0.75\textwidth]{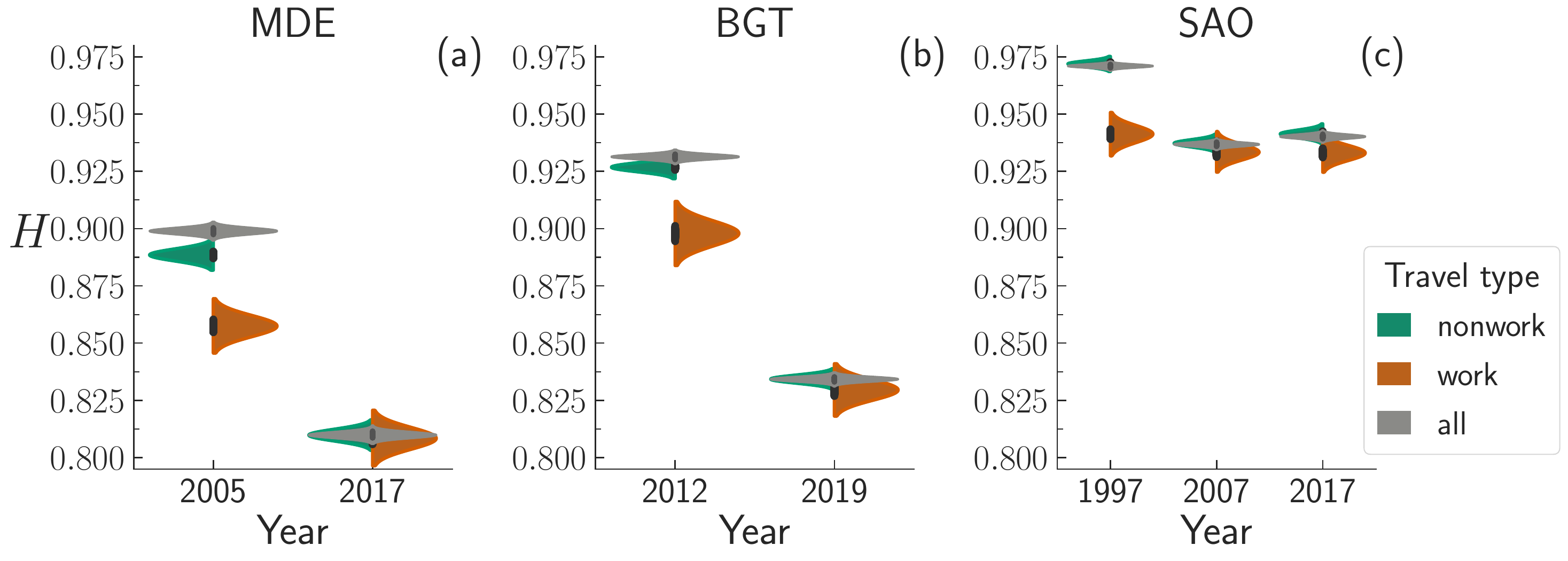}
		\caption{\textbf{Distribution of the bootstrapped mobility diversity, $H$, using the raw data without considering the expansion factors.} For each region and year, we consider the travels made for \all, \work, and \nonwork purposes.}
		\label{fig:sample_purpose}
	\end{figure}
	
	We show in \ref{fig:sample_purpose} Fig the distributions of the mobility diversity of travels made with \work, \nonwork or \all purposes for the regions of \mde, \bgt and \sao.
	Comparing \ref{fig:sample_purpose}~Fig with the results of Fig \ref{M-fig:aggregatedpatterns} (using the expansion factors), we identify that most of our main findings are valid in both samples. First, there is \textcolor{black}{a decrease in mobility diversity, $H$, in the most recent years.} Second, \work travels distributions show smaller values of $H$ than \all and \nonwork travels. Third, the \nonwork purpose of travels also plays a role in the spatial distribution of travels.
	
	The major difference between the results of our data using expansion factor (Fig \ref{M-fig:aggregatedpatterns}) or not (\ref{fig:sample_purpose}~Fig) is the magnitude of the mobility diversity differences between the travel types. We observe that the comparison of \work travels with \nonwork and \all travels are not completely captured by the data sample without expansion factor. Nonetheless, the conclusions drawn from the data using or not the expansion factor remain in general the same. However, we identify differences in the relationship between groups, confirming that the use of the expansion factor is crucial to ensure a fair comparison between groups. Thus, our analyses focus on the data sample using the expansion factor, and when it is necessary, we highlight differences between the results using the expansion factor or not for the case of Medell\'in in 2017.

	\subsection{Mobility diversity distribution by \textit{gender} }
	\label{sec:gendermob}
	
	This section explores the role of gender in mobility diversity by studying the overall $H$ distribution and the travel diversity of men and women. We focus our attention on the work-related travels (\work) in comparison with the diversity produced by the trips made for all the travel purposes. The results obtained and summarised in this section are in agreement with the phenomenology displayed in Fig~\ref{M-fig:genderregions}.
	
\begin{figure}[H]
		\centering
		\includegraphics[width=0.98\textwidth]{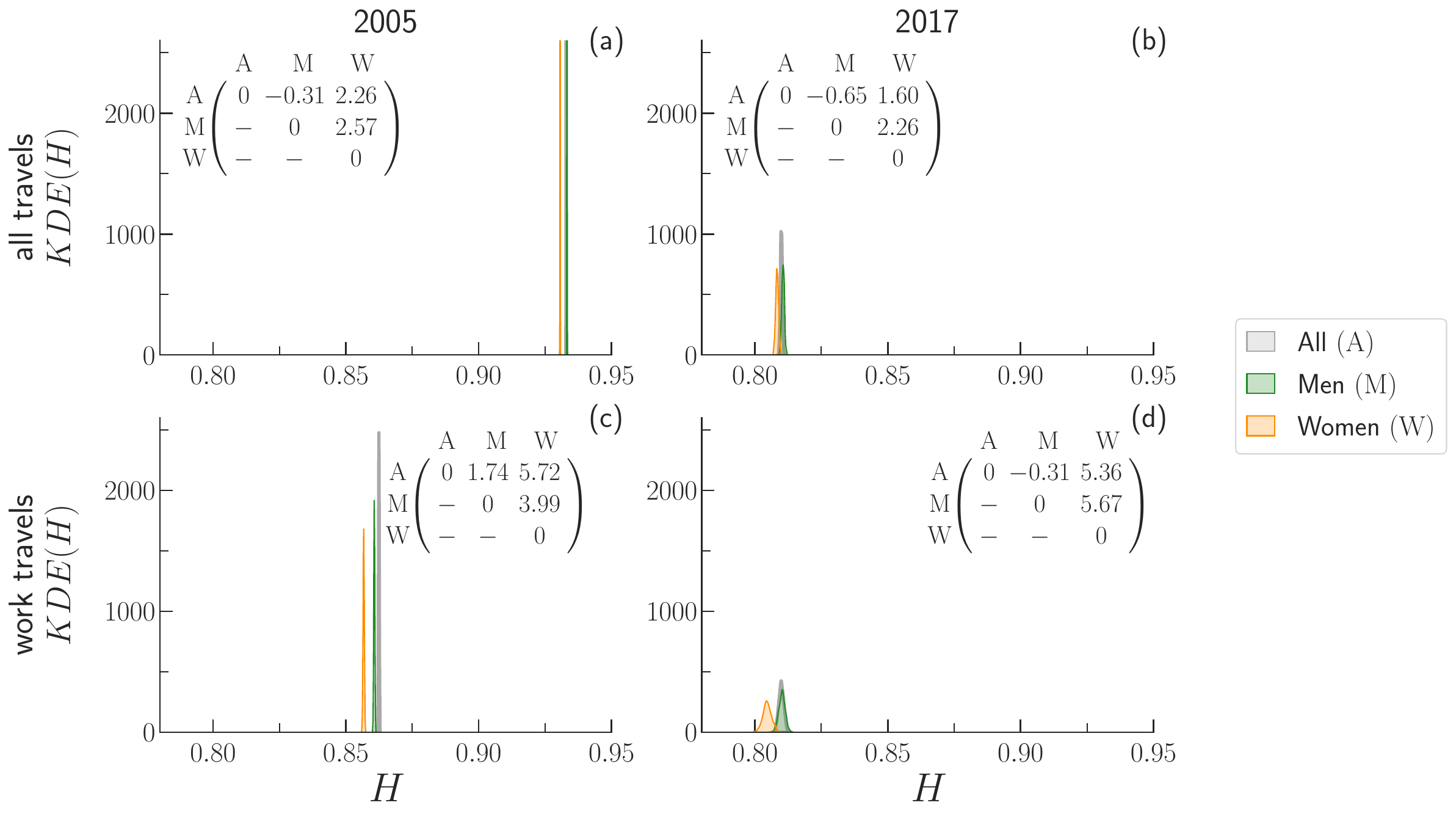}
		\caption{\textbf{Kernel Density Estimation plots of the mobility diversity, $H$, for \all travels (panels \textbf{a,b}) and \work travels (panels \textbf{c,d}) in \mde.} For each travel purpose, we plot the $\kde(H)$ for travels made by men ($M$), women ($W$), and all ($A$) travellers. The matrix appearing in the top left corner of each panel reports the peak-to-peak distance (\ie{} the distance between the median of the distributions) multiplied by a factor of $10^{3}$. The $\kde$s are computed from a distribution of $H$ obtained by bootstrapping 1,000 times \textcolor{black}{60\%} of the available travel records.}
		\label{fig:genderaburra}
	\end{figure}

	For the case of \mde (\ref{fig:genderaburra}~Fig), we observe that men exhibit higher values of $H$ than women. The values of the peak-to-peak distances between the KDEs of $H$ corresponding to \all and \men travels are smaller than the \women counterpart (see \ref{fig:gender_mde_sample_size}~Fig). \textcolor{black}{not a consequence of the fact that men account for the majority of the trips in the datasets compared to women (as shown in \ref{sec:sample_size_effect}~Section)}.

\begin{figure}[H]
		\centering
		\includegraphics[width=0.8\textwidth]{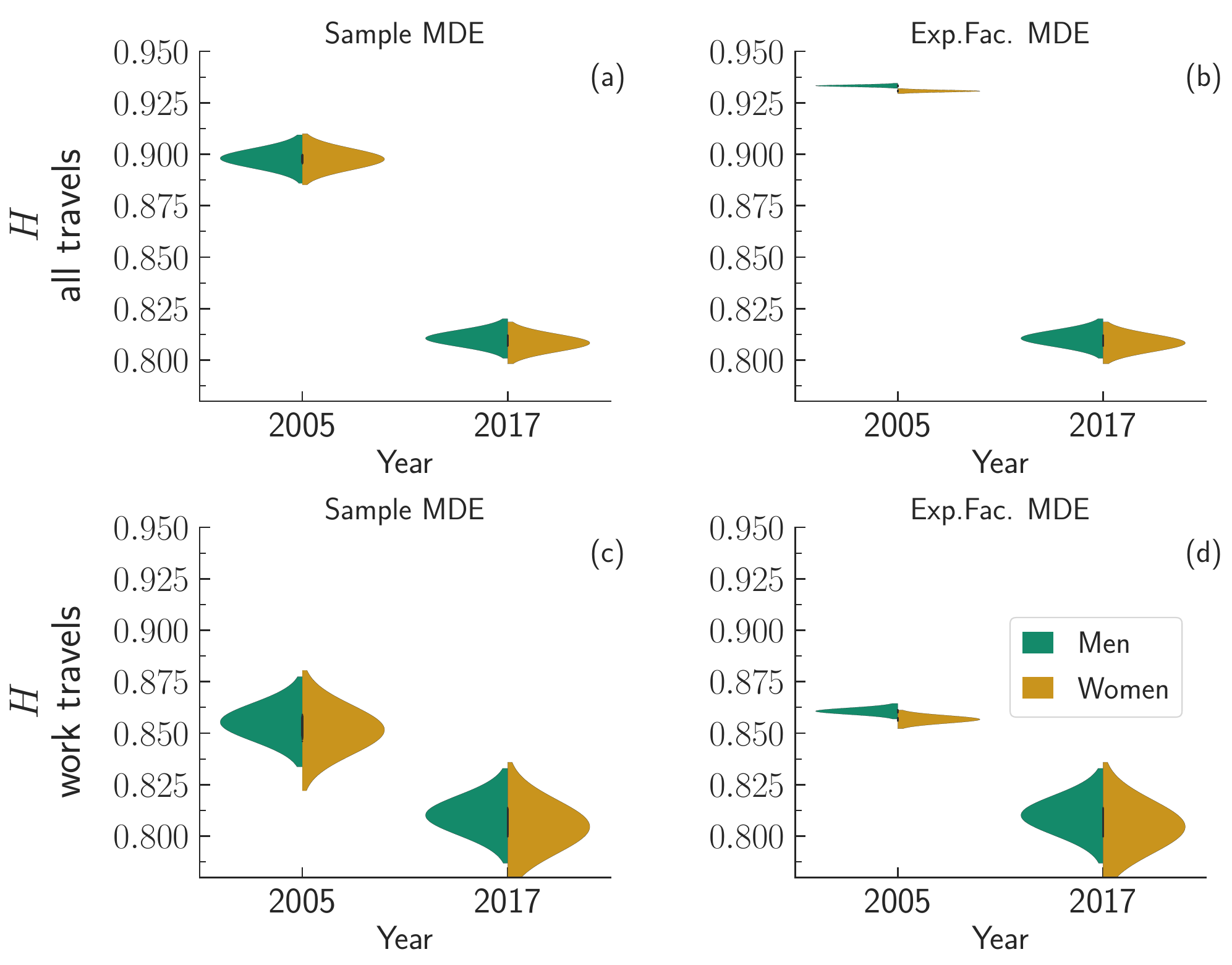}
		\caption{\textbf{Comparing the distributions of the mobility diversity ($H$) for \all travels (panels \textbf{a,b}) and \work travels (panels \textbf{c,d}) within the \mde area.} Panels \textbf{a} and \textbf{c} display the case of raw travel records, whereas panels \textbf{b} and \textbf{d} display the case of travel records obtained using the expansion factors for year 2005.}
		\label{fig:mdeexpanalysis}
	\end{figure}

	Despite the unavailability of expansion factors in \mde 2017, the variation in the mobility diversity over the time is compatible with the ones observed in \bgt and even \sao. Indeed, from the values displayed in Fig \ref{M-fig:aggregatedpatterns} and \ref{fig:mdeexpanalysis} Fig, there is evidence that mobility diversity \textcolor{black}{decreases} over the years. 
	
\begin{figure}[H]
		\includegraphics[width=\textwidth]{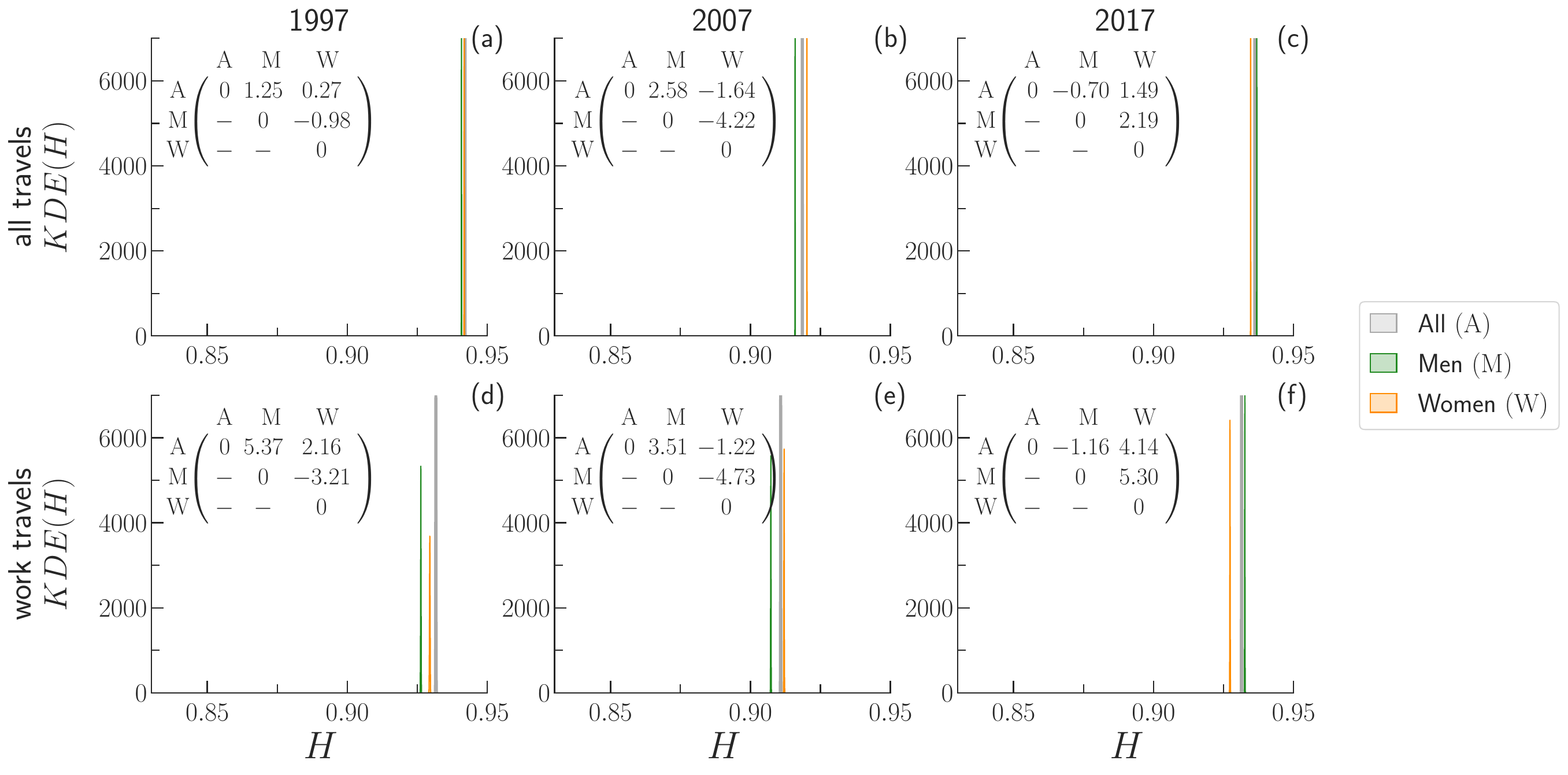}
		\caption{\textbf{Distributions of the mobility diversity, $H$, for \all travels (panels {\bf a}, {\bf b}, and {\bf c}) and \work travels (panels {\bf d},{\bf e}, and {\bf f}) made in \sao.} For each travel purpose, we plot the $\kde(H)$ for travels made by men ($M$), women ($W$), and all travellers ($A$). The matrix appearing in the top left corner of each panel reports the peak-to-peak distance (\ie{} the distance between the median of the distributions) multiplied by a factor of $10^{3}$. The $\kde$s are computed from a distribution of $H$ obtained by bootstrapping 1,000 times \textcolor{black}{60\%} of the available records.}
		\label{fig:kdegendersao}
	\end{figure}

	 \textcolor{black}{In the case of \sao (\ref{fig:kdegendersao}~Fig) we observe that men's mobility tends to display a lower value of $H$ than women's one independently on the travel's purpose. However, such a hierarchy gets inverted in 2017.}

	\newpage
	
\begin{figure}[H]
		\centering
		\includegraphics[width=0.98\textwidth]{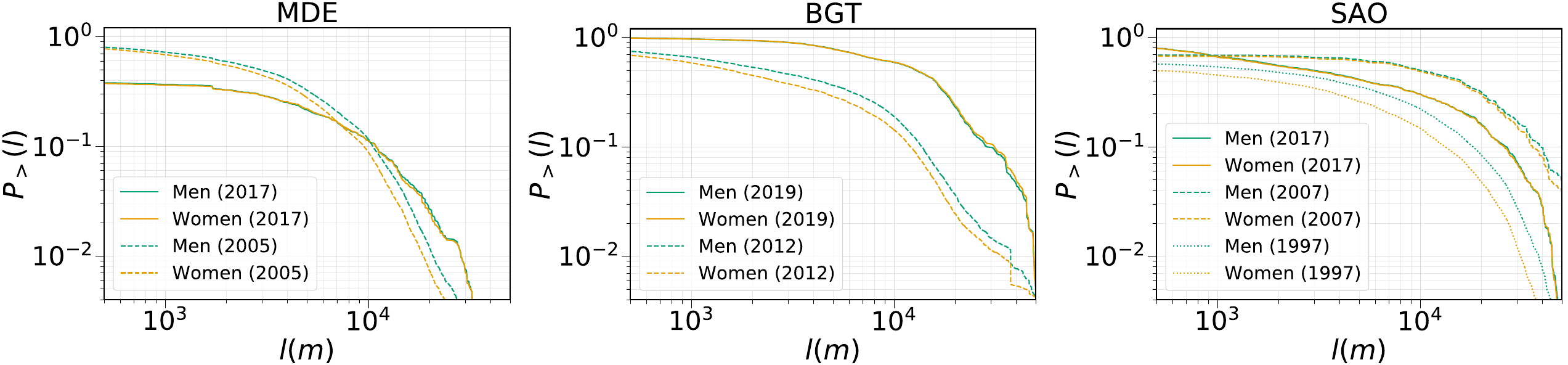}
		\caption{\textbf{Complementary cumulative probability distribution function, $P_{>}(l)$, of the probability of making a travel with a distance between origin and destination zones equal to or greater than $l$.} Each panel refers to a different metropolitan area. }
		\label{fig:traveldistance}
	\end{figure}

	One hallmark of the gender-centred differences in urban mobility is that, on average, women are more likely to perform shorter travels than men. This pattern can also be observed in our data, as shown in \ref{fig:traveldistance}~Fig, and \ref{tab:traveldistance} and \ref{tab:traveldistancetests}~Tables. The travel distance, $l$, is computed as the distance between the centroids of the origin and destination zones. However, the difference in the travel distances distribution does not exclude the chance that travellers (either women or men) can display small values of $H$. The reason is that, in principle, men could have longer travel distance while concentrating their travels in a small number of zones, which is not the case. In this way, the fact that women are more likely to have a shorter travel distance than men would not necessarily impact the mobility diversity of the travels performed by women.
	
\begin{figure}[H]
		\includegraphics[width=0.98\textwidth]{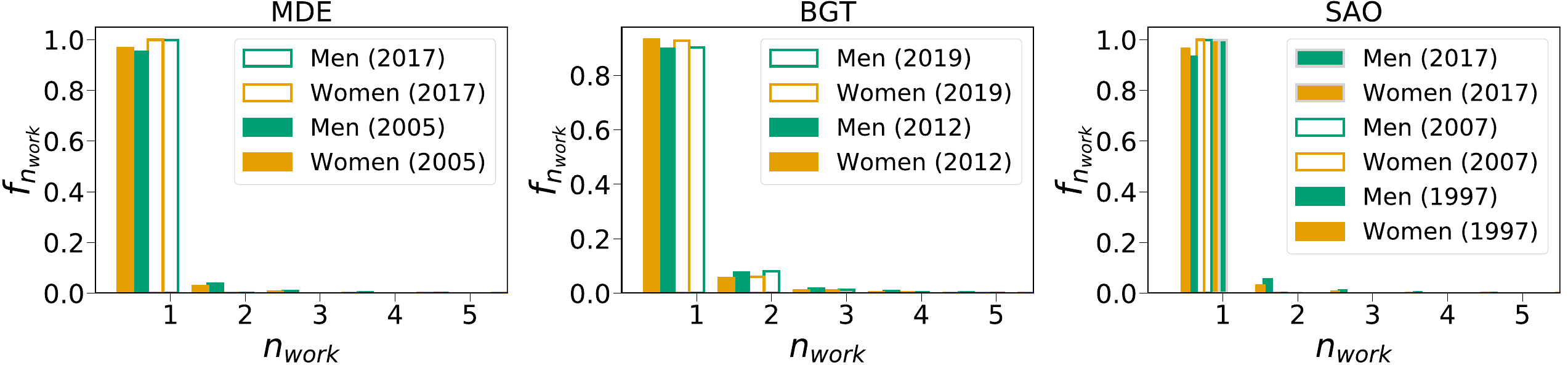}
		\caption{\textbf{Fraction of the number of locations in which an individual works, $f_{n_{work}}$}. The data are disaggregated according to the gender of the travellers.}
		\label{fig:workmorezones}
	\end{figure}

	On the other hand, the fact that women are more likely to move within the same zone could impact the mobility diversity because they are less likely to endeavour to other zones. Women and men display a similar fraction of travels regardless of the travel's purpose (all or to work), and the fact that the origin and destination zones are the same (travels inside a zone) (see  \ref{tab:probabilitysamezones}~Table). The fraction of travels that travellers live and work in the same zone are similar for women and men (see \ref{tab:probabilitysamezones}~Table). Furthermore, women and men work in general in only one zone (see \ref{fig:workmorezones}~Fig), but the latter are slightly more likely to work in more than one zone. Thus, we argue that travels inside zones and the number of workplaces are not impacting differences in the mobility diversity of women and men.
	
	Then, we check whether the majority of the zones are more likely to be visited by men than women for different purposes of travel. \ref{tab:higherfractiontravelsperarea} Table shows the percentage of zones for which there are more travels performed by men than by women for \all, \work and \nonwork purposes. For \all travels, \mde and \sao show a majority of areas being visited by men, and \bgt shows a majority of areas being visited by women. For \work travels, regardless of the region, the majority of the areas are mostly visited by men, and the opposite happens to \nonwork travels. We conclude that women are more likely to be concentrated in a small number of areas to work, and they are also the minority in the majority of the areas.
	
	\begin{table}[H]
		\caption{\textbf{Minimum ($\min(l)$), maximum ($\max(l)$), median ($med(l)$), average ($\avg{l}$), and standard error of the mean ($\varepsilon_l$) of the travel distance $l$ (measured in m) made by men and women in each region and year.}}
		\label{tab:traveldistance}
		\centering
		\resizebox{0.7\linewidth}{!}{
			\begin{tabular}{cclrrrrr}
				\toprule
				City & Year & Gender & \multicolumn{1}{c}{$\min(l)$} & \multicolumn{1}{c}{$\max(l)$} &  \multicolumn{1}{c}{$med(l)$} & \multicolumn{1}{c}{$\avg{l}$} & \multicolumn{1}{c}{$\varepsilon_l$} \\
				\midrule
				\multirow{4}{*}{\mde} & \multirow{2}{*}{2005} & \men & \multirow{2}{*}{102.54} & 59149.04 & 3775.57 & 5025.26  & 3.37 \\
				& & \women & & 58728.50 & 3256.55 & 4463.16 & 3.17 \\\cline{2-8}
				& \multirow{2}{*}{2017} & \men & \multirow{2}{*}{104.58} & 39867.49 &  5349.23 & 7705.08  & 52.37 \\
				& & \women & & 39867.49 & 5631.93 & 7734.20 & 53.62 \\\hline
				\multirow{4}{*}{\bgt} & \multirow{2}{*}{2012} & \men & \multirow{2}{*}{101.35} & 115452.21 & 4207.63 & 6666.98  & 3.49 \\
				& & \women & & 81793.49 & 3008.38 & 5746.65 &  2.99 \\\cline{2-8}
				& \multirow{2}{*}{2019} & \men & \multirow{2}{*}{119.02} & 89115.59 & 12883.82 & 14561.42 &  4.94 \\
				& & \women & & 89115.59 & 12991.28 & 14761.55 &  4.74 \\\hline
				\multirow{6}{*}{\sao} & \multirow{2}{*}{1997} & \men & \multirow{2}{*}{103.91} & 99384.35 & 7081.65 & 10162.40 &  4.05 \\
				& & \women & & 99384.35 & 5297.32 & 8136.43 &  3.88 \\\cline{2-8}
				& \multirow{2}{*}{2007} & \men & \multirow{2}{*}{281.25} & 85235.50 & 17888.80 & 21623.26 &  4.91 \\
				& & \women & & 85235.50 & 17627.78 & 20844.79 & 4.74 \\\cline{2-8}
				& \multirow{2}{*}{2017} & \men & \multirow{2}{*}{130.41} & 49996.82 & 4727.93 & 10015.40 &  3.11 \\
				& & \women & & 62504.24 & 4630.86 & 9952.94 &  3.10 \\
				\bottomrule
			\end{tabular}
		}
	\end{table}

	\begin{table}[H]
	    \centering
		\caption{\textbf{The $p$-values of the Kolmogorov–Smirnov ($KSTest$) and Student $t$ ($TTest$) tests comparing the travel distance performed by \men (M), \women (W) and \all travellers (A).} The symbol $^{***}$ represents that the $p$-value is smaller than 0.001.}\label{tab:traveldistancetests}
		\resizebox{0.92\textwidth}{!}{
			\begin{tabular}{llrrrrrr}
				\toprule
				City & Year & $KSTest(MW)$ & $KSTest(AM)$ & $KSTest(AW)$ &  $TTest(MW)$ & $TTest(AM)$ & $TTest(AW)$ \\
				\midrule
				\multirow{2}{*}{\mde} & 2005 & $^{***}$ & $^{***}$ & $^{***}$ & $^{***}$ & $^{***}$ & $^{***}$ \\
				& 2017 & $^{***}$ & 0.17 & 0.12 & 0.69 & 0.827 & 0.817 \\
				\multirow{2}{*}{\bgt} & 2012 & $^{***}$ & $^{***}$ & $^{***}$ & $^{***}$ & $^{***}$ & $^{***}$ \\
				& 2019 & $^{***}$ & $^{***}$ & $^{***}$ & $^{***}$ & $^{***}$ & $^{***}$ \\
				\multirow{3}{*}{\sao} & 1997 & $^{***}$ & $^{***}$ & $^{***}$ & $^{***}$ & $^{***}$ & $^{***}$ \\
				& 2007 & $^{***}$ & $^{***}$ & $^{***}$ & $^{***}$ & $^{***}$ & $^{***}$ \\
				& 2017 & $^{***}$ & $^{***}$ & $^{***}$ & $^{***}$ & $^{***}$ & $^{***}$ \\
				\bottomrule
			\end{tabular}
		}
	\end{table}

	\begin{table}[H]
		\caption{\textbf{Percentages of the $all$ travels for which the origin and destination zones are the same and are performed by \all (A), \men (M) and \women (W) travellers, $P^{X}_{all} \; X \in \{ A, M, W \}$. The same quantity but for the case of $work$ travels, $P^{X}_{work} \; X \in \{ A, M, W \}$. Finally, we report the percentages of work travels performed by \all (A), \men (M) and \women (W) working in the same zone where they live, $P^{X}_{live=work} \; X \in \{ A, M, W \}$.}}\label{tab:probabilitysamezones}
		\centering
		\resizebox{\textwidth}{!}{
			\begin{tabular}{llrrrrrrrrr}
				\toprule
				City & Year & $P^{A}_{all} (\%)$ & $P^{M}_{all} (\%)$ & $P^{W}_{all} (\%)$ & $P^{A}_{work} (\%)$ & $P^{M}_{work} (\%)$ & $P^{W}_{work} (\%)$ &  $P^{A}_{live=work} (\%)$ &  $P^{M}_{live=work} (\%)$ & $P^{W}_{live=work} (\%)$\\
				\midrule
				\multirow{2}{*}{\mde} & 2005 & 18.34 & 17.52 & 19.20
				& 1.71 & 1.90 & 1.26 
				&  7.76 & 8.02 & 7.38 \\
				
				& 2017 & 18.99 &  16.79 & 21.62
				& 14.39 & 24.31 & 15.73 
				& 22.04 & 26.41 &  17.32 \\
				
				\multirow{2}{*}{\bgt} & 2012 & 28.43  & 25.34  & 30.94
				& 1.92 & 2.60  &  1.51
				&  13.72 & 14.37 & 12.92\\
				
				& 2019 & 1.16  & 1.19  & 1.14
				& 1.71 & 2.06  &  1.58 
				& 10.81 & 10.54 & 11.15\\
				
				\multirow{3}{*}{\sao} & 1997 & 43.83  & 41.05  & 46.90
				& 12.45 & 13.99 &  10.75
				& 23.31 & 21.83 & 25.73 \\
				
				& 2007 & 37.77  & 35.45 & 40.15
				& 11.53 &  12.03 &  11.01  
				& 20.23 & 18.66 & 22.32 \\
				
				& 2017 &  38.69 & 37.75  & 39.63
				& 12.49 &  13.91 & 11.07  
				& 20.63 & 20.23 & 21.11 \\
				\bottomrule
			\end{tabular}
		}
	\end{table}

	\begin{table}[H]
		\caption{\textbf{Percentage of areas for which the fraction of travels performed by men is higher than the the same quantity computed for women for \all, \work and \nonwork travels ($P^{M}_{all,area} > P^{W}_{all,area}$, $P^{M}_{work,area} > P^{W}_{work,area}$ , $P^{M}_{nonwork,area} > P^{W}_{nonwork,area}$).}}\label{tab:higherfractiontravelsperarea}
		\centering
		\resizebox{0.68\textwidth}{!}{
			\begin{tabular}{llccc}
				\toprule
				City & Year & $P^{M}_{all,area} > P^{W}_{all,area}$ & $P^{M}_{work,area} > P^{W}_{work,area}$ & $P^{M}_{nonwork,area} > P^{W}_{nonwork,area}$  \\
				\midrule
				\multirow{2}{*}{\mde} & 2005 & 61.46\% & 79.67\% & 43.20\% \\
				& 2017 & 62.67\% & 91.12\% & 36.40\% \\\hline
				\multirow{2}{*}{\bgt} & 2012 & 28.21\% & 63.79\% & 21.55\% \\
				& 2019 & 26.84\% & 77.09\% & 15.02\% \\\hline
				\multirow{3}{*}{\sao} & 1997 & 63.88\% & 87.20\% & 39.84\% \\
				& 2007 & 46.85\% & 86.12\% & 26.77\% \\
				& 2017 & 50.00\% & 86.69\% & 34.00\% \\
				\bottomrule
			\end{tabular}
		}
	\end{table}

	\subsection{Mobility diversity distribution by \textit{socioeconomic groups}}\label{sec:sociomob}
	
\begin{figure}[H]
		\centering
		\includegraphics[width=0.8\textwidth]{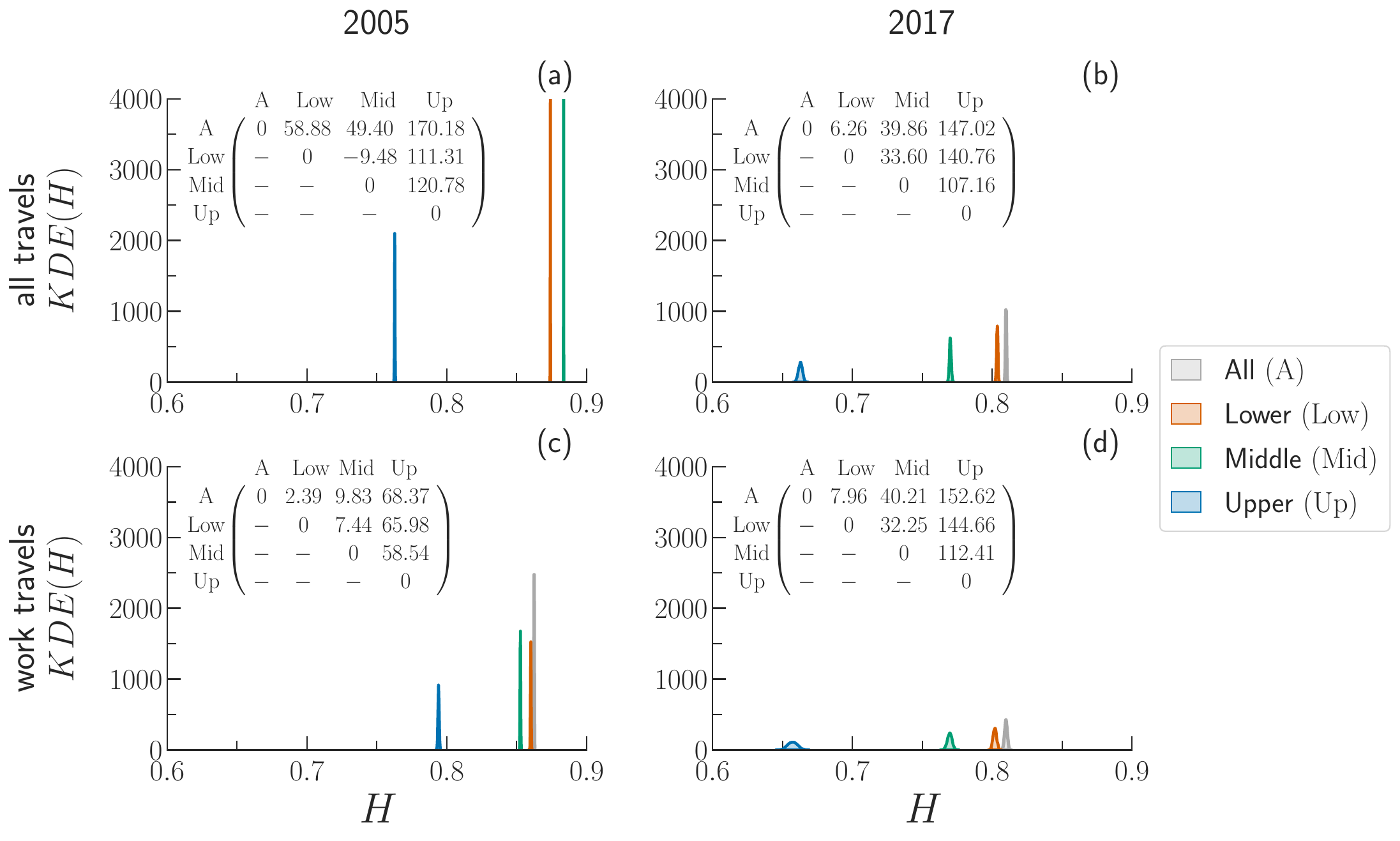}
		\caption{\textbf{KDE plots of the mobility diversity $H$ for \all travels (panels \textbf{a} and {\bf b}), and  \work travels (panels \textbf{c} and {\bf d}) in Medell\'in.} The matrix in the top left corner of each graph reports the peak-to-peak distance between the median of the distribution, multiplied by a factor of $10^{3}$.}
		\label{fig:kdesociomde}
	\end{figure}

\begin{figure}[H]
		\centering
		\includegraphics[width=0.87\textwidth]{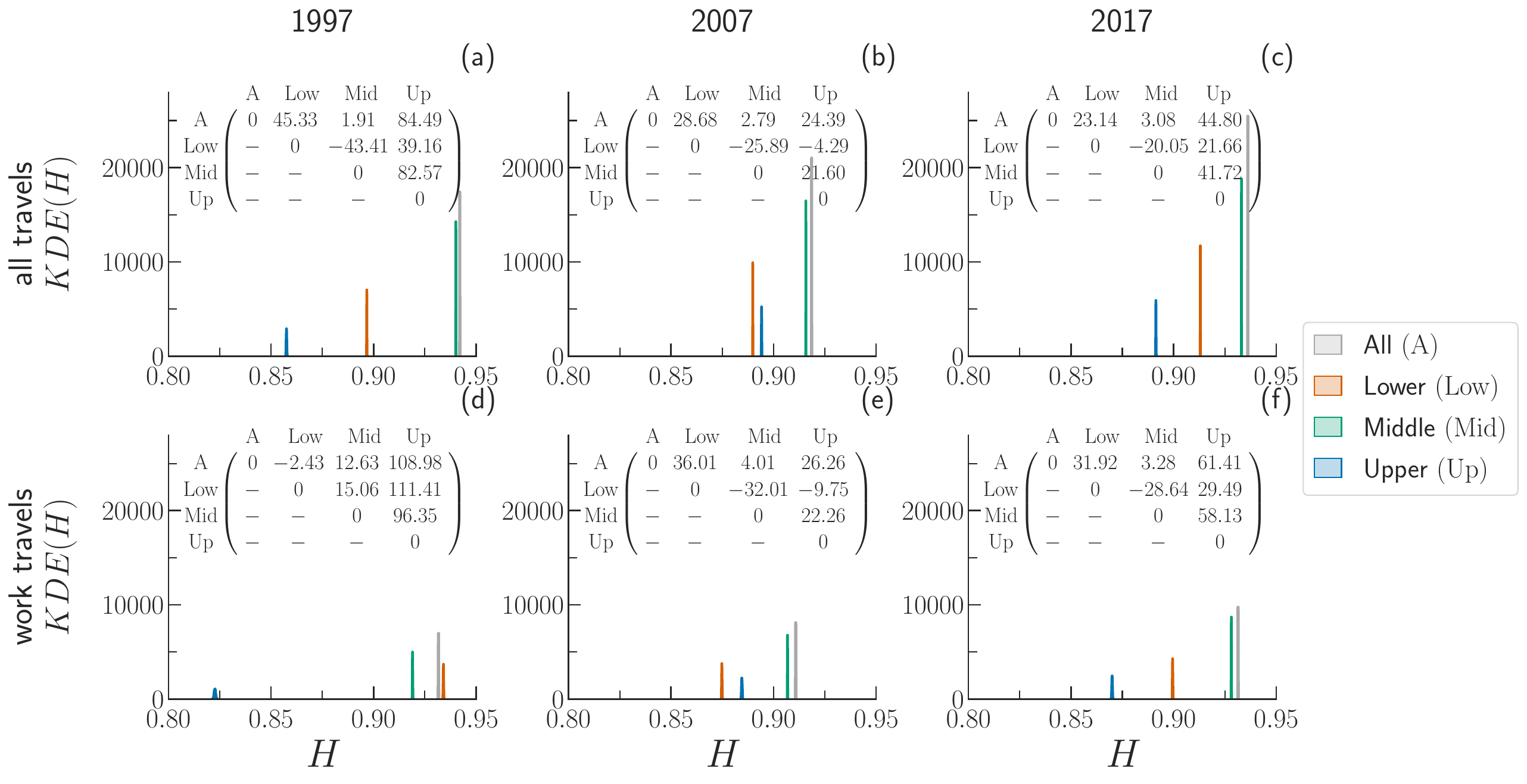}
		\caption{\textbf{KDE plots of the mobility diversity $H$ for \all travels (panels \textbf{a}, {\bf b}, and {\bf c}), and  \work travels (panels \textbf{d}, {\bf e}, and {\bf f}) in S\~ao Paulo.} The matrix in the top left corner of each graph reports the peak-to-peak distance between the median of the distribution, multiplied by a factor of $10^{3}$.}
		\label{fig:kdesociosao}
	\end{figure}

	Similar to the main manuscript, here, we show the distributions of the mobility diversity $H$ for \mde and \sao areas in \ref{fig:kdesociomde} and \ref{fig:kdesociosao} Figs respectively. In \mde, we observe that upper-income travellers display a lower mobility diversity, whereas \textcolor{black}{lower/middle-income travellers} tend to present the highest $H$ values.

	\newpage
	
	In \sao, \textcolor{black}{the mobility diversity varied over the years indicating at the same time increase and decrease across socioeconomic groups.}. We argue here that the impact on the mobility in 2007 could have been largely influenced by the profound economic changes Brazil underwent during that period \cite{de2010impact,ferrari2011brazil}. 
	
	\subsection{Mobility diversity distribution by \textit{gender} and  \textit{socioeconomic groups}}\label{sec:gendersociomob}
	
\begin{figure}[H]
		\includegraphics[width=0.97\textwidth]{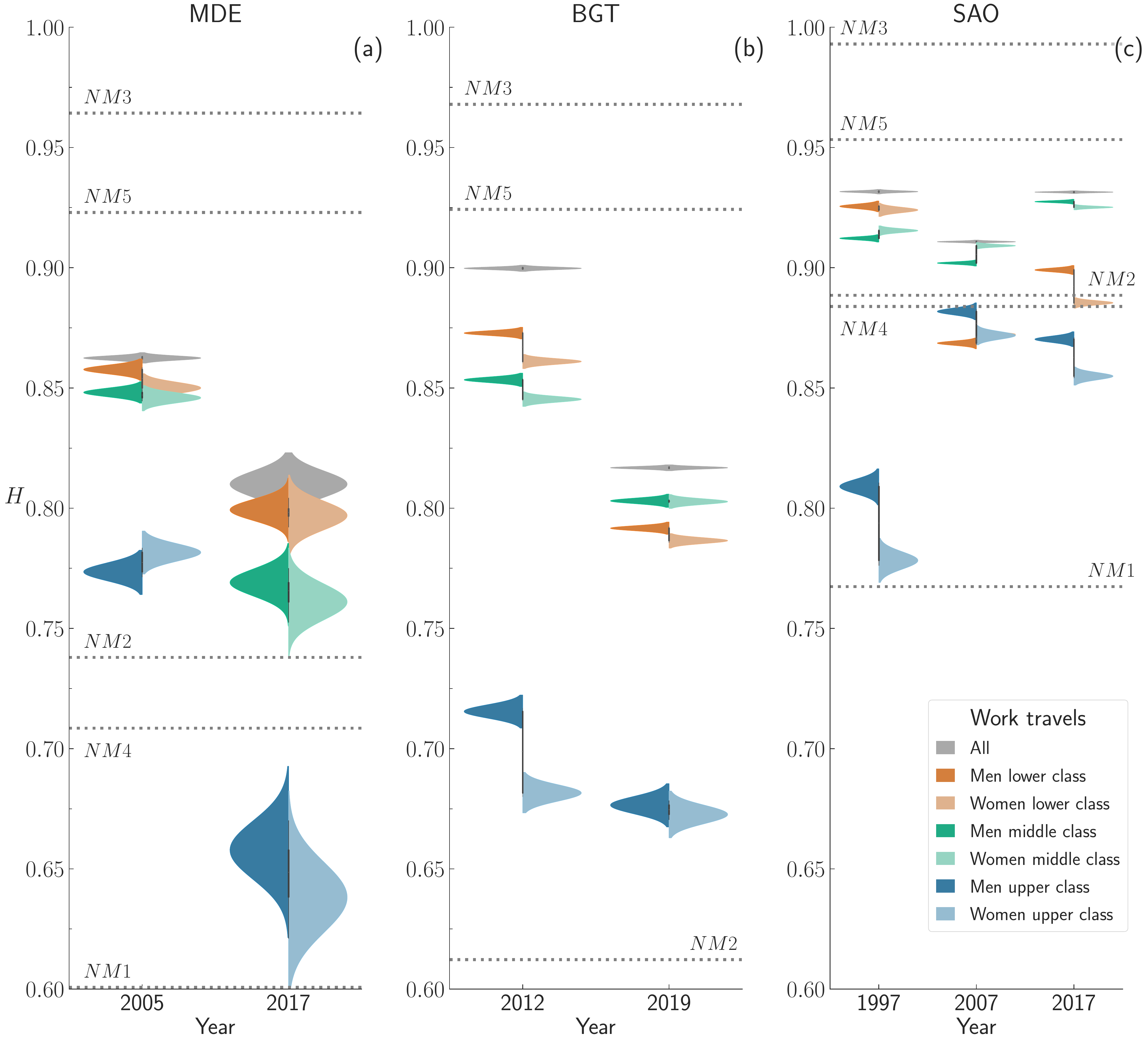}
		\caption{\textbf{Distribution of the mobility diversity, $H$, for travels made by \work purposes by travellers grouped according to their socioeconomic status and gender.} Each column refers to a different region, and for each region, we consider all the available years. For each socioeconomic status (\upperc{}, \middlec{}, and \lowerc{}) a darker hue denotes men travellers, whereas a lighter hue denotes women ones. \textcolor{black}{Dotted grey lines display the values of mobility diversity for each null model (see \ref{sec:null_models}~Section for the details).}}
		\label{fig:workgendersocioregions}
	\end{figure}

	In \ref{fig:workgendersocioregions}~Fig, we display the distributions of $H$ computed for travels made for \work purposes by all combinations of gender and socioeconomic status. As we also see in Fig~\ref{M-fig:allgendersocioregions}, regardless of the purposes, the socioeconomic status shapes the mobility of people considerably, whereas gender exerts a smaller effect. Nonetheless, a marked gender split is also seen in both figures. Within each socioeconomic group, men consistently display higher values of $H$. \textcolor{black}{We also observe that the values of the mobility diversity computed from the travels generated using the null models are different from their empirical counterparts. In \ref{sec:null_models}~Section, we describe each null model, and comment about the corresponding values of mobility diversity.} On average, the gender-centred differences within each socioeconomic group tend to decrease over time, suggesting that a possible gender-level difference in mobility is, indeed, reducing (\ref{tab:genderdifferences}~Table).

	\begin{table}[H]
		\caption{\textbf{Gender differences, $median(H^M_{S}) - median(H^W_{S})$, of the mobility diversity $H$ of travels made for \all and \work purposes by travellers grouped according to their socioeconomic status, $S \in \{ \text{\lowerc{}, \middlec{}, \upperc{}} \}$, and gender, $X \in \{ M, W \}$.} The values report the peak-to-peak distance between the median of the distribution of $H$, multiplied by a factor of $10^{3}$. Negative values (in bold) denote the case $median(H^W_{S}) > median(H^M_{S})$.}
		\label{tab:genderdifferences}
		\centering
		\resizebox{0.65\linewidth}{!}{
			\begin{tabular}{cclrrr}
				\toprule
				City & Year & Purpose & \multicolumn{1}{c}{\lowerc} & \multicolumn{1}{c}{\middlec{}} & \multicolumn{1}{c}{\upperc{}}\\
				\midrule
				\multirow{4}{*}{\mde} 
				& \multirow{2}{*}{2005}
				& \all & 5.83 &  1.96 & 12.10  \\
				& & \work & 7.67  & 2.22  &  \textbf{-8.07} \\
				& \multirow{2}{*}{2017} 
				& \all & 0.25 & 4.51  &  16.15 \\
				& & \work & 2.31 & 8.05  &  19.67 \\
				\hline
				\multirow{4}{*}{\bgt} 
				& \multirow{2}{*}{2012}
				& \all & 6.25 & 4.87  & 15.39  \\
				& & \work &  11.80 & 8.10  &  33.89 \\
				& \multirow{2}{*}{2019} 
				& \all & 2.05 & \textbf{-1.40}  &  0.98 \\
				& & \work & 5.15 & 0.30 & 3.87 \\
				\hline
				\multirow{6}{*}{\sao} 
				& \multirow{2}{*}{1997}
				& \all & 1.69 & \textbf{-0.38}  & 16.05  \\
				& & \work & 1.51 & \textbf{-3.21}  & 30.75  \\
				& \multirow{2}{*}{2007} 
				& \all & \textbf{-4.89} & \textbf{-5.97}  &  3.99 \\
				& & \work & \textbf{-3.64} & \textbf{-7.21} & 9.95    \\
				& \multirow{2}{*}{2017} 
				& \all & 3.48 & 0.69  & 6.51  \\
				& & \work & 13.66 & 2.32  & 15.34  \\
				\bottomrule
			\end{tabular}
		}
	\end{table}

	\section{Statistical verification of the mobility diversity distributions}\label{sec:anova}

	As described in the main manuscript, to account for variations in sample sizes, we employed a bootstrapping strategy to estimate mobility diversity distribution. From these distributions, we used multiple statistical methods to verify the differences in the distributions across groups. 
	The tests we used were the Welch's $t$-test~\cite{delacre2017psychologists}, the ANOVA~\cite{tabachnick2007experimental}, and the Tukey's HSD post hoc test~\cite{brown2005new}. The Welch's $t$-test compares if the distributions of mobility diversity are statistically different from each other. The ANOVA test compares if the averages of the groups' mobility diversity distributions are statistically different, expressing if the result extracted for each element in a group is in fact, different from the other elements. Finally, the Tukey's HSD post hoc test indicates what pairs of groups' means are different. The statistical tests were computed using the following Python packages: pandas, numpy, scipy, statsmodels, and pingouin.

	To evaluate the contribution of the gender and socioeconomic dimensions to the mobility diversity $H$, we first apply the ANOVA one-way and two-way tests to identify whether the distributions present similar average values of $H$. In \ref{tab:anovaalltravelsexp} and \ref{tab:anovaworktravelsexp} Tables, we plot all the values of $F$-statistic and $p$-value of the ANOVA test computed from the mobility diversity, $H$, of travels made for \all and \work purposes by travellers aggregated by gender and socioeconomic status.
	
	First, we test if the values of $H$ from the gender groups are from populations with the same mean values. Considering the \all and \work travels, we can reject the null hypothesis that the mean values of $H$ from men, women, and all travellers are statistically the same because the $p$-values are smaller than 0.01 and the $F$-values are not small. Next, as the ANOVA test does not specify which specific groups differ from each other, we apply the Tukey's HSD post hoc test to discover whether the specific groups hold mutually statistically different. For instance, Tukey's test can tell us if the mean values of $H$ computed for men travellers are not statistically different from the same quantities computed for all travellers but, instead, are statistically different from the women's counterparts.
	
	Applying Tukey's HSD post hoc test, see \ref{tab:mde_2005_tukey} - \ref{tab:sao_2017_tukey} Tables, we observe that the $p$-values from the multi-group means comparisons of the \women, \men and \all distributions of the mobility diversity between the different purpose of travels are smaller than 0.01. The same procedure using ANOVA and Tukey's HSD post hoc tests is applied for the values of $H$ obtained when grouping travellers according to their socioeconomic classes. The $p$-values of the mobility diversity calculated from \all and \work travels using the ANOVA test are all smaller than 0.01, and the $F$-values are even higher than their counterpart computed for the gender-based classification. Using the Tukey's HSD post hoc tests, we can reject that the values of $H$ of the socioeconomic groups are from populations with same mean values.
	
	The two-way ANOVA test is then used to analyse the relationship between gender and socioeconomic status in the measurement of mobility diversity. The values of $F$ are small, hence the outcome of the ANOVA test does not exclude that the mobility diversity, $H$, of travellers belonging to different socioeconomic and gender groups could belong to the same distribution. To exclude such possibility, we decided to apply the Tukey's HSD test. From all the multi-group means comparisons of the distributions of the mobility diversity between different set of travels, see \ref{tab:mde_2005_tukey} - \ref{tab:sao_2017_tukey} Tables, we can not reject that the values of $H$ of the following distributions are from populations with same mean values: \textcolor{black}{\all and \men of \mde in 2017}.
	
	In the section of results of the manuscript, we observe that we can not reject that the distributions of the \work travels are statistically similar (tested by Welch's $t$-test with $p$-value $< 0.01$) in \textcolor{black}{two} cases: %
	\begin{inparaenum}[(i)]
		\item \textcolor{black}{comparing \all and \men of \mde in 2017; and}
		\item \textcolor{black}{comparing \men from the \upperc and travellers of the \upperc{} class of \mde in 2017 regardless of their gender.}
	\end{inparaenum}
	These observations are in agreement with the outcomes of the Tukey's test. 
	
	In general, we observe that the null hypotheses of gender groups and socioeconomic groups displaying similar mean values of mobility diversity can be rejected. When gender and socioeconomic dimensions are combined, for the majority of the cases, we can reject that the distributions are obtained from populations with the same mean values. Thus, each gender and socioeconomic group alone or taken together display different distributions of mobility diversity. Such a difference in the distributions of $H$ means that gender and socioeconomic differences are based on how each group explores the space available.

	\begin{table}[H]
		\caption{\textbf{$F$ Statistic of the ANOVA Test computed from the mobility diversity of all travels.} All the $p$-values are smaller than 0.001.}\label{tab:anovaalltravelsexp}
		\resizebox{\linewidth}{!}{
			\begin{tabular}{l|rr|rr|rrr}
				\toprule
				Location & \multicolumn{2}{c|}{\mde} & \multicolumn{2}{c|}{\bgt} & \multicolumn{3}{c}{\sao} \\
				Years & \multicolumn{1}{c}{2005} & \multicolumn{1}{c|}{2017} & \multicolumn{1}{c}{2012} & \multicolumn{1}{c|}{2019} & \multicolumn{1}{c}{1997} & \multicolumn{1}{c}{2007} & \multicolumn{1}{c}{2017} \\
				\midrule\midrule
				Gender Groups & 573569 & 4690 & 6023260 & 93689 & 537431 & 6456450 & 2296951 \\
				Socioeconomic Groups & 391579690 & 4127072 &  1138264618 & 35079596 & 301501736 & 177736269 & 228956044 \\
				Combined Groups & 8604 & 3527 & 5711 & 3060 & 5145 & 8739 & 5984 \\
				\bottomrule
			\end{tabular}
		}
	\end{table}
	
	\begin{table}[H]
		\caption{\textbf{$F$ Statistic of the ANOVA Test computed from the mobility diversity of work travels.} All the $p$-values are smaller than 0.001.}\label{tab:anovaworktravelsexp}
		\resizebox{\linewidth}{!}{
			\begin{tabular}{l|rr|rr|rrr}
				\toprule
				Location & \multicolumn{2}{c|}{\mde} & \multicolumn{2}{c|}{\bgt} & \multicolumn{3}{c}{\sao} \\
				Years & \multicolumn{1}{c}{2005} & \multicolumn{1}{c|}{2017} & \multicolumn{1}{c}{2012} & \multicolumn{1}{c|}{2019} & \multicolumn{1}{c}{1997} & \multicolumn{1}{c}{2007} & \multicolumn{1}{c}{2017} \\
				\midrule\midrule
				Gender Groups & 186411 & 4826 & 6164761 & 742155 & 1551884 & 1377470 & 2164971 \\
				Socioeconomic Groups & 7272494 & 762876 & 66412851 & 20475418 & 36892298 & 37295014 & 75000823 \\
				Combined Groups & 3420 & 3826 & 4705 & 3406 & 2522 & 9095 & 6249 \\
				\bottomrule
			\end{tabular}
		}
	\end{table}

	%
	%
	%
	\begin{table}[H]
		\caption{\textbf{Multi-group means comparisons of the distributions of the mobility diversity between different set of travels in \mde{} 2005 using Tukey’s HSD test.} The values presented are multiplied by $10^{2}$. The $^{***}$ symbol denotes a $p$-value smaller than 0.001. We highlight the cells of groups having $p$-values higher than 0.001.}
		\label{tab:mde_2005_tukey}
		\centering
		\resizebox{0.70\linewidth}{!}{%
			\begin{tabular}{l|l|r|rr|c}
				\toprule
				\multicolumn{1}{c|}{Travels} & \multicolumn{1}{c|}{Groups} & \multicolumn{1}{c|}{Mean} & \multicolumn{2}{c|}{95\% Confidence interval} &\multicolumn{1}{c}{Adjusted} \\
				& & \multicolumn{1}{c|}{difference} & Lower bound & Upper bound &\multicolumn{1}{c}{$p$-value} \\
				\midrule
				\multirow{30}{*}{\all} & (\all) $\times$ (\men) & 0.0308 & 0.0302 & 0.0314 & \multirow{30}{*}{$^{***}$} \\
				& (\all) $\times$ (\women) & -0.2263  &  -0.2269 &-0.2257  & \\
				& (\men) $\times$ (\women) & -0.2571 &   -0.2577& -0.2565   & \\
				& (\all) $\times$ (\lowerc{}) & 86.472 &  86.4707 & 86.4733  & \\
				& (\all) $\times$ (\middlec{}) & 87.4194 &  87.4181 & 87.4207  & \\
				& (\all) $\times$ (\upperc{}) & 75.3408 & 75.3395 & 75.3421  & \\
				& (\lowerc{}) $\times$ (\middlec{}) & 0.9475 &  0.9462 &  0.9488 & \\
				& (\lowerc{}) $\times$ (\upperc{}) & -11.1312 &  -11.1325& -11.1299   & \\
				& (\middlec{}) $\times$ (\upperc{}) & -12.0786 &  -12.0799& -12.0773  & \\
				& (\all) $\times$ (\men-\lowerc{}) & -5.7453 &  -5.7475 & -5.7431  & \\
				& (\all) $\times$ (\men-\middlec{}) & -4.9728 &  -4.975&  -4.9706  & \\
				& (\all) $\times$ (\men-\upperc{}) & -16.8361 &  -16.8383& -16.8339  & \\
				& (\all) $\times$ (\women-\lowerc{}) & -6.3292 &  -6.3314  & -6.327  & \\
				& (\all) $\times$ (\women-\middlec{}) & -5.1697 &  -5.1719 & -5.1675   & \\
				& (\all) $\times$ (\women-\upperc{}) & -18.047 &  -18.0492& -18.0448  & \\
				& (\men-\lowerc{}) $\times$ (\men-\middlec{}) & 0.7725 &  0.7703 &  0.7747  & \\
				& (\men-\lowerc{}) $\times$ (\men-\upperc{}) & -11.0908 &  -11.093 &-11.0886  & \\
				& (\men-\lowerc{}) $\times$ (\women-\lowerc{}) & -0.5839 & -0.5861 & -0.5817  & \\
				& (\men-\lowerc{}) $\times$ (\women-\middlec{}) & 0.5756 &  0.5734  & 0.5778  & \\
				& (\men-\lowerc{}) $\times$ (\women-\upperc{}) & -12.3017 &  -12.3039& -12.2995  & \\
				& (\men-\middlec{}) $\times$ (\men-\upperc{}) & -11.8633 &  -11.8655 &-11.8611  & \\
				& (\men-\middlec{}) $\times$ (\women-\lowerc{}) & -1.3564 &  -1.3586 & -1.3542  & \\
				& (\men-\middlec{}) $\times$ (\women-\middlec{})& -0.1969 &  -0.1991 & -0.1947   & \\
				& (\men-\middlec{}) $\times$ (\women-\upperc{}) & -13.0742 &  -13.0764 & -13.072  & \\
				& (\men-\upperc{}) $\times$ (\women-\lowerc{}) & 10.5069 &  10.5047 & 10.5091  & \\
				& (\men-\upperc{}) $\times$ (\women-\middlec{}) & 11.6664 & 11.6642 & 11.6686  & \\
				& (\men-\upperc{}) $\times$ (\women-\upperc{}) & -1.2109 &  -1.2131 & -1.2087   & \\
				& (\women-\lowerc{}) $\times$ (\women-\middlec{}) & 1.1595 &  1.1573 &  1.1617  & \\
				& (\women-\lowerc{}) $\times$ (\women-\upperc{}) & -11.7178 &  -11.72 &-11.7156  & \\
				& (\women-\middlec{}) $\times$ (\women-\upperc{}) & -12.8773 &  -12.8795 & -12.8751  & \\
				\midrule
				\multirow{29}{*}{\work} & (\all) $\times$ (\men) & -0.1733 &  -0.1755 & -0.1711  & \multirow{30}{*}{$^{***}$} \\
				& (\all) $\times$ (\women) & -0.5707  & -0.5729 & -0.5685 & \\
				& (\men) $\times$ (\women)  & -0.3974 &  -0.3996 &  -0.3952& \\
				& (\all) $\times$ (\lowerc{})  & 85.1415 & 85.1383 & 85.1447 & \\
				& (\all) $\times$ (\middlec{}) & 84.3991 & 84.3959  & 84.4024 & \\
				& (\all) $\times$ (\upperc{})  & 78.5477 & 78.5445 & 78.551 & \\
				& (\lowerc{}) $\times$ (\middlec{})  & -0.7424 & -0.7456 & -0.7392 & \\
				& (\lowerc{}) $\times$ (\upperc{})  & -6.5938 & -6.597 & -6.5905 & \\
				& (\middlec{}) $\times$ (\upperc{})  & -5.8514 & -5.8546 & -5.8482 & \\
				& (\all) $\times$ (\men-\lowerc{})  & -0.4769  & -0.4823 & -0.4715 & \\
				& (\all) $\times$ (\men-\middlec{})  & -1.4315 & -1.4369 & -1.4261 & \\
				& (\all) $\times$ (\men-\upperc{})  & -8.8917 & -8.8971 & -8.8863 & \\
				& (\all) $\times$ (\women-\lowerc{})  & -1.2447 & -1.2501 & -1.2393 & \\
				& (\all) $\times$ (\women-\middlec{})  & -1.6542 & -1.6596 & -1.6488  & \\
				& (\all) $\times$ (\women-\upperc{})  & -8.084 & -8.0894 & -8.0786   & \\
				& (\men-\lowerc{}) $\times$ (\men-\middlec{})  & -0.9547 & -0.9601 & -0.9493   & \\
				& (\men-\lowerc{}) $\times$ (\men-\upperc{})  & -8.4148 & -8.4202 & -8.4094   & \\
				& (\men-\lowerc{}) $\times$ (\women-\lowerc{})  & -0.7678 & -0.7733 &  -0.7624  & \\
				& (\men-\lowerc{}) $\times$ (\women-\middlec{})  & -1.1773 & -1.1827 &  -1.1719   & \\
				& (\men-\lowerc{}) $\times$ (\women-\upperc{})  & -7.6072 &  -7.6126 &  -7.6018   & \\
				& (\men-\middlec{}) $\times$ (\men-\upperc{})  & -7.4601 & -7.4656 & -7.4547 &   \\
				& (\men-\middlec{}) $\times$ (\women-\lowerc{})  & 0.1868 & 0.1814 &  0.1922 &   \\
				& (\men-\middlec{}) $\times$ (\women-\middlec{})  & -0.2226 & -0.228 & -0.2172    & \\
				& (\men-\middlec{}) $\times$ (\women-\upperc{})  & -6.6525 & -6.6579&  -6.6471   & \\
				& (\men-\upperc{}) $\times$ (\women-\lowerc{})  & 7.647 & 7.6416 &  7.6524 &   \\
				& (\men-\upperc{}) $\times$ (\women-\middlec{})  & 7.2375 & 7.2321 &  7.2429 &   \\
				& (\men-\upperc{}) $\times$ (\women-\upperc{})  & 0.8076 & 0.8022&    0.813 &   \\
				& (\women-\lowerc{}) $\times$ (\women-\middlec{}) & -0.4095 & -0.4149 &  -0.404   & \\
				& (\women-\lowerc{}) $\times$ (\women-\upperc{})  & -6.8393 & -6.8447&  -6.8339   & \\
				& (\women-\middlec{}) $\times$ (\women-\upperc{})  & -6.4299 & -6.4353 & -6.4245   & \\
				\bottomrule
			\end{tabular}%
		}
	\end{table}

	\begin{table}[H]
		\caption{\textbf{Multi-group means comparisons of the distributions of the mobility diversity between different set of travels in \mde 2017 using the Tukey’s HSD test.} See the caption of \ref{tab:mde_2005_tukey}~Table for the description of each column, and the notation.}
		\label{tab:mde_2017_tukey}
		\centering
		\resizebox{0.70\linewidth}{!}{
			\begin{tabular}{l|l|r|rr|c}
				\toprule
				\multicolumn{1}{c|}{Travels} & \multicolumn{1}{c|}{Groups} & \multicolumn{1}{c|}{Mean} & \multicolumn{2}{c|}{95\% Confidence interval} &\multicolumn{1}{c}{Adjusted} \\
				& & \multicolumn{1}{c|}{difference} & Lower bound & Upper bound &\multicolumn{1}{c}{$p$-value} \\
				\midrule
				\multirow{30}{*}{\all} & (\all) $\times$ (\men) &  0.0637 & 0.0586 & 0.0688 & \multirow{30}{*}{$^{***}$} \\
				& (\all) $\times$ (\women) & -0.1613 & -0.1664 & -0.1562 &  \\
				& (\men) $\times$ (\women) & -0.225 & -0.2301 & -0.2199 &  \\
				& (\all) $\times$ (\lowerc{})& 79.5541 &  79.5448 & 79.5633  &  \\
				& (\all) $\times$ (\middlec{}) & 76.1984 &  76.1891 & 76.2076  &  \\
				& (\all) $\times$ (\upperc{})& 65.4818 &  65.4725 &  65.491  &  \\
				& (\lowerc{}) $\times$ (\middlec{}) & -3.3557 &  -3.3649 & -3.3465 &  \\
				& (\lowerc{}) $\times$ (\upperc{}) & -14.0723 & -14.0815 & -14.0631   &  \\
				& (\middlec{}) $\times$ (\upperc{})& -10.7166 & -10.7258& -10.7074 &  \\
				& (\all) $\times$ (\men\lowerc{})& -0.7026 &  -0.7195 & -0.6858 &  \\
				& (\all) $\times$ (\men-\middlec{}) & -3.8696 &  -3.8864 & -3.8528  &  \\
				& (\all) $\times$ (\men-\upperc{}) & -14.1376 & -14.1544& -14.1208  &  \\
				& (\all) $\times$ (\women\lowerc{}) & -0.7279 &  -0.7447 & -0.7111  &  \\
				& (\all) $\times$ (\women-\middlec{}) &  -4.3216 &  -4.3384 & -4.3048 &  \\
				& (\all) $\times$ (\women-\upperc{}) & -15.7535 & -15.7703& -15.7367  &  \\
				& (\men\lowerc{}) $\times$ (\men-\middlec{}) & -3.1669 &  -3.1838 & -3.1501  &  \\
				& (\men\lowerc{}) $\times$ (\men-\upperc{}) & -13.4349 & -13.4517& -13.4181 &  \\
				& (\men\lowerc{}) $\times$ (\women\lowerc{}) & -0.0252 &   -0.042 & -0.0084 &  \\
				& (\men\lowerc{}) $\times$ (\women-\middlec{}) & -3.6189 &  -3.6357&  -3.6021 &  \\
				& (\men\lowerc{}) $\times$ (\women-\upperc{}) & -15.0508 & -15.0676 & -15.034   &  \\
				& (\men-\middlec{}) $\times$ (\men-\upperc{}) & -10.268 & -10.2848& -10.2512  &  \\
				& (\men-\middlec{}) $\times$ (\women\lowerc{}) & 3.1417 &   3.1249  & 3.1585 &  \\
				& (\men-\middlec{}) $\times$ (\women-\middlec{}) & -0.452 &  -0.4688 & -0.4352 &  \\
				& (\men-\middlec{}) $\times$ (\women-\upperc{})& -11.8839 & -11.9007& -11.8671  &  \\
				& (\men-\upperc{}) $\times$ (\women\lowerc{}) & 13.4097 &  13.3929&  13.4265  &  \\
				& (\men-\upperc{}) $\times$ (\women-\middlec{})& 9.816 &   9.7992&   9.8328 &  \\
				& (\men-\upperc{}) $\times$ (\women-\upperc{}) & -1.6159 &  -1.6327 & -1.5991 &  \\
				& (\women\lowerc{}) $\times$ (\women-\middlec{}) &  -3.5937 &  -3.6105 & -3.5769  &  \\
				& (\women\lowerc{}) $\times$ (\women-\upperc{}) & -15.0256 & -15.0424 &-15.0088  &  \\
				& (\women-\middlec{}) $\times$ (\women-\upperc{}) & -11.4319 & -11.4487 &-11.4151  &  \\
				\midrule
				\rowcolor{lightgray}
				\multirow{30}{*}{\work} & (\all) $\times$ (\men) &  0.0109 & -0.0025 & 0.0244   &  0.1356  \\
				& (\all) $\times$ (\women) & -0.5492  & -0.5626 & -0.5357 & \multirow{29}{*}{$^{***}$} \\
				& (\men) $\times$ (\women) & -0.5601  & -0.5735&  -0.5467  &  \\
				& (\all) $\times$ (\lowerc{}) & 79.3821 &  79.3596 &  79.4046 &  \\
				& (\all) $\times$ (\middlec{}) & 76.1565 &   76.134 &   76.179  &  \\
				& (\all) $\times$ (\upperc{}) & 64.9359 &  64.9134 &  64.9584  &  \\
				& (\lowerc{}) $\times$ (\middlec{}) &-3.2256 &  -3.2481 &  -3.2031  &  \\
				& (\lowerc{}) $\times$ (\upperc{}) & -14.4462 & -14.4686 & -14.4237  &  \\
				& (\middlec{}) $\times$ (\upperc{}) & -11.2206 &  -11.243&  -11.1981  &  \\
				& (\all) $\times$ (\men\lowerc{}) & -1.0741 &  -1.1158  & -1.0323  &  \\
				& (\all) $\times$ (\men-\middlec{}) & -4.0935 &  -4.1352 &  -4.0517 &  \\
				& (\all) $\times$ (\men-\upperc{}) & -15.223 & -15.2647 & -15.1813   &  \\
				& (\all) $\times$ (\women\lowerc{}) &-1.306 &  -1.3477 &  -1.2643   &  \\
				& (\all) $\times$ (\women-\middlec{}) & -4.8986 &  -4.9403&   -4.8569  &  \\
				& (\all) $\times$ (\women-\upperc{}) & -17.1903 &  -17.232&  -17.1486   &  \\
				& (\men\lowerc{}) $\times$ (\men-\middlec{}) & -3.0194 &  -3.0611  & -2.9777    &  \\
				& (\men\lowerc{}) $\times$ (\men-\upperc{}) & -14.149 & -14.1907 & -14.1072  &  \\
				& (\men\lowerc{}) $\times$ (\women\lowerc{}) & -0.2319 &  -0.2736 &  -0.1902   &  \\
				& (\men\lowerc{}) $\times$ (\women-\middlec{}) & -3.8245 &  -3.8662 &  -3.7828  &  \\
				& (\men\lowerc{}) $\times$ (\women-\upperc{}) & -16.1162 &  -16.158&  -16.0745   &  \\
				& (\men-\middlec{}) $\times$ (\men-\upperc{}) & -11.1296 & -11.1713&  -11.0878  &  \\
				& (\men-\middlec{}) $\times$ (\women\lowerc{}) &2.7875 &   2.7458&    2.8292  &  \\
				& (\men-\middlec{}) $\times$ (\women-\middlec{}) & -0.8051 &  -0.8468&   -0.7634  &  \\
				& (\men-\middlec{}) $\times$ (\women-\upperc{}) & -13.0968 & -13.1386 & -13.0551  &  \\
				& (\men-\upperc{}) $\times$ (\women\lowerc{}) & 13.917 &  13.8753  & 13.9588  &  \\
				& (\men-\upperc{}) $\times$ (\women-\middlec{}) & 10.3244 &  10.2827  & 10.3662  &  \\
				& (\men-\upperc{}) $\times$ (\women-\upperc{})& -1.9673 &   -2.009 &  -1.9256  &  \\
				& (\women\lowerc{}) $\times$ (\women-\middlec{}) & -3.5926 &  -3.6343 &  -3.5509  &  \\
				& (\women\lowerc{}) $\times$ (\women-\upperc{}) & -15.8843 &  -15.926&  -15.8426  &  \\
				& (\women-\middlec{}) $\times$ (\women-\upperc{}) & -12.2917 & -12.3335 &   -12.25  &  \\
				\bottomrule
			\end{tabular}
		}
	\end{table}
	
	\begin{table}[H]
		\caption{\textbf{Multi-group means comparisons of the distributions of the mobility diversity between different set of travels in \bgt 2012 using Tukey’s HSD test.} See the caption of \ref{tab:mde_2005_tukey}~Table for the description of each column, and the notation.}
		\label{tab:bgt_2012_tukey}
		\centering
		\resizebox{0.70\linewidth}{!}{
			\begin{tabular}{l|l|r|rr|c}
				\toprule
				\multicolumn{1}{c|}{Travels} & \multicolumn{1}{c|}{Groups} & \multicolumn{1}{c|}{Mean} & \multicolumn{2}{c|}{95\% Confidence interval} &\multicolumn{1}{c}{Adjusted} \\
				& & \multicolumn{1}{c|}{difference} & Lower bound & Upper bound &\multicolumn{1}{c}{$p$-value} \\
				\midrule
				\multirow{30}{*}{\all} & (\all) $\times$ (\men) &   -0.0497 &   -0.05 &-0.0494  & \multirow{30}{*}{$^{***}$} \\
				& (\all) $\times$ (\women) & -0.4077 &  -0.408 &-0.4074  &  \\
				& (\men) $\times$ (\women) & -0.358 & -0.3583 & -0.3577  &  \\
				& (\all) $\times$ (\lowerc{})& 87.0619 &  87.0609 & 87.0629  &  \\
				& (\all) $\times$ (\middlec{}) & 89.0512 &  89.0502 & 89.0522   &  \\
				& (\all) $\times$ (\upperc{})& 71.4057 &  71.4046 & 71.4067  &  \\
				& (\lowerc{}) $\times$ (\middlec{}) & 1.9893 &   1.9883 &  1.9903 &  \\
				& (\lowerc{}) $\times$ (\upperc{}) & -15.6563 & -15.6573 &-15.6552 &  \\
				& (\middlec{}) $\times$ (\upperc{})& -17.6455 & -17.6466& -17.6445   &  \\
				& (\all) $\times$ (\men\lowerc{})& -6.1861 &  -6.1878  &-6.1844 &  \\
				& (\all) $\times$ (\men-\middlec{}) & -4.2557 &  -4.2574  & -4.254  &  \\
				& (\all) $\times$ (\men-\upperc{}) &  -21.8737 & -21.8754 & -21.872   &  \\
				& (\all) $\times$ (\women\lowerc{}) & -6.812 &  -6.8137 & -6.8103    &  \\
				& (\all) $\times$ (\women-\middlec{}) & -4.743 &  -4.7446 & -4.7413 &  \\
				& (\all) $\times$ (\women-\upperc{}) & -23.4137 & -23.4154 & -23.412  &  \\
				& (\men\lowerc{}) $\times$ (\men-\middlec{}) & 1.9304 &   1.9287  & 1.9321  &  \\
				& (\men\lowerc{}) $\times$ (\men-\upperc{}) &-15.6875 & -15.6892& -15.6858  &  \\
				& (\men\lowerc{}) $\times$ (\women\lowerc{}) &  -0.6258 &  -0.6275 & -0.6241  &  \\
				& (\men\lowerc{}) $\times$ (\women-\middlec{}) & 1.4432 &   1.4415 &  1.4449  &  \\
				& (\men\lowerc{}) $\times$ (\women-\upperc{}) & -17.2275 & -17.2292 &-17.2258  &  \\
				& (\men-\middlec{}) $\times$ (\men-\upperc{}) & -17.618 & -17.6196 &-17.6163  &  \\
				& (\men-\middlec{}) $\times$ (\women\lowerc{}) & -2.5562 &  -2.5579 & -2.5545   &  \\
				& (\men-\middlec{}) $\times$ (\women-\middlec{}) & -0.4872 &  -0.4889 & -0.4855  &  \\
				& (\men-\middlec{}) $\times$ (\women-\upperc{})& -19.1579 & -19.1596& -19.1563   &  \\
				& (\men-\upperc{}) $\times$ (\women\lowerc{}) & 15.0617 &    15.06 & 15.0634 &  \\
				& (\men-\upperc{}) $\times$ (\women-\middlec{})& 17.1307 &   17.129 & 17.1324 &  \\
				& (\men-\upperc{}) $\times$ (\women-\upperc{}) &-1.54 &  -1.5417 & -1.5383  &  \\
				& (\women\lowerc{}) $\times$ (\women-\middlec{}) & 2.069 &   2.0673  & 2.0707  &  \\
				& (\women\lowerc{}) $\times$ (\women-\upperc{}) & -16.6017 & -16.6034  &  -16.6    &  \\
				& (\women-\middlec{}) $\times$ (\women-\upperc{}) & -18.6707 & -18.6724 & -18.669  &  \\
				\midrule
				\multirow{30}{*}{\work} & (\all) $\times$ (\men) &  -0.9601 &-0.9612 & -0.9589   & \multirow{30}{*}{$^{***}$} \\
				& (\all) $\times$ (\women) & -1.5977 &-1.5989 & -1.5966  &  \\
				& (\men) $\times$ (\women) & -0.6377 &-0.6388 & -0.6365  &  \\
				& (\all) $\times$ (\lowerc{}) & 88.1965 & 88.1941 &  88.1989  &  \\
				& (\all) $\times$ (\middlec{}) & 85.972 & 85.9696 &  85.9743  &  \\
				& (\all) $\times$ (\upperc{}) & 72.908 & 72.9056 &  72.9104  &  \\
				& (\lowerc{}) $\times$ (\middlec{}) & -2.2246 &  -2.227 &  -2.2222  &  \\
				& (\lowerc{}) $\times$ (\upperc{}) & -15.2885 &-15.2909&  -15.2861  &  \\
				& (\middlec{}) $\times$ (\upperc{}) & -13.0639 &-13.0663&  -13.0615  &  \\
				& (\all) $\times$ (\men\lowerc{}) & -2.6924 & -2.6965 &  -2.6884  &  \\
				& (\all) $\times$ (\men-\middlec{}) & -4.6356 & -4.6396 &  -4.6315 &  \\
				& (\all) $\times$ (\men-\upperc{}) & -18.4256 &-18.4297&  -18.4216  &  \\
				& (\all) $\times$ (\women\lowerc{}) & -3.8732 & -3.8773  & -3.8692 &  \\
				& (\all) $\times$ (\women-\middlec{}) & -5.4457 & -5.4497 &  -5.4416 &  \\
				& (\all) $\times$ (\women-\upperc{}) & -21.8152 &-21.8192 & -21.8111  &  \\
				& (\men\lowerc{}) $\times$ (\men-\middlec{}) & -1.9431 & -1.9472  & -1.9391 &  \\
				& (\men\lowerc{}) $\times$ (\men-\upperc{}) & -15.7332 &-15.7373 & -15.7292  &  \\
				& (\men\lowerc{}) $\times$ (\women\lowerc{}) & -1.1808 & -1.1849  & -1.1768 &  \\
				& (\men\lowerc{}) $\times$ (\women-\middlec{}) & -2.7533 & -2.7573  & -2.7492   &  \\
				& (\men\lowerc{}) $\times$ (\women-\upperc{}) & -19.1227 &-19.1268&  -19.1187 &  \\
				& (\men-\middlec{}) $\times$ (\men-\upperc{}) & -13.7901 &-13.7941  & -13.786 &  \\
				& (\men-\middlec{}) $\times$ (\women\lowerc{}) & 0.7623 &  0.7583  &  0.7664 &  \\
				& (\men-\middlec{}) $\times$ (\women-\middlec{}) & -0.8101 & -0.8142  & -0.8061 &  \\
				& (\men-\middlec{}) $\times$ (\women-\upperc{}) & -17.1796 &-17.1836&  -17.1755  &  \\
				& (\men-\upperc{}) $\times$ (\women\lowerc{}) & 14.5524 & 14.5483&   14.5564  &  \\
				& (\men-\upperc{}) $\times$ (\women-\middlec{}) & 12.98 & 12.9759  &  12.984 &  \\
				& (\men-\upperc{}) $\times$ (\women-\upperc{})& -3.3895 & -3.3936&   -3.3855  &  \\
				& (\women\lowerc{}) $\times$ (\women-\middlec{}) & -1.5724 & -1.5765 &  -1.5684   &  \\
				& (\women\lowerc{}) $\times$ (\women-\upperc{}) & -17.9419 & -17.946&  -17.9379  &  \\
				& (\women-\middlec{}) $\times$ (\women-\upperc{}) & -16.3695 & -16.3735&  -16.3654  &  \\
				\bottomrule
			\end{tabular}
		}
	\end{table}

	\begin{table}[H]
		\caption{\textbf{Multi-group means comparisons of the distributions of the mobility diversity between different set of travels in \bgt 2019 using Tukey’s HSD test.} See the caption of \ref{tab:mde_2005_tukey}~Table for the description of each column, and the notation.}
		\label{tab:bgt_2019_tukey}
		\centering
		\resizebox{0.70\linewidth}{!}{
			\begin{tabular}{l|l|r|rr|c}
				\toprule
				\multicolumn{1}{c|}{Travels} & \multicolumn{1}{c|}{Groups} & \multicolumn{1}{c|}{Mean} & \multicolumn{2}{c|}{95\% Confidence interval} &\multicolumn{1}{c}{Adjusted} \\
				& & \multicolumn{1}{c|}{difference} & Lower bound & Upper bound &\multicolumn{1}{c}{$p$-value} \\
				\midrule
				\multirow{30}{*}{\all} & (\all) $\times$ (\men) &   -0.0376 & -0.038 & -0.0373  & \multirow{30}{*}{$^{***}$} \\
				& (\all) $\times$ (\women) & -0.0561 & -0.0564 &-0.0557  &  \\
				& (\men) $\times$ (\women) &  -0.0184 & -0.0188 & -0.0181  &  \\
				& (\all) $\times$ (\lowerc{})& 81.0821 & 81.0813 &  81.083  &  \\
				& (\all) $\times$ (\middlec{}) & 81.8183 & 81.8174 & 81.8191  &  \\
				& (\all) $\times$ (\upperc{})& 78.0621 & 78.0612 &  78.063   &  \\
				& (\lowerc{}) $\times$ (\middlec{}) & 0.7361 &  0.7353 &   0.737  &  \\
				& (\lowerc{}) $\times$ (\upperc{}) & -3.0201 & -3.0209 & -3.0192  &  \\
				& (\middlec{}) $\times$ (\upperc{})&-3.7562 &  -3.757 & -3.7553 &  \\
				& (\all) $\times$ (\men\lowerc{})& -0.8819 & -0.8835 & -0.8804  &  \\
				& (\all) $\times$ (\men-\middlec{}) & -0.3397 & -0.3413 & -0.3382   &  \\
				& (\all) $\times$ (\men-\upperc{}) & -4.8074 & -4.8089 & -4.8059  &  \\
				& (\all) $\times$ (\women\lowerc{}) & -1.0872 & -1.0888 & -1.0857  &  \\
				& (\all) $\times$ (\women-\middlec{}) & -0.1996 & -0.2012&  -0.1981  &  \\
				& (\all) $\times$ (\women-\upperc{}) & -4.9056 & -4.9071&   -4.904 &  \\
				& (\men\lowerc{}) $\times$ (\men-\middlec{}) & 0.5422 &  0.5407 &  0.5437  &  \\
				& (\men\lowerc{}) $\times$ (\men-\upperc{}) & -3.9255 &  -3.927 & -3.9239  &  \\
				& (\men\lowerc{}) $\times$ (\women\lowerc{}) & -0.2053 & -0.2068 & -0.2038   &  \\
				& (\men\lowerc{}) $\times$ (\women-\middlec{}) & 0.6823 &  0.6808 &  0.6838 &  \\
				& (\men\lowerc{}) $\times$ (\women-\upperc{}) & -4.0236 & -4.0252 & -4.0221  &  \\
				& (\men-\middlec{}) $\times$ (\men-\upperc{}) & -4.4677 & -4.4692 & -4.4661  &  \\
				& (\men-\middlec{}) $\times$ (\women\lowerc{}) & -0.7475 &  -0.749 &  -0.746  &  \\
				& (\men-\middlec{}) $\times$ (\women-\middlec{}) & 0.1401 &  0.1386 &  0.1416   &  \\
				& (\men-\middlec{}) $\times$ (\women-\upperc{})& -4.5658 & -4.5674&  -4.5643    &  \\
				& (\men-\upperc{}) $\times$ (\women\lowerc{}) & 3.7202 &  3.7186 &  3.7217   &  \\
				& (\men-\upperc{}) $\times$ (\women-\middlec{})&  4.6078 &  4.6062 &  4.6093  &  \\
				& (\men-\upperc{}) $\times$ (\women-\upperc{}) & -0.0982 & -0.0997 & -0.0966   &  \\
				& (\women\lowerc{}) $\times$ (\women-\middlec{}) &0.8876 &  0.8861 &  0.8891    &  \\
				& (\women\lowerc{}) $\times$ (\women-\upperc{}) & -3.8183 & -3.8199 & -3.8168  &  \\
				& (\women-\middlec{}) $\times$ (\women-\upperc{}) & -4.706 & -4.7075 & -4.7044  &  \\
				\midrule
				\multirow{30}{*}{\work} & (\all) $\times$ (\men) &  -0.3398 & -0.3409 & -0.3387  & \multirow{30}{*}{$^{***}$} \\
				& (\all) $\times$ (\women) &  -0.523 & -0.5242 & -0.5219  &  \\
				& (\men) $\times$ (\women) & -0.1833 & -0.1844&  -0.1822   &  \\
				& (\all) $\times$ (\lowerc{}) & 78.9254 & 78.9227 & 78.9281  &  \\
				& (\all) $\times$ (\middlec{}) &  80.3745 & 80.3718&  80.3772 &  \\
				& (\all) $\times$ (\upperc{}) &  71.1352 & 71.1325 & 71.1379  &  \\
				& (\lowerc{}) $\times$ (\middlec{}) &  1.4491 &  1.4465 &  1.4518  &  \\
				& (\lowerc{}) $\times$ (\upperc{}) &  -7.7902 & -7.7929 & -7.7875 &  \\
				& (\middlec{}) $\times$ (\upperc{}) & -9.2393 &  -9.242 & -9.2366  &  \\
				& (\all) $\times$ (\men\lowerc{}) &  -2.5151 &  -2.5197 &  -2.5104  &  \\
				& (\all) $\times$ (\men-\middlec{}) &  -1.3778 &  -1.3825 &  -1.3731 &  \\
				& (\all) $\times$ (\men-\upperc{}) &   -14.0293 &  -14.034 & -14.0246 &  \\
				& (\all) $\times$ (\women\lowerc{}) &  -3.0307 &  -3.0354  &  -3.026   &  \\
				& (\all) $\times$ (\women-\middlec{}) & -1.4085 &  -1.4132  & -1.4038 &  \\
				& (\all) $\times$ (\women-\upperc{}) &  -14.4171 & -14.4218&  -14.4125  &  \\
				& (\men\lowerc{}) $\times$ (\men-\middlec{}) &   1.1372 &   1.1326  &  1.1419 &  \\
				& (\men\lowerc{}) $\times$ (\men-\upperc{}) & -11.5143 & -11.5189 & -11.5096 &  \\
				& (\men\lowerc{}) $\times$ (\women\lowerc{}) & -0.5156 &  -0.5203 &  -0.5109  &  \\
				& (\men\lowerc{}) $\times$ (\women-\middlec{}) &   1.1066 &   1.1019  &  1.1113  &  \\
				& (\men\lowerc{}) $\times$ (\women-\upperc{}) &  -11.9021 & -11.9068&  -11.8974  &  \\
				& (\men-\middlec{}) $\times$ (\men-\upperc{}) &  -12.6515 & -12.6562&  -12.6468  &  \\
				& (\men-\middlec{}) $\times$ (\women\lowerc{}) &  -1.6529 &  -1.6576  & -1.6482   &  \\
				& (\men-\middlec{}) $\times$ (\women-\middlec{}) &  -0.0307 &  -0.0354  &  -0.026 &  \\
				& (\men-\middlec{}) $\times$ (\women-\upperc{}) & -13.0393 &  -13.044 & -13.0346 &  \\
				& (\men-\upperc{}) $\times$ (\women\lowerc{}) & 10.9986 &  10.9939  & 11.0033    &  \\
				& (\men-\upperc{}) $\times$ (\women-\middlec{}) & 12.6208 &  12.6161  & 12.6255   &  \\
				& (\men-\upperc{}) $\times$ (\women-\upperc{})& -0.3878 &  -0.3925 &  -0.3831  &  \\
				& (\women\lowerc{}) $\times$ (\women-\middlec{}) & 1.6222 &   1.6175  &  1.6269  &  \\
				& (\women\lowerc{}) $\times$ (\women-\upperc{}) & -11.3865 & -11.3911 & -11.3818  &  \\
				& (\women-\middlec{}) $\times$ (\women-\upperc{}) & -13.0087 & -13.0133  & -13.004 &  \\
				\bottomrule
			\end{tabular}
		}
	\end{table}

	\begin{table}[H]
		\caption{\textbf{Multi-group means comparisons of the distributions of the mobility diversity between different set of travels in \sao 1997 using Tukey’s HSD test.} See the caption of \ref{tab:mde_2005_tukey}~Table for the description of each column, and the notation.}
		\label{tab:sao_1997_tukey}
		\centering
		\resizebox{0.70\linewidth}{!}{
			\begin{tabular}{l|l|r|rr|c}
				\toprule
				\multicolumn{1}{c|}{Travels} & \multicolumn{1}{c|}{Groups} & \multicolumn{1}{c|}{Mean} & \multicolumn{2}{c|}{95\% Confidence interval} &\multicolumn{1}{c}{Adjusted} \\
				& & \multicolumn{1}{c|}{difference} & Lower bound & Upper bound &\multicolumn{1}{c}{$p$-value} \\
				\midrule
				\multirow{30}{*}{\all} & (\all) $\times$ (\men) &  -0.1254 & -0.1257   &-0.1251  & \multirow{30}{*}{$^{***}$} \\
				& (\all) $\times$ (\women) &  -0.0272 & -0.0275  & -0.0269  &  \\
				& (\men) $\times$ (\women) & 0.0981 &  0.0978  &  0.0984   &  \\
				& (\all) $\times$ (\lowerc{})& 88.7281 & 88.7272   &88.7289   &  \\
				& (\all) $\times$ (\middlec{}) & 93.0693 & 93.0684   &93.0702   &  \\
				& (\all) $\times$ (\upperc{})& 84.8127 & 84.8118   &84.8136  &  \\
				& (\lowerc{}) $\times$ (\middlec{}) & 4.3413 &  4.3404    &4.3422  &  \\
				& (\lowerc{}) $\times$ (\upperc{}) & -3.9154 & -3.9163   &-3.9145  &  \\
				& (\middlec{}) $\times$ (\upperc{})& -8.2567 & -8.2575   &-8.2558  &  \\
				& (\all) $\times$ (\men\lowerc{})&  -4.627 & -4.6286  & -4.6255   &  \\
				& (\all) $\times$ (\men-\middlec{}) & -0.3233 & -0.3248  & -0.3218  &  \\
				& (\all) $\times$ (\men-\upperc{}) & -8.3645 &  -8.366   &-8.3629 &  \\
				& (\all) $\times$ (\women\lowerc{}) & -4.7965 & -4.7981   & -4.795 &  \\
				& (\all) $\times$ (\women-\middlec{}) & -0.2844 &  -0.286  & -0.2829  &  \\
				& (\all) $\times$ (\women-\upperc{}) & -9.9697 & -9.9712  & -9.9681  &  \\
				& (\men\lowerc{}) $\times$ (\men-\middlec{}) & 4.3037 &  4.3022   & 4.3053  &  \\
				& (\men\lowerc{}) $\times$ (\men-\upperc{}) & -3.7374 & -3.7389  & -3.7359  &  \\
				& (\men\lowerc{}) $\times$ (\women\lowerc{}) & -0.1695 &  -0.171   & -0.168   &  \\
				& (\men\lowerc{}) $\times$ (\women-\middlec{}) &  4.3426 &  4.3411   & 4.3441  &  \\
				& (\men\lowerc{}) $\times$ (\women-\upperc{}) & -5.3426 & -5.3442   &-5.3411 &  \\
				& (\men-\middlec{}) $\times$ (\men-\upperc{}) & -8.0412 & -8.0427   &-8.0396 &  \\
				& (\men-\middlec{}) $\times$ (\women\lowerc{}) & -4.4732 & -4.4748   &-4.4717 &  \\
				& (\men-\middlec{}) $\times$ (\women-\middlec{}) & 0.0388 &  0.0373    &0.0404  &  \\
				& (\men-\middlec{}) $\times$ (\women-\upperc{})& -9.6464 & -9.6479   &-9.6448  &  \\
				& (\men-\upperc{}) $\times$ (\women\lowerc{}) & 3.5679 &  3.5664   & 3.5695  &  \\
				& (\men-\upperc{}) $\times$ (\women-\middlec{})& 8.08 &  8.0785   & 8.0815  &  \\
				& (\men-\upperc{}) $\times$ (\women-\upperc{}) & -1.6052 & -1.6068   &-1.6037  &  \\
				& (\women\lowerc{}) $\times$ (\women-\middlec{}) & 4.5121 &  4.5105   & 4.5136  &  \\
				& (\women\lowerc{}) $\times$ (\women-\upperc{}) & -5.1731 & -5.1747   &-5.1716  &  \\
				& (\women-\middlec{}) $\times$ (\women-\upperc{}) & -9.6852 & -9.6868   &-9.6837  &  \\
				\midrule
				\multirow{30}{*}{\work} & (\all) $\times$ (\men) &  -0.5365 & -0.5373   &-0.5357   & \multirow{30}{*}{$^{***}$} \\
				& (\all) $\times$ (\women) & -0.2166 & -0.2175   &-0.2158  &  \\
				& (\men) $\times$ (\women) & 0.3199 &   0.319   & 0.3207  &  \\
				& (\all) $\times$ (\lowerc{}) & 92.4728 &  92.4706   & 92.4749 &  \\
				& (\all) $\times$ (\middlec{}) & 90.9687 &  90.9666   & 90.9709   &  \\
				& (\all) $\times$ (\upperc{}) & 81.3344 &  81.3323   & 81.3366  &  \\
				& (\lowerc{}) $\times$ (\middlec{}) & -1.504 &  -1.5062  &  -1.5019 &  \\
				& (\lowerc{}) $\times$ (\upperc{}) & -11.1383 & -11.1405   &-11.1362  &  \\
				& (\middlec{}) $\times$ (\upperc{}) & -9.6343 &  -9.6364   & -9.6321  &  \\
				& (\all) $\times$ (\men\lowerc{}) & -0.6123 &  -0.6164   & -0.6082  &  \\
				& (\all) $\times$ (\men-\middlec{}) & -1.9421 &  -1.9462    &-1.9381  &  \\
				& (\all) $\times$ (\men-\upperc{}) & -12.2619 & -12.2659   &-12.2578  &  \\
				& (\all) $\times$ (\women\lowerc{}) & -0.7635 &  -0.7676    &-0.7594  &  \\
				& (\all) $\times$ (\women-\middlec{}) &-1.6211 &  -1.6252   & -1.6171    &  \\
				& (\all) $\times$ (\women-\upperc{}) & -15.3371 & -15.3412   & -15.333   &  \\
				& (\men\lowerc{}) $\times$ (\men-\middlec{}) & -1.3298 &  -1.3339  &  -1.3258  &  \\
				& (\men\lowerc{}) $\times$ (\men-\upperc{}) & -11.6496 & -11.6536   &-11.6455   &  \\
				& (\men\lowerc{}) $\times$ (\women\lowerc{}) & -0.1512 &  -0.1553  &  -0.1471  &  \\
				& (\men\lowerc{}) $\times$ (\women-\middlec{}) &-1.0088 &  -1.0129   & -1.0048  &  \\
				& (\men\lowerc{}) $\times$ (\women-\upperc{}) & -14.7248 & -14.7289  & -14.7207  &  \\
				& (\men-\middlec{}) $\times$ (\men-\upperc{}) & -10.3197 & -10.3238  & -10.3156  &  \\
				& (\men-\middlec{}) $\times$ (\women\lowerc{}) & 1.1786 &   1.1745   &  1.1827  &  \\
				& (\men-\middlec{}) $\times$ (\women-\middlec{}) &  0.321 &   0.3169  &   0.3251  &  \\
				& (\men-\middlec{}) $\times$ (\women-\upperc{}) & -13.395 &  -13.399   &-13.3909  &  \\
				& (\men-\upperc{}) $\times$ (\women\lowerc{}) &  11.4983 &  11.4943   & 11.5024   &  \\
				& (\men-\upperc{}) $\times$ (\women-\middlec{}) & 10.6407 &  10.6366   & 10.6448 &  \\
				& (\men-\upperc{}) $\times$ (\women-\upperc{})& -3.0752 &  -3.0793   & -3.0712  &  \\
				& (\women\lowerc{}) $\times$ (\women-\middlec{}) & -0.8576 &  -0.8617   & -0.8535  &  \\
				& (\women\lowerc{}) $\times$ (\women-\upperc{}) & -14.5736 & -14.5777  & -14.5695  &  \\
				& (\women-\middlec{}) $\times$ (\women-\upperc{}) & -13.716 &   -13.72  & -13.7119  &  \\
				\bottomrule
			\end{tabular}
		}
	\end{table}

	\begin{table}[H]
		\caption{\textbf{Multi-group means comparisons of the distributions of the mobility diversity between different set of travels in \sao 2007 using Tukey’s HSD test.} See the caption of \ref{tab:mde_2005_tukey}~Table for the description of each column, and the notation.}
		\label{tab:sao_2007_tukey}
		\centering
		\resizebox{0.70\linewidth}{!}{
			\begin{tabular}{l|l|r|rr|c}
				\toprule
				\multicolumn{1}{c|}{Travels} & \multicolumn{1}{c|}{Groups} & \multicolumn{1}{c|}{Mean} & \multicolumn{2}{c|}{95\% Confidence interval} &\multicolumn{1}{c}{Adjusted} \\
				& & \multicolumn{1}{c|}{difference} & Lower bound & Upper bound &\multicolumn{1}{c}{$p$-value} \\
				\midrule
				\multirow{30}{*}{\all} & (\all) $\times$ (\men) &  -0.2582 & -0.2584 & -0.2579  & \multirow{30}{*}{$^{***}$} \\
				& (\all) $\times$ (\women) & 0.164 &  0.1637  & 0.1642  &  \\
				& (\men) $\times$ (\women) & 0.4221 &  0.4219 &  0.4224  &  \\
				& (\all) $\times$ (\lowerc{})& 88.067 & 88.0665 & 88.0675  &  \\
				& (\all) $\times$ (\middlec{}) & 90.6562 & 90.6557 & 90.6568  &  \\
				& (\all) $\times$ (\upperc{})& 88.4965 &  88.496 &  88.497  &  \\
				& (\lowerc{}) $\times$ (\middlec{}) & 2.5893 &  2.5887 &  2.5898   &  \\
				& (\lowerc{}) $\times$ (\upperc{}) & 0.4295 &   0.429  &   0.43  &  \\
				& (\middlec{}) $\times$ (\upperc{})& -2.1598 & -2.1603 & -2.1593   &  \\
				& (\all) $\times$ (\men\lowerc{})& -3.2646 & -3.2655 & -3.2637  &  \\
				& (\all) $\times$ (\men-\middlec{}) & -0.6351 &  -0.636 & -0.6343  &  \\
				& (\all) $\times$ (\men-\upperc{}) &  -2.562 & -2.5628 & -2.5611  &  \\
				& (\all) $\times$ (\women\lowerc{}) & -2.775 & -2.7759 & -2.7741  &  \\
				& (\all) $\times$ (\women-\middlec{}) & -0.0381 &  -0.039 & -0.0373  &  \\
				& (\all) $\times$ (\women-\upperc{}) & -2.9615 & -2.9624 & -2.9606 &  \\
				& (\men\lowerc{}) $\times$ (\men-\middlec{}) & 2.6294 &  2.6286&   2.6303   &  \\
				& (\men\lowerc{}) $\times$ (\men-\upperc{}) & 0.7026 &  0.7018&   0.7035  &  \\
				& (\men\lowerc{}) $\times$ (\women\lowerc{}) & 0.4896 &  0.4887 &  0.4905   &  \\
				& (\men\lowerc{}) $\times$ (\women-\middlec{}) & 3.2265 &  3.2256 &  3.2273  &  \\
				& (\men\lowerc{}) $\times$ (\women-\upperc{}) &  0.3031 &  0.3022 &   0.304 &  \\
				& (\men-\middlec{}) $\times$ (\men-\upperc{}) & -1.9268 & -1.9277 & -1.9259 &  \\
				& (\men-\middlec{}) $\times$ (\women\lowerc{}) & -2.1398 & -2.1407 &  -2.139  &  \\
				& (\men-\middlec{}) $\times$ (\women-\middlec{}) & 0.597 &  0.5961 &  0.5979   &  \\
				& (\men-\middlec{}) $\times$ (\women-\upperc{})& -2.3264 & -2.3272&  -2.3255  &  \\
				& (\men-\upperc{}) $\times$ (\women\lowerc{}) & -0.213 & -0.2139 & -0.2121 &  \\
				& (\men-\upperc{}) $\times$ (\women-\middlec{})& 2.5238 &  2.5229  & 2.5247  &  \\
				& (\men-\upperc{}) $\times$ (\women-\upperc{}) & -0.3996 & -0.4004 & -0.3987  &  \\
				& (\women\lowerc{}) $\times$ (\women-\middlec{}) &2.7368 &   2.736 &  2.7377 &  \\
				& (\women\lowerc{}) $\times$ (\women-\upperc{}) & -0.1865 & -0.1874 &  -0.1857  &  \\
				& (\women-\middlec{}) $\times$ (\women-\upperc{}) & -2.9234 & -2.9242&  -2.9225  &  \\
				\midrule
				\multirow{30}{*}{\work} & (\all) $\times$ (\men) &  -0.3503 & -0.3509 & -0.3497  & \multirow{30}{*}{$^{***}$} \\
				& (\all) $\times$ (\women) & 0.1219 &  0.1213 &  0.1226   &  \\
				& (\men) $\times$ (\women) & 0.4722 &  0.4716 &  0.4729   &  \\
				& (\all) $\times$ (\lowerc{}) &  86.5667 & 86.5655 & 86.5679 &  \\
				& (\all) $\times$ (\middlec{}) & 89.7687 & 89.7674 & 89.7699  &  \\
				& (\all) $\times$ (\upperc{}) & 87.5448 & 87.5436  & 87.546  &  \\
				& (\lowerc{}) $\times$ (\middlec{}) & 3.202 &  3.2007 &  3.2032  &  \\
				& (\lowerc{}) $\times$ (\upperc{}) & 0.9781 &  0.9768 &  0.9793 &  \\
				& (\middlec{}) $\times$ (\upperc{}) & -2.2239 & -2.2251 & -2.2227 &  \\
				& (\all) $\times$ (\men\lowerc{}) & -4.2172 & -4.2194 & -4.2151   &  \\
				& (\all) $\times$ (\men-\middlec{}) & -0.8862 & -0.8884 & -0.8841  &  \\
				& (\all) $\times$ (\men-\upperc{}) & -2.8996 & -2.9017 & -2.8975   &  \\
				& (\all) $\times$ (\women\lowerc{}) & -3.8529 &  -3.855 & -3.8508  &  \\
				& (\all) $\times$ (\women-\middlec{}) & -0.1643 & -0.1665 & -0.1622   &  \\
				& (\all) $\times$ (\women-\upperc{}) & -3.8952 & -3.8974&  -3.8931  &  \\
				& (\men\lowerc{}) $\times$ (\men-\middlec{}) & 3.331 &  3.3289&   3.3331   &  \\
				& (\men\lowerc{}) $\times$ (\men-\upperc{}) & 1.3176 &  1.3155 &  1.3198  &  \\
				& (\men\lowerc{}) $\times$ (\women\lowerc{}) &  0.3643 &  0.3622 &  0.3665 &  \\
				& (\men\lowerc{}) $\times$ (\women-\middlec{}) & 4.0529 &  4.0508 &   4.055  &  \\
				& (\men\lowerc{}) $\times$ (\women-\upperc{}) & 0.322 &  0.3199 &  0.3241   &  \\
				& (\men-\middlec{}) $\times$ (\men-\upperc{}) & -2.0134 & -2.0155 & -2.0112    &  \\
				& (\men-\middlec{}) $\times$ (\women\lowerc{}) & -2.9667 & -2.9688 & -2.9645  &  \\
				& (\men-\middlec{}) $\times$ (\women-\middlec{}) & 0.7219 &  0.7198 &   0.724   &  \\
				& (\men-\middlec{}) $\times$ (\women-\upperc{}) & -3.009 & -3.0111 & -3.0069 &  \\
				& (\men-\upperc{}) $\times$ (\women\lowerc{}) &  -0.9533 & -0.9554 & -0.9512    &  \\
				& (\men-\upperc{}) $\times$ (\women-\middlec{}) & 2.7353 &  2.7331  & 2.7374   &  \\
				& (\men-\upperc{}) $\times$ (\women-\upperc{})& -0.9956 & -0.9978 & -0.9935  &  \\
				& (\women\lowerc{}) $\times$ (\women-\middlec{}) & 3.6886 &  3.6864 &  3.6907  &  \\
				& (\women\lowerc{}) $\times$ (\women-\upperc{}) & -0.0423 & -0.0445 & -0.0402   &  \\
				& (\women-\middlec{}) $\times$ (\women-\upperc{}) & -3.7309 &  -3.733&  -3.7288  &  \\
				\bottomrule
			\end{tabular}
		}
	\end{table}

	\begin{table}[H]
		\caption{\textbf{Multi-group means comparisons of the distributions of the mobility diversity between different set of travels in \sao 2017 using Tukey’s HSD test.} See the caption of \ref{tab:mde_2005_tukey}~Table for the description of each column, and the notation.}
		\label{tab:sao_2017_tukey}
		\centering
		\resizebox{0.70\linewidth}{!}{
			\begin{tabular}{l|l|r|rr|c}
				\toprule
				\multicolumn{1}{c|}{Travels} & \multicolumn{1}{c|}{Groups} & \multicolumn{1}{c|}{Mean} & \multicolumn{2}{c|}{95\% Confidence interval} &\multicolumn{1}{c}{Adjusted} \\
				& & \multicolumn{1}{c|}{difference} & Lower bound & Upper bound &\multicolumn{1}{c}{$p$-value} \\
				\midrule
				\multirow{30}{*}{\all} & (\all) $\times$ (\men) &  0.0704 &  0.0702 & 0.0707  & \multirow{30}{*}{$^{***}$} \\
				& (\all) $\times$ (\women) & -0.1485 & -0.1488& -0.1483  &  \\
				& (\men) $\times$ (\women) & -0.219 & -0.2192 &-0.2187 &  \\
				& (\all) $\times$ (\lowerc{})& 90.3673 & 90.3668 &90.3678  &  \\
				& (\all) $\times$ (\middlec{}) &  92.373 & 92.3725 &92.3734  &  \\
				& (\all) $\times$ (\upperc{})& 88.2008 & 88.2004 &88.2013  &  \\
				& (\lowerc{}) $\times$ (\middlec{}) &  2.0057 &  2.0052 & 2.0061   &  \\
				& (\lowerc{}) $\times$ (\upperc{}) & -2.1665 & -2.1669 & -2.166  &  \\
				& (\middlec{}) $\times$ (\upperc{})& -4.1721 & -4.1726 &-4.1717  &  \\
				& (\all) $\times$ (\men\lowerc{})& -2.3309 & -2.3317 &-2.3301 &  \\
				& (\all) $\times$ (\men-\middlec{}) & -0.3343 & -0.3351 &-0.3335   &  \\
				& (\all) $\times$ (\men-\upperc{}) & -4.4809 & -4.4817 &-4.4801  &  \\
				& (\all) $\times$ (\women\lowerc{}) & -2.6799 & -2.6807 &-2.6791 &  \\
				& (\all) $\times$ (\women-\middlec{}) & -0.4041 & -0.4049 &-0.4033  &  \\
				& (\all) $\times$ (\women-\upperc{}) & -5.132 & -5.1328 &-5.1312 &  \\
				& (\men\lowerc{}) $\times$ (\men-\middlec{}) & 1.9966 &  1.9958 &  1.9974   &  \\
				& (\men\lowerc{}) $\times$ (\men-\upperc{}) & -2.15 & -2.1508 &-2.1492  &  \\
				& (\men\lowerc{}) $\times$ (\women\lowerc{}) & -0.3489 & -0.3497 &-0.3481 &  \\
				& (\men\lowerc{}) $\times$ (\women-\middlec{}) & 1.9268 &   1.926 & 1.9276 &  \\
				& (\men\lowerc{}) $\times$ (\women-\upperc{}) & -2.8011 & -2.8019 &-2.8003  &  \\
				& (\men-\middlec{}) $\times$ (\men-\upperc{}) & -4.1466 & -4.1474 &-4.1458 &  \\
				& (\men-\middlec{}) $\times$ (\women\lowerc{}) & -2.3456 & -2.3464 &-2.3448  &  \\
				& (\men-\middlec{}) $\times$ (\women-\middlec{}) & -0.0698 & -0.0706 & -0.069  &  \\
				& (\men-\middlec{}) $\times$ (\women-\upperc{})& -4.7977 & -4.7985 &-4.7969  &  \\
				& (\men-\upperc{}) $\times$ (\women\lowerc{}) &1.801 &  1.8002 & 1.8018  &  \\
				& (\men-\upperc{}) $\times$ (\women-\middlec{})& 4.0768 &   4.076  &4.0776  &  \\
				& (\men-\upperc{}) $\times$ (\women-\upperc{}) & -0.6511 & -0.6519 &-0.6503  &  \\
				& (\women\lowerc{}) $\times$ (\women-\middlec{}) & 2.2757 &  2.2749 & 2.2765   &  \\
				& (\women\lowerc{}) $\times$ (\women-\upperc{}) & -2.4522 &  -2.453 &-2.4514  &  \\
				& (\women-\middlec{}) $\times$ (\women-\upperc{}) & -4.7279 & -4.7287& -4.7271  &  \\
				\midrule
				\multirow{30}{*}{\work} & (\all) $\times$ (\men) &  0.116 &  0.1154 & 0.1165  & \multirow{30}{*}{$^{***}$} \\
				& (\all) $\times$ (\women) &-0.4142 & -0.4147 &-0.4136 &  \\
				& (\men) $\times$ (\women) & -0.5301 & -0.5307 &-0.5296  &  \\
				& (\all) $\times$ (\lowerc{}) & 89.0198 & 89.0187 &89.0209   &  \\
				& (\all) $\times$ (\middlec{}) & 91.8841 &  91.883 &91.8851  &  \\
				& (\all) $\times$ (\upperc{}) & 86.0713 & 86.0702 &86.0724  &  \\
				& (\lowerc{}) $\times$ (\middlec{}) & 2.8643 &  2.8632 & 2.8654  &  \\
				& (\lowerc{}) $\times$ (\upperc{}) & -2.9485 & -2.9496 &-2.9474  &  \\
				& (\middlec{}) $\times$ (\upperc{}) & -5.8127 & -5.8138& -5.8117  &  \\
				& (\all) $\times$ (\men\lowerc{}) & -3.2379 & -3.2398 &-3.2359   &  \\
				& (\all) $\times$ (\men-\middlec{}) & -0.3976 & -0.3995 &-0.3956  &  \\
				& (\all) $\times$ (\men-\upperc{}) & -6.1201 & -6.1221 &-6.1182  &  \\
				& (\all) $\times$ (\women\lowerc{}) & -4.6039 & -4.6058 &-4.6019 &  \\
				& (\all) $\times$ (\women-\middlec{}) & -0.6298 & -0.6318 &-0.6279   &  \\
				& (\all) $\times$ (\women-\upperc{}) & -7.6548 & -7.6567 &-7.6528  &  \\
				& (\men\lowerc{}) $\times$ (\men-\middlec{}) & 2.8403 &  2.8384 & 2.8422  &  \\
				& (\men\lowerc{}) $\times$ (\men-\upperc{}) &-2.8822 & -2.8842& -2.8803  &  \\
				& (\men\lowerc{}) $\times$ (\women\lowerc{}) & -1.366 &  -1.368& -1.3641 &  \\
				& (\men\lowerc{}) $\times$ (\women-\middlec{}) & 2.608 &  2.6061  &  2.61  &  \\
				& (\men\lowerc{}) $\times$ (\women-\upperc{}) & -4.4169 & -4.4189  &-4.415  &  \\
				& (\men-\middlec{}) $\times$ (\men-\upperc{}) & -5.7225 & -5.7245 &-5.7206   &  \\
				& (\men-\middlec{}) $\times$ (\women\lowerc{}) & -4.2063 & -4.2083 &-4.2044 &  \\
				& (\men-\middlec{}) $\times$ (\women-\middlec{}) &  -0.2323 & -0.2342 &-0.2303  &  \\
				& (\men-\middlec{}) $\times$ (\women-\upperc{}) & -7.2572 & -7.2592 &-7.2553  &  \\
				& (\men-\upperc{}) $\times$ (\women\lowerc{}) & 1.5162 &  1.5143 & 1.5182 &  \\
				& (\men-\upperc{}) $\times$ (\women-\middlec{}) & 5.4903 &  5.4883 & 5.4922 &  \\
				& (\men-\upperc{}) $\times$ (\women-\upperc{})& -1.5347 & -1.5366 &-1.5327  &  \\
				& (\women\lowerc{}) $\times$ (\women-\middlec{}) & 3.9741 &  3.9721  & 3.976    &  \\
				& (\women\lowerc{}) $\times$ (\women-\upperc{}) & -3.0509 & -3.0528 & -3.049 &  \\
				& (\women-\middlec{}) $\times$ (\women-\upperc{}) &-7.025 & -7.0269 & -7.023  &  \\
				\bottomrule
			\end{tabular}
		}
	\end{table}
	
	\newpage

	\section{Effects of changing the sample size}
	\label{sec:sample_size_effect}
	
	We present here the analysis of the effects of changing the sample size on the mobility diversity's value. Our goal is to check whether our main results/conclusions depend on the sample size. Specifically, we investigate the saturation of $H$ as we increase the size of the sample used in the bootstraps of travels made by different gender and socioeconomic groups. Rather than using samples \textcolor{black}{made using 60\% of all the travels available for each group, we consider samples with a size ranging from 1,000 travels up to the maximum possible size for each region, year, and purpose of travel (the values of the maximum sample size possible are detailed in \ref{tab:alltravelsexp} and \ref{tab:worktravelsexp} Tables).} 
	
\begin{figure}[H]
		\centering
		\includegraphics[width=0.8\textwidth]{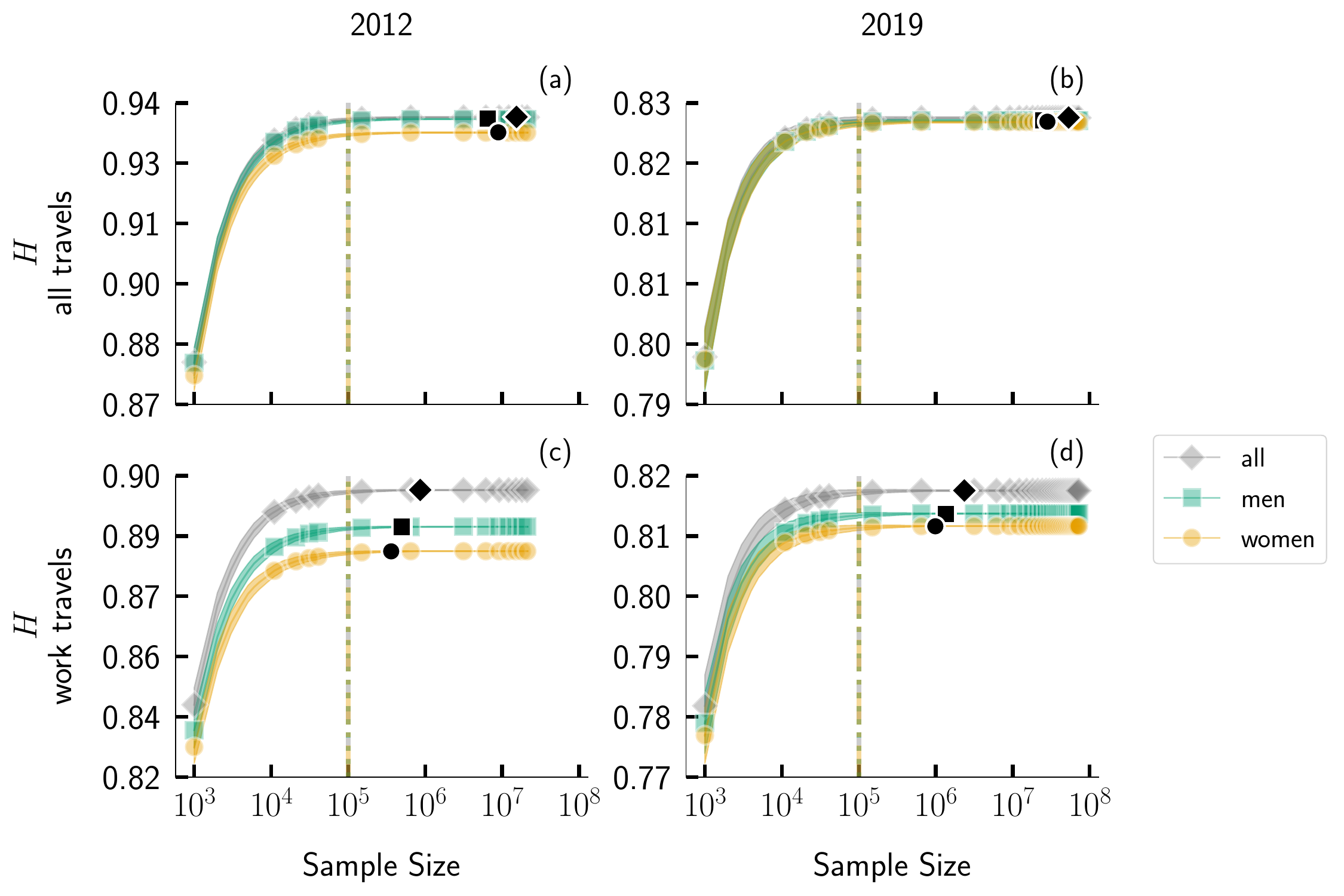}
		\caption{\textbf{Values of the mobility diversity, $H$, for different sample's sizes for travels made by travellers grouped by gender in \bgt.} We consider either \all travels (panels {\bf a} and {\bf b}), or \work travels (panels {\bf c} and {\bf d}) only. The shaded area accounts for the standard deviation of the values obtained from averaging the results over 1,000 realisations. Each column accounts for a different year. The vertical lines denotes the size from which the values of mobility diversity stabilise. The black symbols correspond to the same quantity obtained using a sample's size equal to \textcolor{black}{60\%} of all the travels available.}
		\label{fig:gender_bgt_sample_size}
	\end{figure}

\begin{figure}[H]
		\centering
		\includegraphics[width=0.97\textwidth]{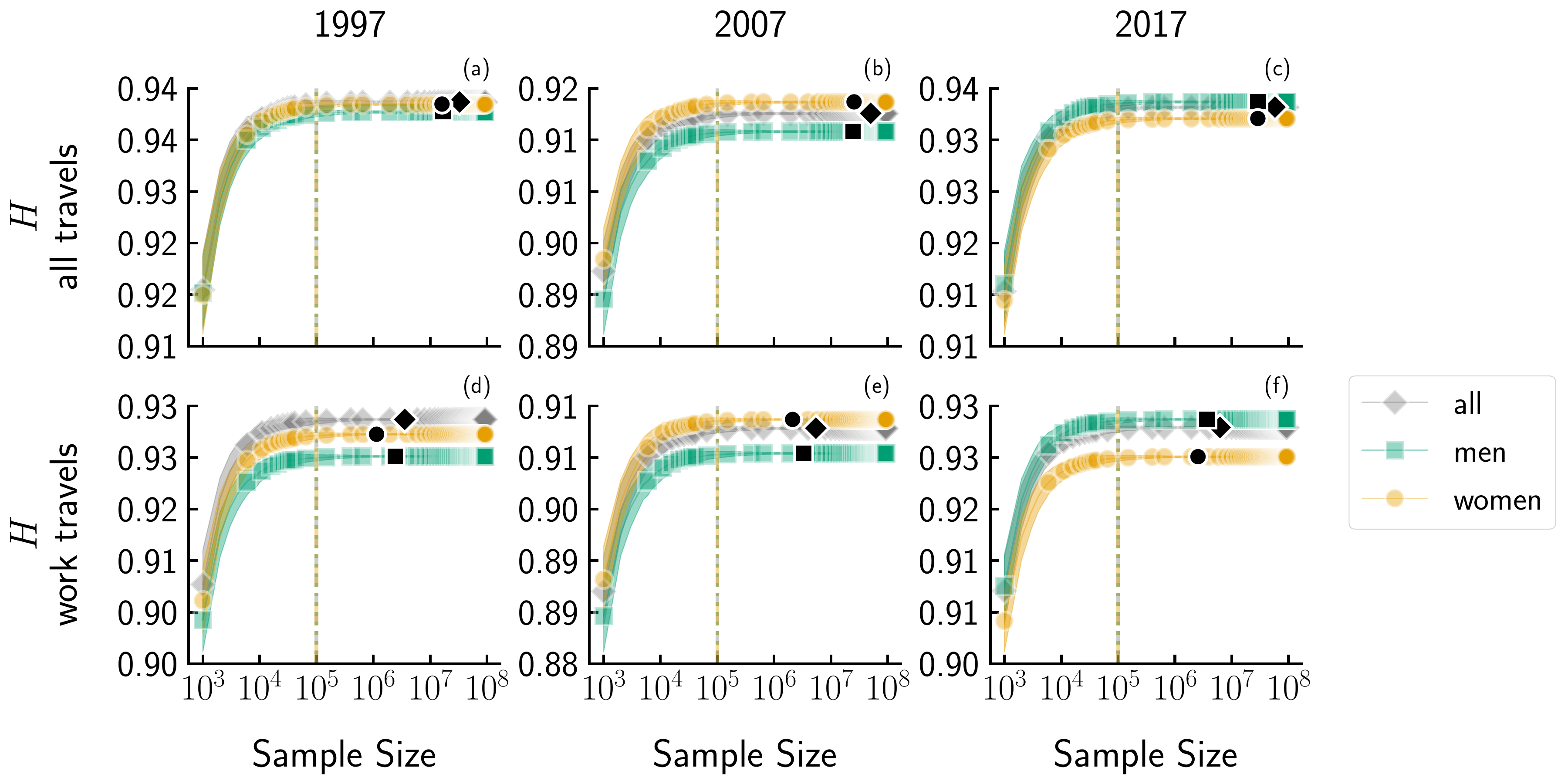}
		\caption{\textbf{Values of the mobility diversity, $H$, for different sample's sizes for travels made by travellers grouped by gender in \sao.} We consider either \all travels (panels {\bf a}, {\bf b}, and {\bf c}), or \work travels (panels {\bf d}, {\bf e}, and {\bf f}) only. See the caption of \ref{fig:gender_bgt_sample_size}~Fig for the description of the notation.}
		\label{fig:gender_sao_sample_size}
	\end{figure}

\begin{figure}[H]
		\centering
		\includegraphics[width=0.8\textwidth]{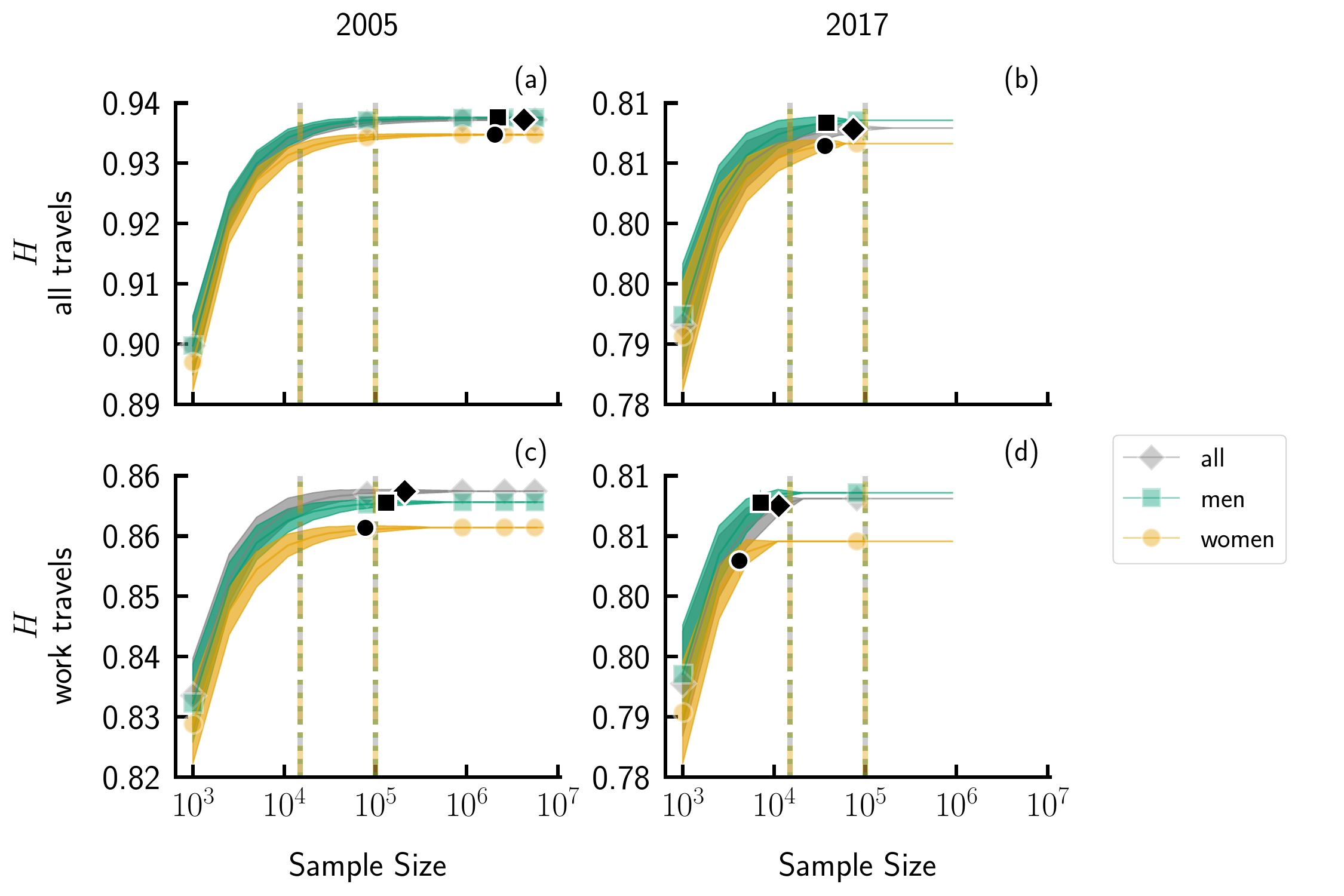}
		\caption{\textbf{Values of the mobility diversity, $H$, for different sample's sizes for travels made by travellers grouped by gender in \mde{}.} We consider either \all travels (panels {\bf a} and {\bf b}), or \work travels (panels {\bf c} and {\bf d}) only. See the caption of \ref{fig:gender_bgt_sample_size}~Fig for the description of the notation.}
		\label{fig:gender_mde_sample_size}
	\end{figure}

	\textcolor{black}{We begin our analysis by considering the effects of sample's size on the travels grouped by gender only. \ref{fig:gender_bgt_sample_size}--\ref{fig:gender_mde_sample_size}~Figs display the values of $H$ as we increase the size of the sample for \bgt{}, \sao{}, and \mde{}. We observe a saturation of the values of $H$ as soon as the sample's size exceeds \textcolor{black}{$10^{4}$ travels for \mde, and $10^{5}$ travels for \bgt and \sao, instead. Such a value is smaller than 60\% of all the travels, indicated by the black symbols appearing in} \ref{fig:gender_bgt_sample_size}--\ref{fig:gender_mde_sample_size}~Figs). Nevertheless, men continue to consistently show higher values of $H$ than women.}
	
	%
	%
	
	%
	%

	%
	%

\begin{figure}[H]
		\centering
		\includegraphics[width=0.8\textwidth]{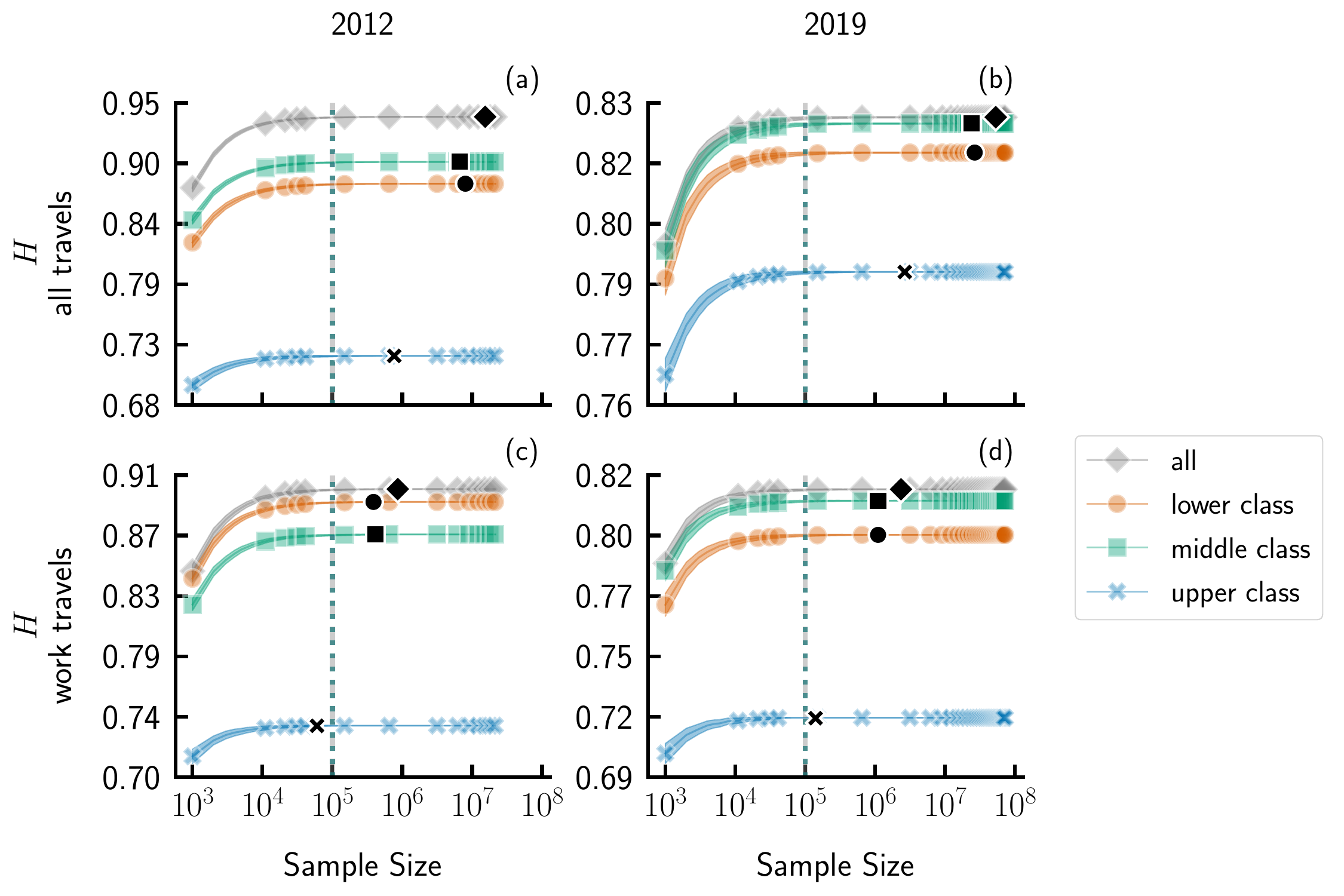}
		\caption{V\textbf{alues of the mobility diversity, $H$, for different sample's sizes for travels made by travellers grouped by socioeconomic class in \bgt{}.} We consider either \all travels (panels {\bf a} and {\bf b}), or \work travels (panels {\bf c} and {\bf d}) only. See the caption of \ref{fig:gender_bgt_sample_size}~Fig for the description of the notation.}
		\label{fig:socio_bgt_sample_size}
	\end{figure}

\begin{figure}[H]
		\centering
		\includegraphics[width=0.97\textwidth]{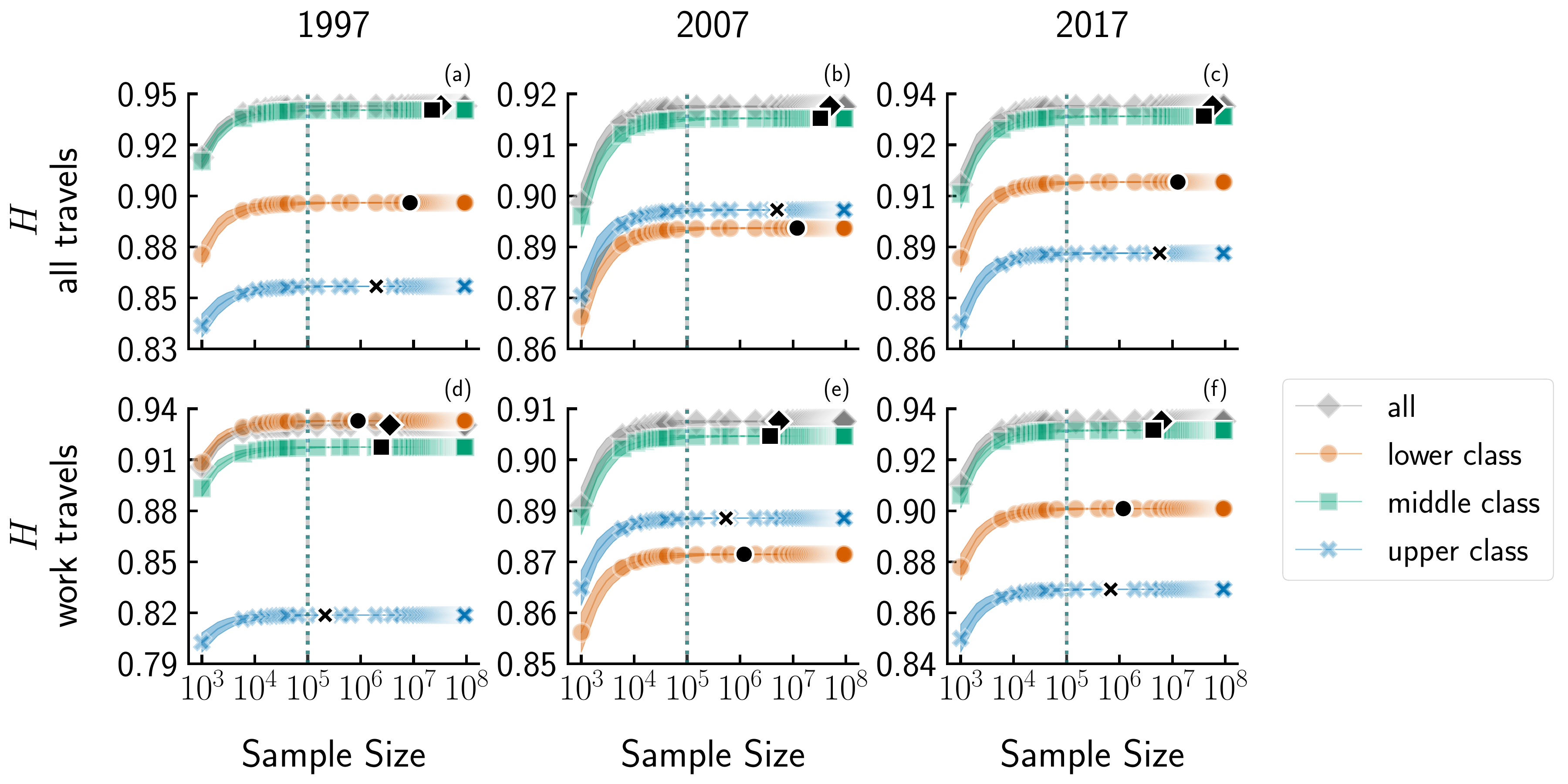}
		\caption{\textbf{Values of the mobility diversity, $H$, for different sample's sizes for travels made by travellers grouped by socioeconomic class in \sao{}.} We consider either \all travels (panels {\bf a}, {\bf b}, and {\bf c}), or \work travels (panels {\bf d}, {\bf e}, and {\bf f}) only. See the caption of \ref{fig:gender_bgt_sample_size}~Fig for the description of the notation.}
		\label{fig:socio_sao_sample_size}
	\end{figure}

\begin{figure}[H]
		\centering
		\includegraphics[width=0.8\textwidth]{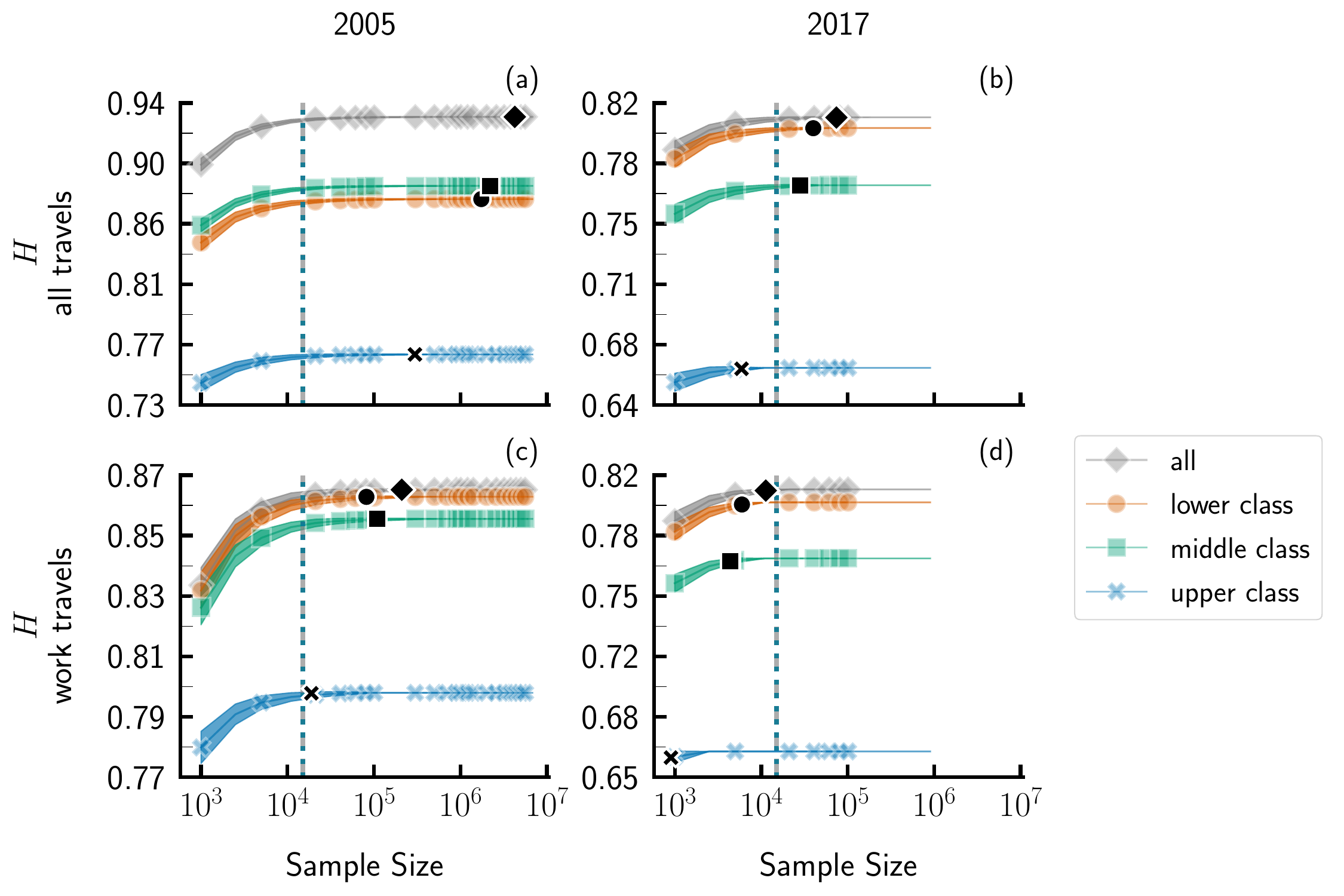}
		\caption{\textbf{Values of the mobility diversity, $H$, for different sample's sizes for travels made by travellers grouped by socioeconomic class in \mde{}.} We consider either \all travels (panels {\bf a} and {\bf b}), or \work travels (panels {\bf c} and {\bf d}) only. See the caption of \ref{fig:gender_bgt_sample_size}~Fig for the description of the notation.}
		\label{fig:socio_mde_sample_size}
	\end{figure}

	\textcolor{black}{After studying the effects of sample's size on gender alone, we can repeat the same analysis for the socioeconomic factor. \ref{fig:socio_bgt_sample_size} -- \ref{fig:socio_mde_sample_size} Figs show the effects of changing the sample's for the three socioeconomic groups considered. Overall, the phenomenology observed remains more or less the same as the gender only case. Such an analogy lead us to conclude that despite the presence of some differences between the empirical (complete) results and the sampled ones, our conclusions are not affected by the choice of sample's size.}

	%
	%

	%
	%

	%
	%
	
\begin{figure}[H]
		\includegraphics[width=0.97\textwidth]{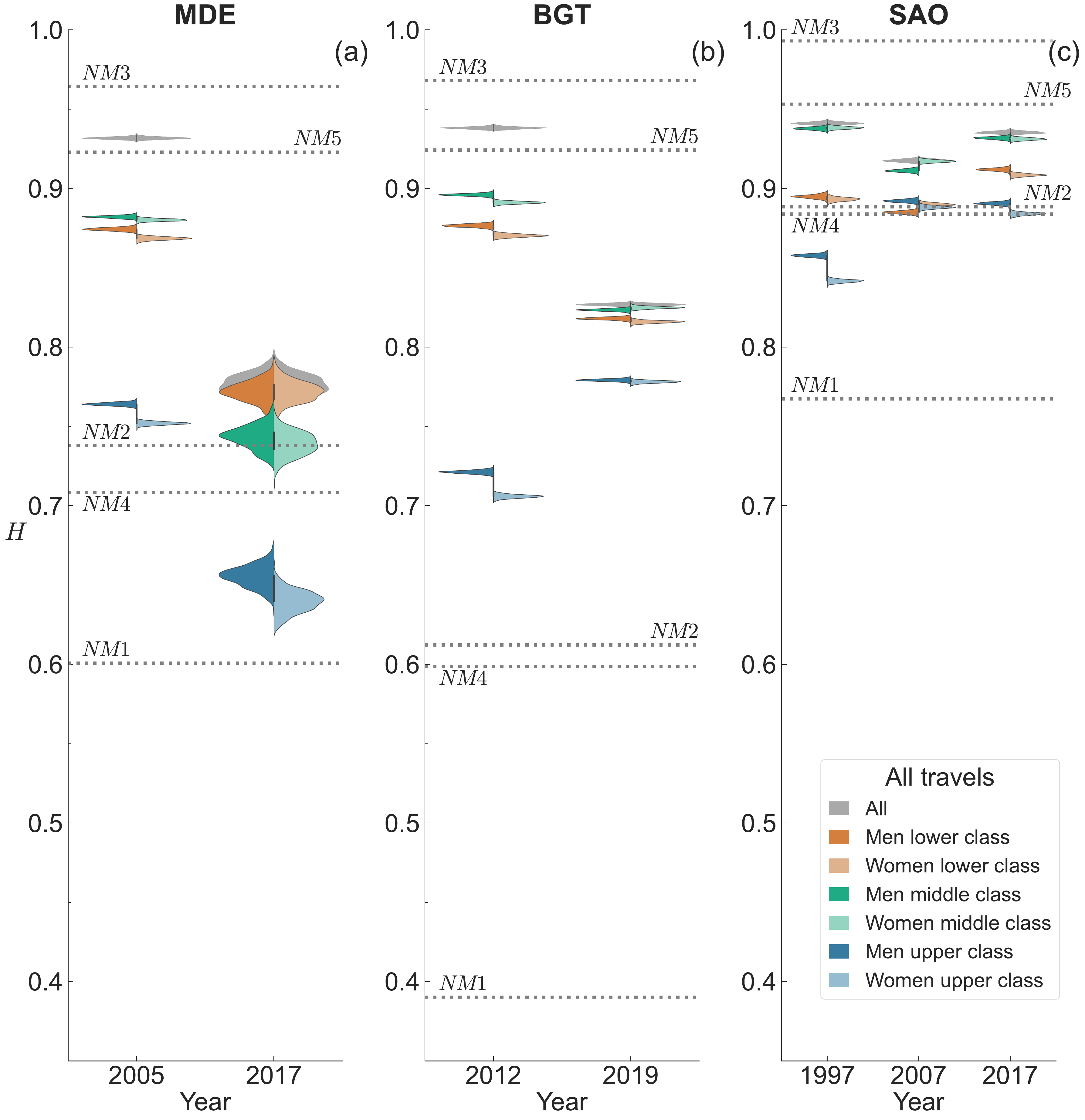}
		\caption{\textbf{Distribution of the mobility diversity, $H$, for \textcolor{black}{25,000 travels (except \mde for 2017 that is 550 travels)} made by \all purposes by travellers grouped according to their socioeconomic status and gender.} Each column refers to a different region, and for each region, we consider all the available years. For each socioeconomic status (\upperc{}, \middlec{}, and \lowerc{}) darker hue denotes men travellers, whereas lighter hue denotes women ones. The dotted lines denote the values of $H$ computed using travels generated by each null model.}
		\label{fig:allgendersocioregions_maxsample}
	\end{figure}

\begin{figure}[H]
		\includegraphics[width=0.97\textwidth]{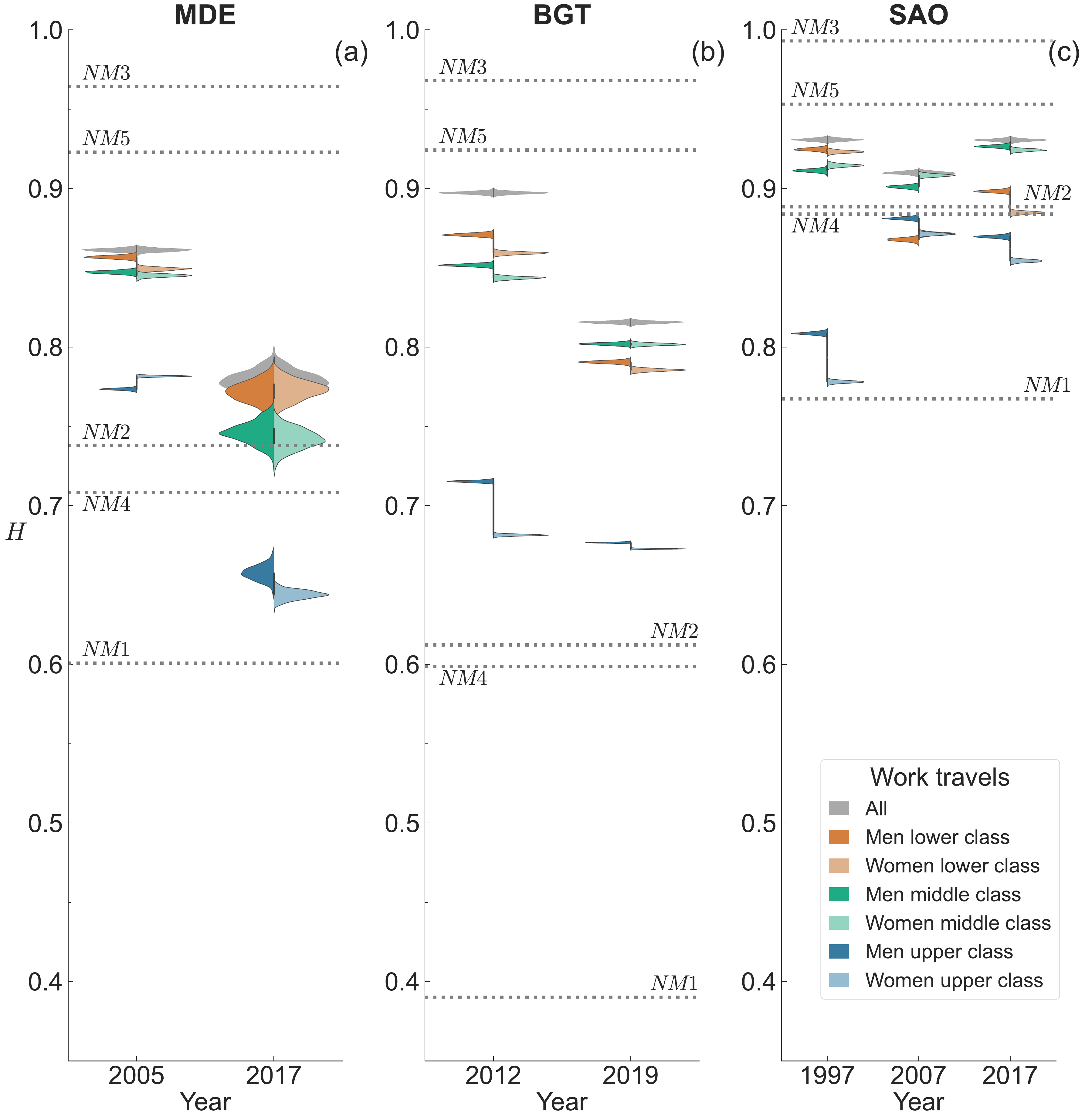}
		\caption{\textbf{Distribution of the mobility diversity, $H$, for the maximum number of travels in the sample size range made by \work purposes by travellers grouped according to their socioeconomic status and gender.} Each column refers to a different region, and for each region, we consider all the available years. For each socioeconomic status (\upperc{}, \middlec{}, and \lowerc{}) darker hue denotes men travellers, whereas lighter hue denotes women ones. The dotted lines denote the values of $H$ computed using travels generated by each null model.}
		\label{fig:workgendersocioregions_maxsample}
	\end{figure}

	Finally, we analyse the effects of changing the sample's size for travels grouped according to gender and socioeconomic status simultaneously. In \ref{fig:allgendersocioregions_maxsample} and \ref{fig:workgendersocioregions_maxsample}~Figs, we report the violin plots of $H$ \textcolor{black}{computed with the values of sample's size across groups equal to 25,000 travels (except \mde for 2017 that we had to establish 550 travels)}. Eyeballing at these figures does not highlight any qualitative difference with the hierarchies displayed in Fig.~\ref{M-fig:allgendersocioregions} and \ref{fig:workgendersocioregions}~Fig Notwithstanding, the values of $H$ displayed in ~\ref{fig:allgendersocioregions_maxsample} and \ref{fig:workgendersocioregions_maxsample}~Figs are \textcolor{black}{approximately the same. This is true due to the fact that the values of $H$ saturate when the sample size exceeds $10^4$/$10^5$ (\mde / \bgt and \sao), thus confirming the robustness of our results.}
	
	%
	%
	
	%
	%

\begin{figure}[H]
		\centering
		\includegraphics[width=0.8\textwidth]{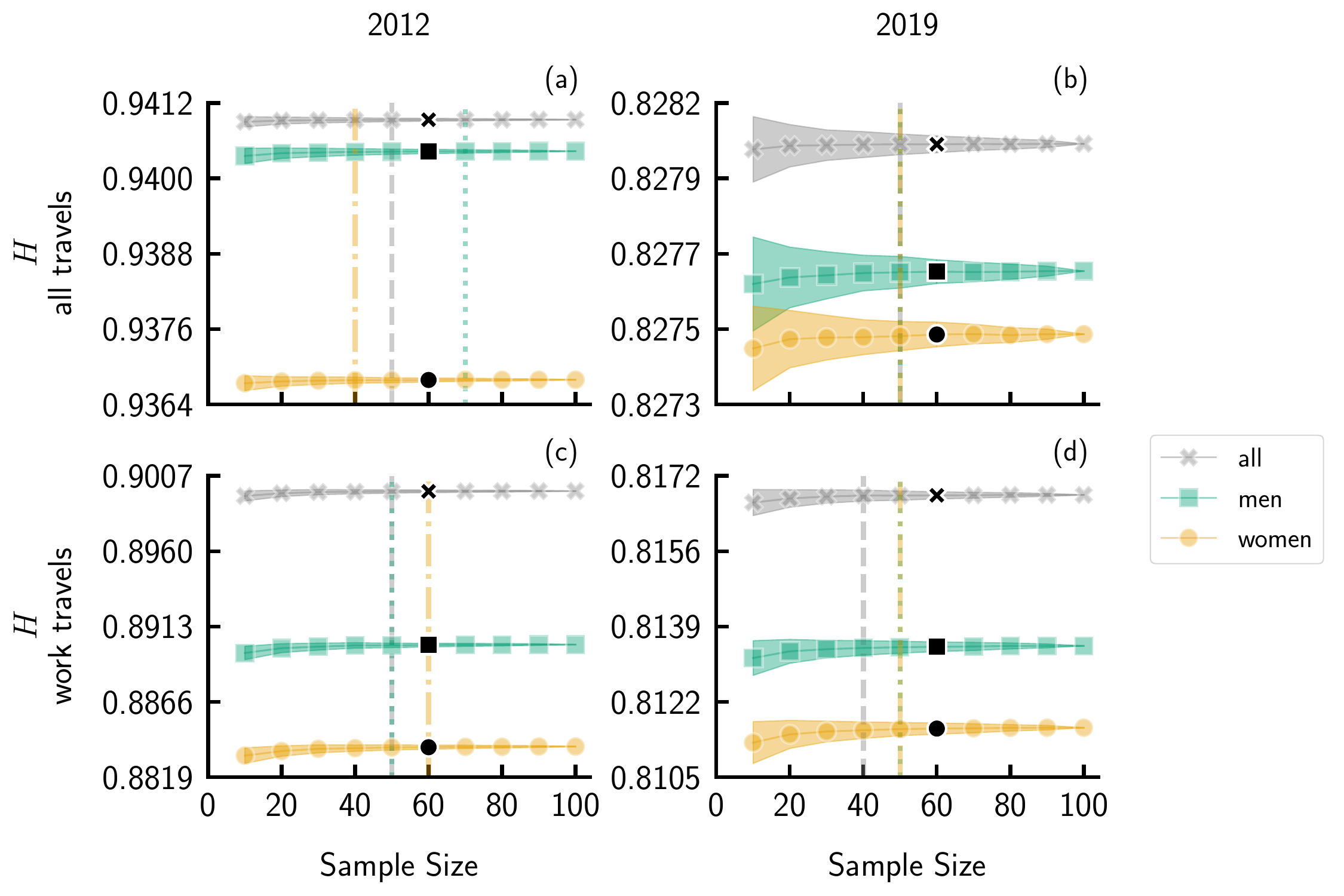}
		\caption{\textbf{Values of the mobility diversity, $H$, for different sample's sizes (in percentage) of travels made by travellers grouped by gender in \bgt{}.} We consider either \all travels (panels {\bf a} and {\bf b}), or \work travels (panels {\bf c} and {\bf d}) only. The shaded area accounts for the standard deviation of the values obtained from averaging the results over 1,000 realisations. Each column accounts for a different year, and the dashed lines represent the saturation of the values of the mobility diversity.}
		\label{fig:percentage_effect}
	\end{figure}

	\textcolor{black}{We performed also additional analyses on the impact of choosing a specific percentage of travels per group. We observe that even when we use just 10\% of the whole data, the differences between the groups remain pretty clear. However, the stagnation of the values of mobility diversity is reached -- on average, -- for samples using at least \textcolor{black}{60\%} of the travels. As an example, we show in  \ref{fig:percentage_effect} Fig the values of $H$ as one increases the size of the sample for \bgt{} area.} \textcolor{black}{Summing up, we argue that the validity of our results and conclusions seems not affected by the sample's size.}
	
	\section{Null models}
	\label{sec:null_models}
	
\begin{figure}[H]
		\centering
		\includegraphics[width=0.98\textwidth]{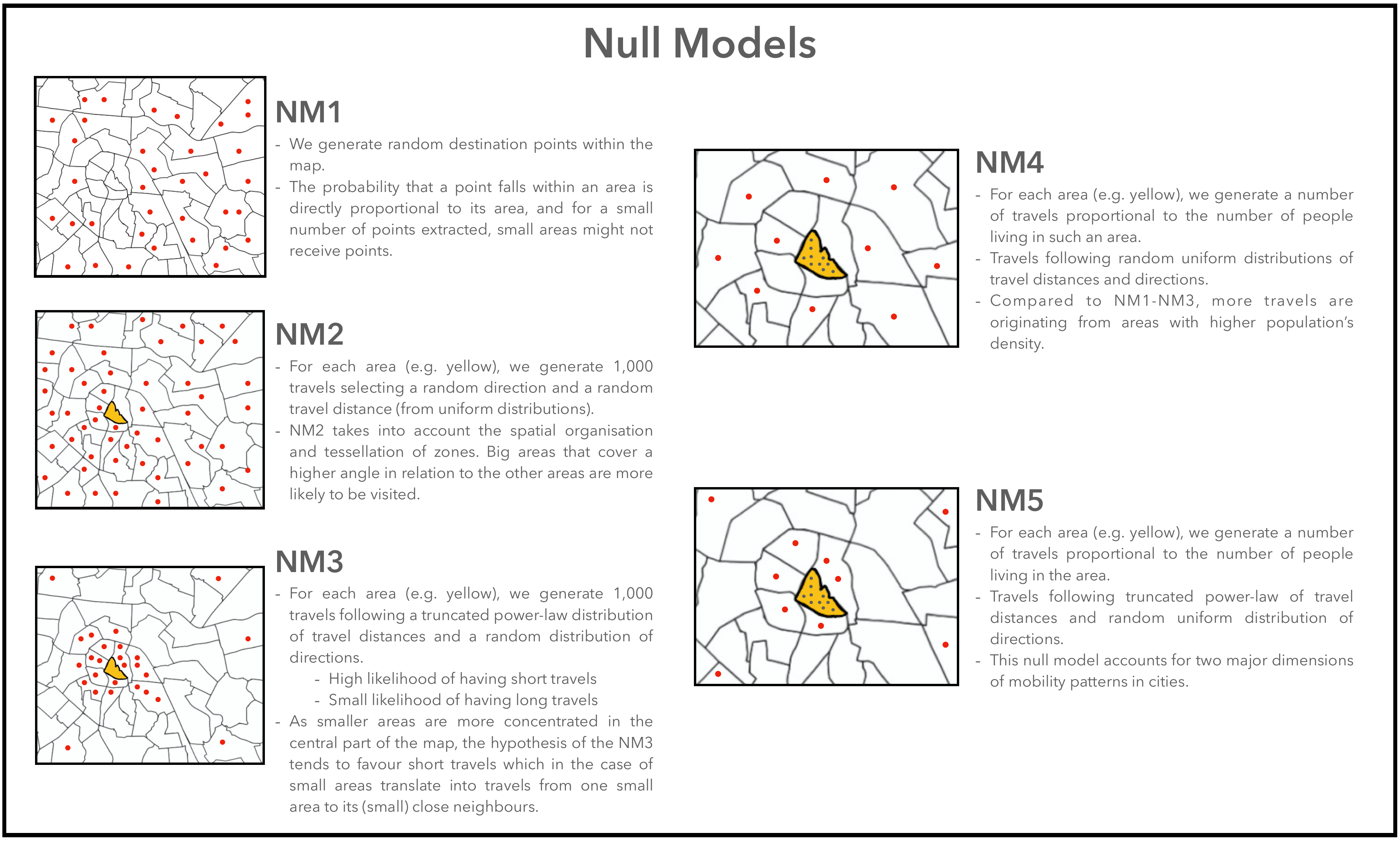}
		\caption{\textbf{Schematic summary of the main features of the null models considered.} For each null model, we list its main properties. Red dots appearing in the maps denote the travels' destinations.}
		\label{fig:null_models_diagram}
	\end{figure}

	In this section we present a comparison between the values of $H$ computed using the data, and the same quantities obtained using a null model. In particular, we focus our attention on three aspects: $i$) the tessellation of the urban area into zones of different size, $ii$) the dis-homogeneity of the distance of travels, and $iii$) the population living in a zone. The combination of these ingredients gives rise to five different null models $\text{\nullm{$x$}}$ with $x \in \{1,\ldots,5\}$, each accounting for one -- or more, -- of the aforementioned aspects; with \nullm{1} being the least realistic model, and \nullm{5} the most realistic one. \ref{fig:null_models_diagram}~Fig contains a schematic summary of the characteristics of all the null models.
	
	For each null model, we generate a number of travels equal to 1,000 multiplied for the number of zones (\eg{} for \sao{} we generate 248,000 travels). The number of people living in a given zone is computed in relation to the density of people resident within each zone. The travel distance, instead, is computed either uniformly at random or extracted from a truncated power-law distribution within the range between 100 and 60,000 meters \cite{Barbosa2018}. Such extremes correspond to the minimum and maximum distances observed in our data. If the destination point of a travel falls outside the urban area, we continue to extract a new point until its position falls within the urban area. If both the travel origin and destination zones coincide, we keep the travel. After generating the travels, we compute first the probability that their destinations fall within a certain zone, $i$, using Eq~\eqref{M-eq:probability}, and then the value of $H$ using Eq~\eqref{M-eq:mobdiv}. Finally, we average the results over 1,000 realisations.
	
	%
	%
	
	\textcolor{black}{\nullm{1} is the most naive model, and accounts only for the tessellation of the urban area. Specifically, the model corresponds to a random scattering of points (\ie{} travels' destinations) over the urban area. According to this model, the probability that a travel ends in a given zone is only proportional to the zone's area. \nullm{2} expands slightly \nullm{1} by taking into account the travel distance limits (100 meters until 60,000 meters). \nullm{3} is similar to \nullm{2} with the exception that the travel's distance is extracted from a truncated power-law function. Such a difference translates into the presence of more short range travels which, in turn, corresponds to more travels towards neighbour zones if the origin zone is small, and more travels within the same zone if the origin zone is big, instead. \nullm{4} is similar to \nullm{2} but it accounts for the heterogeneity in the density of people living in each zone. This means that the number of travels starting from a given zone is proportional to the number of people living in it. Finally, \nullm{5} is the most realistic model and differentiates from \nullm{4} because the travel distance follows a truncated power-law distribution.}
	
	\textcolor{black}{\ref{tab:null_models_summary} and \ref{tab:null_models_summary_groups_together}~Tables display the values of $H$ computed for the spatial distributions of destinations in all the areas, years, purpose, and by the groups considered in our study. In addition, the tables contain also the values of $H$ corresponding to the destinations' distributions generated by each null model. We performed both the Kolmogorov-Smirnov and Welch's $t$-test on each combination (pair) of empirical and synthetic distributions of $H$ and, we have found that all of the combinations reject the null hypothesis of the aforementioned tests with a $p$-value smaller than 0.001. The sole violation of the previous statement occurs for the case of \nullm{4} versus $H_{\text{\upperc{}}}$ of \work{} travels made in \sao{} during 2007.}

	\textcolor{black}{In \ref{sec:sample_size_effect} Section, \ref{fig:allgendersocioregions_maxsample} and \ref{fig:workgendersocioregions_maxsample}~Figs display the violin plots of $H$ computed using the highest possible value of sample's size across groups. In this section, instead, we can observe in \ref{tab:null_models_summary_groups_together}~Table how much the median values of $H$ for each group differ from the same quantity computed via the null models. The difference between these values indicates that taken together gender and socioeconomic status exert a remarkable effect on the value of the mobility diversity.}
	
	%
	%
	
	\begin{table}[H]
		\centering
		\caption{\textbf{Summary of the values of the mobility diversity, $H$, of empirical data and null models.} For each area, year, and purpose of travel we report the values of $H$ computed for all travels ($H_{\text{\all{}}}$), gender ($H_{\text{\men{}}}$, $H_{\text{\women{}}}$), and socioeconomic status ($H_{\text{\lowerc{}}}$, $H_{\text{\middlec{}}}$, $H_{\text{\upperc{}}}$). We report also the value of $H$ computed using different null models (\nullm{$x$} with $x \in \{1,\ldots,5\}$) averaged over 1,000 realisations.}
		\label{tab:null_models_summary}
		\resizebox{0.95\linewidth}{!}{
			\begin{tabular}{c|c|c|c|c|c|c|c|c|c|c|c|c|c}
				\toprule
				\multirow{2}{*}{Area} & \multirow{2}{*}{Year} & Travel & \multirow{2}{*}{$H_{\text{\all{}}}$} & \multirow{2}{*}{$H_{\text{\men{}}}$} & \multirow{2}{*}{$H_{\text{\women{}}}$} & \multirow{2}{*}{$H_{\text{\lowerc{}}}$} &  \multirow{2}{*}{$H_{\text{\middlec{}}}$} & \multirow{2}{*}{$H_{\text{\upperc{}}}$} & \multicolumn{5}{c}{$\avg{H}$}\\
				\cline{10-14}
				& & Type & & & & & & & \nullm{1} & \nullm{2} & \nullm{3} & \nullm{4} & \nullm{5} \\
				
				\midrule
				
				\multirow{4}{*}{\mde{}}
				& \multirow{2}{*}{2005} & \all{} & 0.9329 & 0.9332 & 0.9306 & 0.8740 & 0.8835 & 0.7627 & \multirow{2}{*}{0.6008} & \multirow{2}{*}{0.7379} & \multirow{2}{*}{0.9643} & \multirow{2}{*}{0.7020} & \multirow{2}{*}{0.8847}\\
				& & \work{} & 0.8624 & 0.8607 & 0.8567 & 0.8600 & 0.8526 & 0.7941 & & & & & \\
				\cline{2-14}
				
				& \multirow{2}{*}{2017} & \all{} & 0.8099 & 0.8106 & 0.8083 & 0.8036 & 0.7700 & 0.6629 & \multirow{2}{*}{0.6008} & \multirow{2}{*}{0.7379} & \multirow{2}{*}{0.9643} & \multirow{2}{*}{0.7085} & \multirow{2}{*}{0.9230}\\
				& & \work{} & 0.8100 & 0.8101 & 0.8045 & 0.8019 & 0.7696 & 0.6574 & & & & & \\
				\hline

				\multirow{4}{*}{\bgt{}}
				& \multirow{2}{*}{2012} & \all{} & 0.9409 & 0.9404 & 0.9368 & 0.8800 & 0.8999 & 0.7234 & \multirow{2}{*}{0.3902} & \multirow{2}{*}{0.6123} & \multirow{2}{*}{0.9680} & \multirow{2}{*}{0.9261} & \multirow{2}{*}{0.9485}\\
				& & \work{} & 0.8997 & 0.8901 & 0.8837 & 0.8909 & 0.8687 & 0.7380 & & & & & \\
				\cline{2-14}
				
				& \multirow{2}{*}{2019} & \all{} & 0.8280 & 0.8276 & 0.8274 & 0.8191 & 0.8264 & 0.7889 & \multirow{2}{*}{0.3902} & \multirow{2}{*}{0.6123} & \multirow{2}{*}{0.9680} & \multirow{2}{*}{0.5988} & \multirow{2}{*}{0.9243}\\
				& & \work{} & 0.8168 & 0.8134 & 0.8115 & 0.7974 & 0.8119 & 0.7195 & & & & & \\
				\hline

				\multirow{6}{*}{\sao{}}
				& \multirow{2}{*}{1997} & \all{} & 0.9420 & 0.9407 & 0.9417 & 0.8967 & 0.9401 & 0.8575 & \multirow{2}{*}{0.7674} & \multirow{2}{*}{0.8885} & \multirow{2}{*}{0.9930} & \multirow{2}{*}{0.8760} & \multirow{2}{*}{0.9483}\\
				& & \work{} & 0.9316 & 0.9262 & 0.9294 & 0.9340 & 0.9190 & 0.8226 & & & & & \\
				\cline{2-14}
				
				& \multirow{2}{*}{2007} & \all{} & 0.9185 & 0.9159 & 0.9201 & 0.8898 & 0.9157 & 0.8941 & \multirow{2}{*}{0.7674} & \multirow{2}{*}{0.8885} & \multirow{2}{*}{0.9930} & \multirow{2}{*}{0.8862} & \multirow{2}{*}{0.9446}\\
				& & \work{} & 0.9107 & 0.9072 & 0.9120 & 0.8747 & 0.9067 & 0.8845 & & & & & \\
				\cline{2-14}
				
				& \multirow{2}{*}{2017} & \all{} & 0.9361 & 0.9368 & 0.9346 & 0.9130 & 0.9330 & 0.8913 & \multirow{2}{*}{0.7674} & \multirow{2}{*}{0.8885} & \multirow{2}{*}{0.9930} & \multirow{2}{*}{0.8839} & \multirow{2}{*}{0.9533}\\
				& & \work{} & 0.9314 & 0.9325 & 0.9272 & 0.8995 & 0.9281 & 0.8700 & & & & & \\
				
				\bottomrule
			\end{tabular}
		}
	\end{table}

	\begin{table}[H]
		\centering
		\caption{\textbf{Summary of the values of the mobility diversity, $H$, of empirical data and null models.} For each area, year, and purpose of travel we report the values of $H$ computed for the travels performed by each group considering the gender and socioeconomic class together: Men Lower class ($H_{\text{M,L}}$), Men Middle class ($H_{\text{M,M}}$), Men Upper class ($H_{\text{M,U}}$), Men Lower class ($H_{\text{W,L}}$), Women Middle class ($H_{\text{W,M}}$) and Women Upper class ($H_{\text{W,U}}$). We report also the value of $H$ computed using different null models (\nullm{$x$} with $x \in \{1,\ldots,5\}$) averaged over 1,000 realisations.}
		\label{tab:null_models_summary_groups_together}
		\resizebox{0.95\linewidth}{!}{
			\begin{tabular}{c|c|c|c|c|c|c|c|c|c|c|c|c|c}
				\toprule
				\multirow{2}{*}{Area} & \multirow{2}{*}{Year} & Travel & \multirow{2}{*}{$H_{\text{M,L}}$} & \multirow{2}{*}{$H_{\text{M,M}}$} & \multirow{2}{*}{$H_{\text{M,U}}$} & \multirow{2}{*}{$H_{\text{W,L}}$} &  \multirow{2}{*}{$H_{\text{W,M}}$} & \multirow{2}{*}{$H_{\text{W,U}}$} & \multicolumn{5}{c}{$\avg{H}$}\\
				\cline{10-14}
				& & Type & & & & & & & \nullm{1} & \nullm{2} & \nullm{3} & \nullm{4} & \nullm{5} \\
				
				\midrule
				
				\multirow{4}{*}{\mde{}}
				& \multirow{2}{*}{2005} & \all{} & 0.8754 & 0.8832 & 0.7645 & 0.8696 & 0.8812 & 0.7524 & \multirow{2}{*}{0.6008} & \multirow{2}{*}{0.7379} & \multirow{2}{*}{0.9643} & \multirow{2}{*}{0.7020} & \multirow{2}{*}{0.8847}\\
				& & \work{} & 0.8576 & 0.8481 & 0.7735 & 0.8499 & 0.8458 & 0.7815 & & & & & \\
				\cline{2-14}
				
				& \multirow{2}{*}{2017} & \all{} & 0.8029 & 0.7712 & 0.6685 & 0.8026 & 0.7667 & 0.6524 & \multirow{2}{*}{0.6008} & \multirow{2}{*}{0.7379} & \multirow{2}{*}{0.9643} & \multirow{2}{*}{0.7085} & \multirow{2}{*}{0.9230}\\
				& & \work{} & 0.7992 & 0.7690 & 0.6577 & 0.7969 & 0.7610 & 0.6381 & & & & & \\
				\hline

				\multirow{4}{*}{\bgt{}}
				& \multirow{2}{*}{2012} & \all{} & 0.8790 & 0.8983 & 0.7221 & 0.8727 & 0.8934 & 0.7067 & \multirow{2}{*}{0.3902} & \multirow{2}{*}{0.6123} & \multirow{2}{*}{0.9680} & \multirow{2}{*}{0.9261} & \multirow{2}{*}{0.9485}\\
				& & \work{} & 0.8728 & 0.8533 & 0.7154 & 0.8609 & 0.8452 & 0.6815 & & & & & \\
				\cline{2-14}
				
				& \multirow{2}{*}{2019} & \all{} & 0.8192 & 0.8246 & 0.7799 & 0.8171 & 0.8260 & 0.7789 & \multirow{2}{*}{0.3902} & \multirow{2}{*}{0.6123} & \multirow{2}{*}{0.9680} & \multirow{2}{*}{0.5988} & \multirow{2}{*}{0.9243}\\
				& & \work{} & 0.7916 & 0.8030 & 0.6765 & 0.7865 & 0.8027 & 0.6726 & & & & & \\
				\hline

				\multirow{6}{*}{\sao{}}
				& \multirow{2}{*}{1997} & \all{} & 0.8957 & 0.9387 & 0.8583 & 0.8940 & 0.9391 & 0.8423 & \multirow{2}{*}{0.7674} & \multirow{2}{*}{0.8885} & \multirow{2}{*}{0.9930} & \multirow{2}{*}{0.8760} & \multirow{2}{*}{0.9483}\\
				& & \work{} & 0.9255 & 0.9122 & 0.8090 & 0.9239 & 0.9154 & 0.7782 & & & & & \\
				\cline{2-14}
				
				& \multirow{2}{*}{2007} & \all{} & 0.8858 & 0.9121 & 0.8929 & 0.8907 & 0.9181 & 0.8889 & \multirow{2}{*}{0.7674} & \multirow{2}{*}{0.8885} & \multirow{2}{*}{0.9930} & \multirow{2}{*}{0.8862} & \multirow{2}{*}{0.9446}\\
				& & \work{} & 0.8686 & 0.9019 & 0.8817 & 0.8722 & 0.9091 & 0.8718 & & & & & \\
				\cline{2-14}
				
				& \multirow{2}{*}{2017} & \all{} & 0.9128 & 0.9328 & 0.8913 & 0.9093 & 0.9321 & 0.8848 & \multirow{2}{*}{0.7674} & \multirow{2}{*}{0.8885} & \multirow{2}{*}{0.9930} & \multirow{2}{*}{0.8839} & \multirow{2}{*}{0.9533}\\
				& & \work{} & 0.8990 & 0.9274 & 0.8702 & 0.8853 & 0.9251 & 0.8548 & & & & & \\
				
				\bottomrule
			\end{tabular}
		}
	\end{table}

    \newpage
    
    \section{Effects of endogenous and residential based travels on mobility diversity}
    \label{sec:endo_resid_travels}
    
    
    \textcolor{black}{In this section, we seek to understand whether the empirical values of mobility diversity stem from the spatial segregation of men and women or whether they are the byproduct of intrinsic differences affecting how men and women move. To answer to such a conundrum, we investigate the role that \emph{endogenous} and \emph{residential based} travels exert on the value of mobility diversity removing these travels one at a time from the set of travels made by a particular group of travellers, $X$, selected according to their gender or socioeconomic status (or combinations of them).}
    
    \textcolor{black}{More specifically, endogenous travels are those for which the origin and the destination zones coincide (from a network perspective, these travels correspond to self-loops). Residential based travels, instead, correspond to those travels whose destination zone coincides with the zone where the traveller lives. It is worth noting, however, that considering as residential travels only those for which the destination zone coincides with the zone where the traveller lives does not eliminate completely the residential mobility, as travels with origin zone coinciding with the traveller's living zone do not satisfy such a classification.}
    
    \textcolor{black}{To quantify the amount of endogenous mobility one can look at \ref{tab:probabilitysamezones}~Table. In particular, we notice that the average percentage of travels (made for \all{} purposes) for which the origin and destination zones are the same, $\avg{P^A_{all}}$, is around 27\%. The same quantity computed for \work{} travels, $\avg{P^A_{work}}$, is approximately 8\%. Finally, the percentage of work travels whose destination zone (\ie{} where the traveller works) coincides with the zone where the traveller lives, $P^{A}_{live=work}$, oscillates between 8\% and 23\%. \ref{tab:probabilitysamezoneslive} Table summarises, instead, the percentages of residential travels made for either \all or \work purposes. As done for the endogenous mobility, the average percentage of residential travels made for \all purposes, $\avg{P^{A}_{dest=live}}$, is approximately 22\%, whereas the same quantity computed for \work travels (\ie, the traveller home and work zones coincide), $\avg{P^{A}_{live=work}}$, is approximately 17\%.}
    
    %
    %
	\begin{table}[H]
		\caption{\textbf{Percentages of the travels made for $all$ purposes performed by traveller of type $X$ (\ie \all (A), \men (M) and \women (W)) having as destination zone the same zone where the traveller lives, $P^{X}_{dest=live}$.} Column $P^{X}_{live=work}$ denotes the same quantity computed for $work$ travels.}
		\label{tab:probabilitysamezoneslive}
		\centering
		\resizebox{\textwidth}{!}{
			\begin{tabular}{ll|rrr|rrr}
				\toprule
				City & Year &  $P^{A}_{dest=live} (\%)$ &  $P^{M}_{dest=live} (\%)$ & $P^{W}_{dest=live} (\%)$ &  $P^{A}_{live=work} (\%)$ &  $P^{M}_{live=work} (\%)$ & $P^{W}_{live=work} (\%)$\\
				\midrule
				\multirow{2}{*}{\mde} & 2005 & 16.31 & 15.58 & 17.11 & 7.76 & 8.02 & 7.38 \\
				                      & 2017 & 17.07 & 15.01 & 19.54 & 22.04 & 26.41 & 17.32 \\
				\multirow{2}{*}{\bgt} & 2012 & 18.85 & 16.99 & 20.37 & 13.72 & 14.37 & 12.92 \\
				                      & 2019 & 4.84 & 4.35 & 5.34 & 10.81 & 10.54 & 11.15 \\
				\multirow{3}{*}{\sao} & 1997 & 34.75 & 32.06 & 37.77 & 23.31 & 21.83 & 25.73 \\
				                      & 2007 & 31.48 & 29.57 & 33.45 & 20.23 & 18.66 & 22.32 \\
				                      & 2017 & 32.31 & 34.62 & 33.98 & 20.63 & 20.23 & 21.11 \\
				\bottomrule
			\end{tabular}
		}
	\end{table}

\begin{figure}[H]
    \centering
    \includegraphics[width=0.9\textwidth]{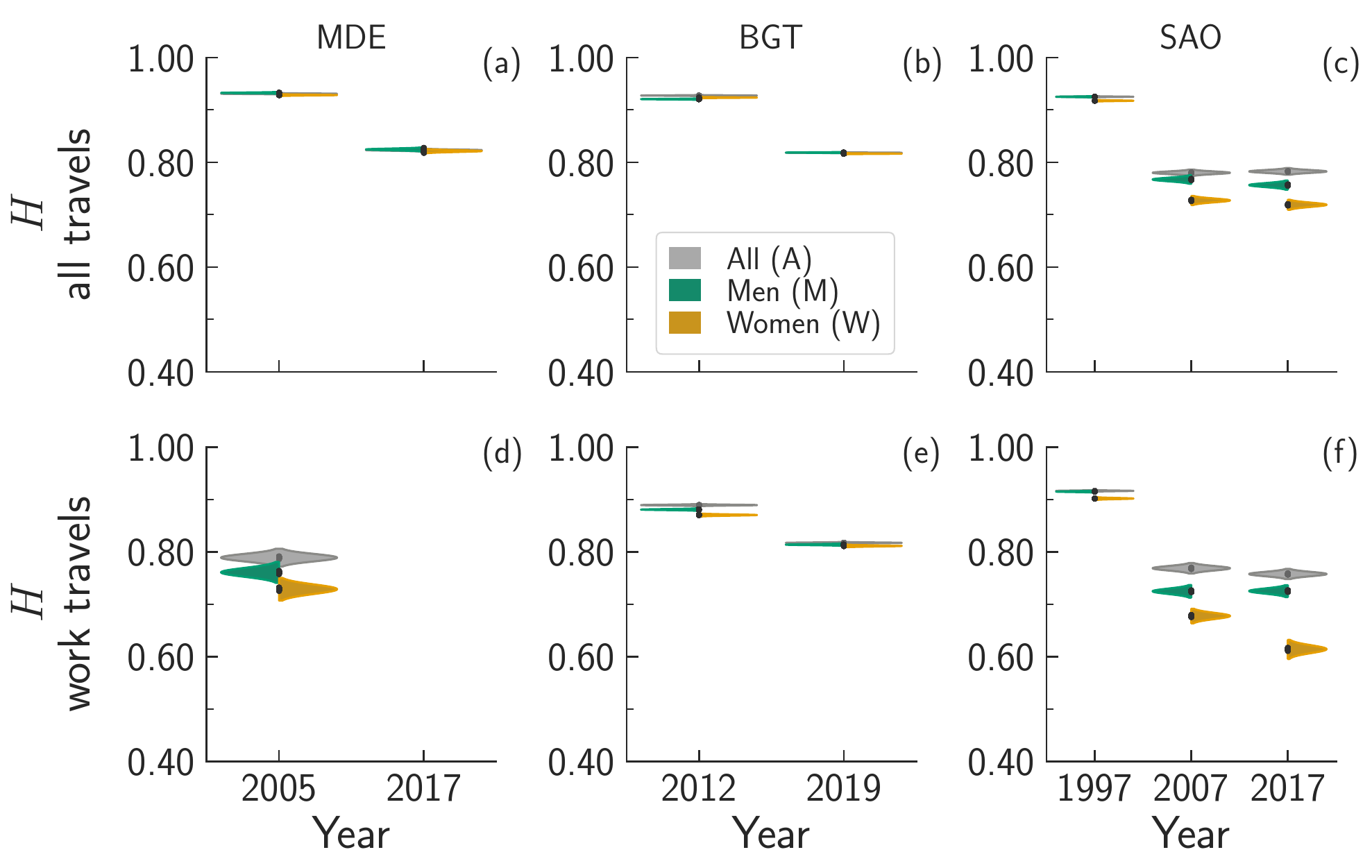}
    \caption{\textbf{Violin plots of the bootstrapped mobility diversity, $H$, for travels having different origin and destination zones}. The top row (panels a-c) accounts for travels made for \all purposes, whereas the bottom row (panels d-e) displays the results for \work travels. The data for year 2017 in the \mde area are missing as they are too scant.}
    \label{fig:genderregions_noselfloop}
    \end{figure}

\begin{figure}[H]
    \centering
    \includegraphics[width=0.9\textwidth]{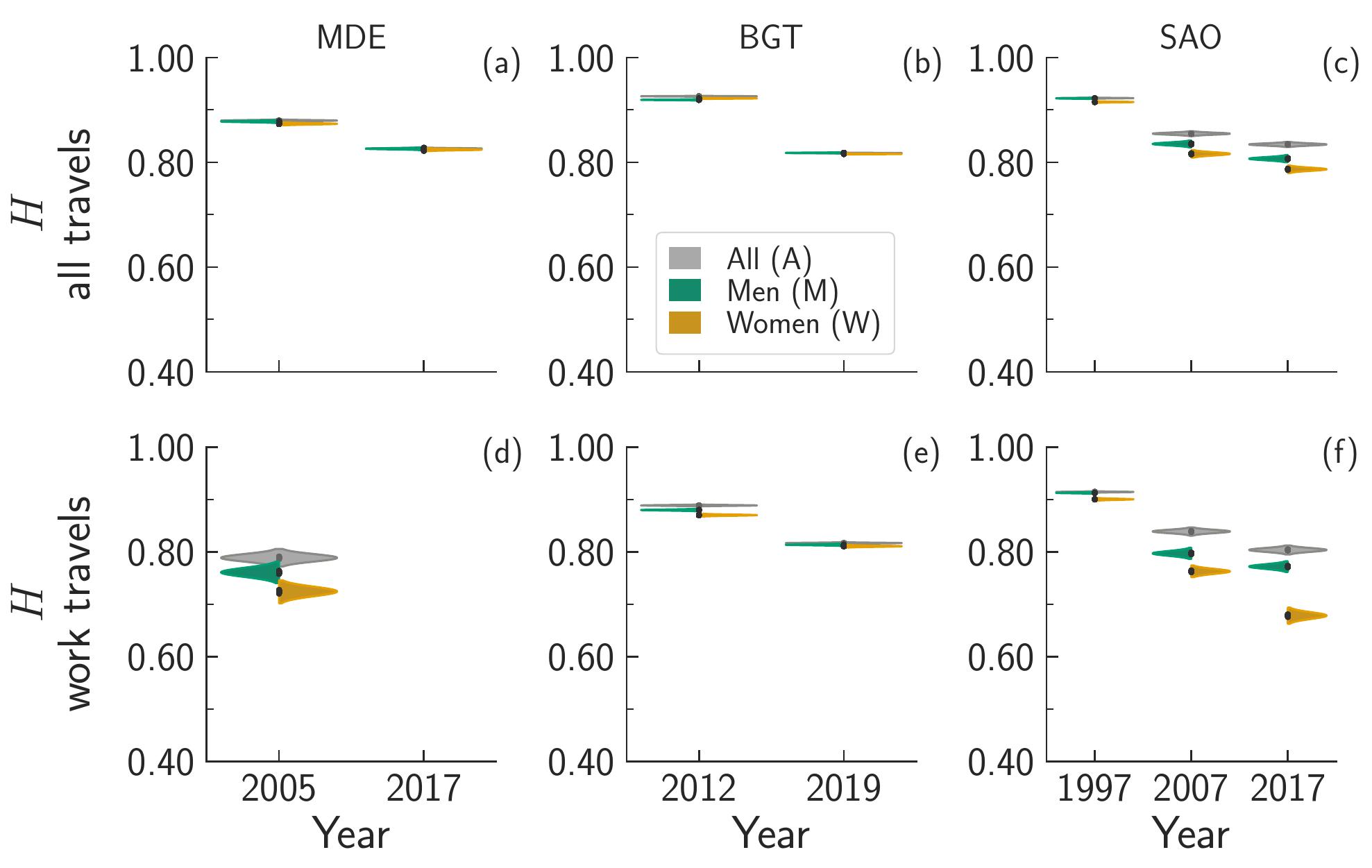}
    \caption{\textbf{Violin plots of the bootstrapped mobility diversity, $H$, for travels whose destination zone does not coincide with the zone where the traveller lives.} See the caption of \ref{fig:genderregions_noselfloop}~Fig for the notation's details and other information.}
    \label{fig:genderregions_nores}
    \end{figure}

    \textcolor{black}{After removing either the endogenous or the residential travels, we quantify the effects of gender by computing $H$ on the set of the remaining travels. \ref{fig:genderregions_noselfloop}~Fig portrays the violin plots of $H$ computed for the sets of non-endogenous travels made either for \all (panels a-c), or \work (panels d-e) purpose. The visual comparison of the average values of $H$ displayed in Fig~\ref{M-fig:genderregions} and \ref{fig:genderregions_noselfloop}~Fig highlights in general a decrease of $H$ in the latter, especially for the \sao{'s} area, as well as a starker gender difference. Moreover, the removal of endogenous travels affects more the diversity of \work travels than that of \all purposes travels. We repeat the comparison for the case of travels excluding the residential ones (\ref{fig:genderregions_nores}~Fig). As for the non-endogenous travels case, we do observe a general decrease of the values of $H$ together with an amplification of the gender differences.}
    %
    %
    %
    %
    %

    \newpage
    
\begin{figure}[H]
		\centering
		\includegraphics[width=0.9\textwidth]{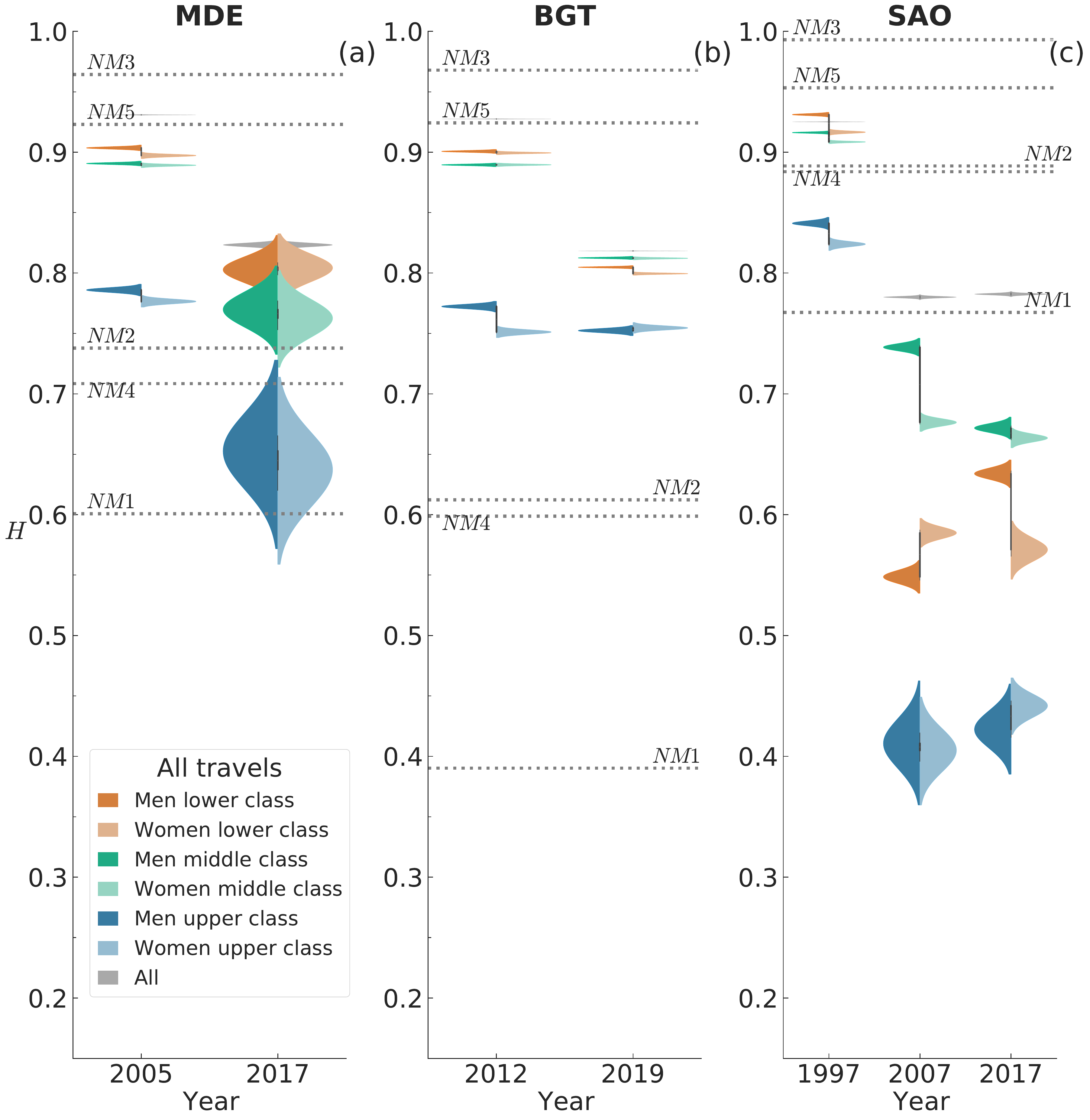}
		\caption{\textbf{Violin plots of the mobility diversity, $H$, of travels made for \all purposes having different origin and destination zones made by travellers grouped according to their socioeconomic status and gender.} Each plot refers to a different region and, for each region, we consider all the available years. For each socioeconomic status (\upperc, \middlec, and \lowerc) a darker hue denotes men travellers, whereas lighter hue denotes women ones. Dotted lines in grey denote the values of $H$ computed from travels generated using null models $NMx$ with $x \in \{ 1, \ldots, 5 \}$ (see \ref{sec:null_models}~Section).}
		\label{fig:noselfloops_allgendersocioregions}
	\end{figure}

\begin{figure}[H]
		\centering
		\includegraphics[width=0.9\textwidth]{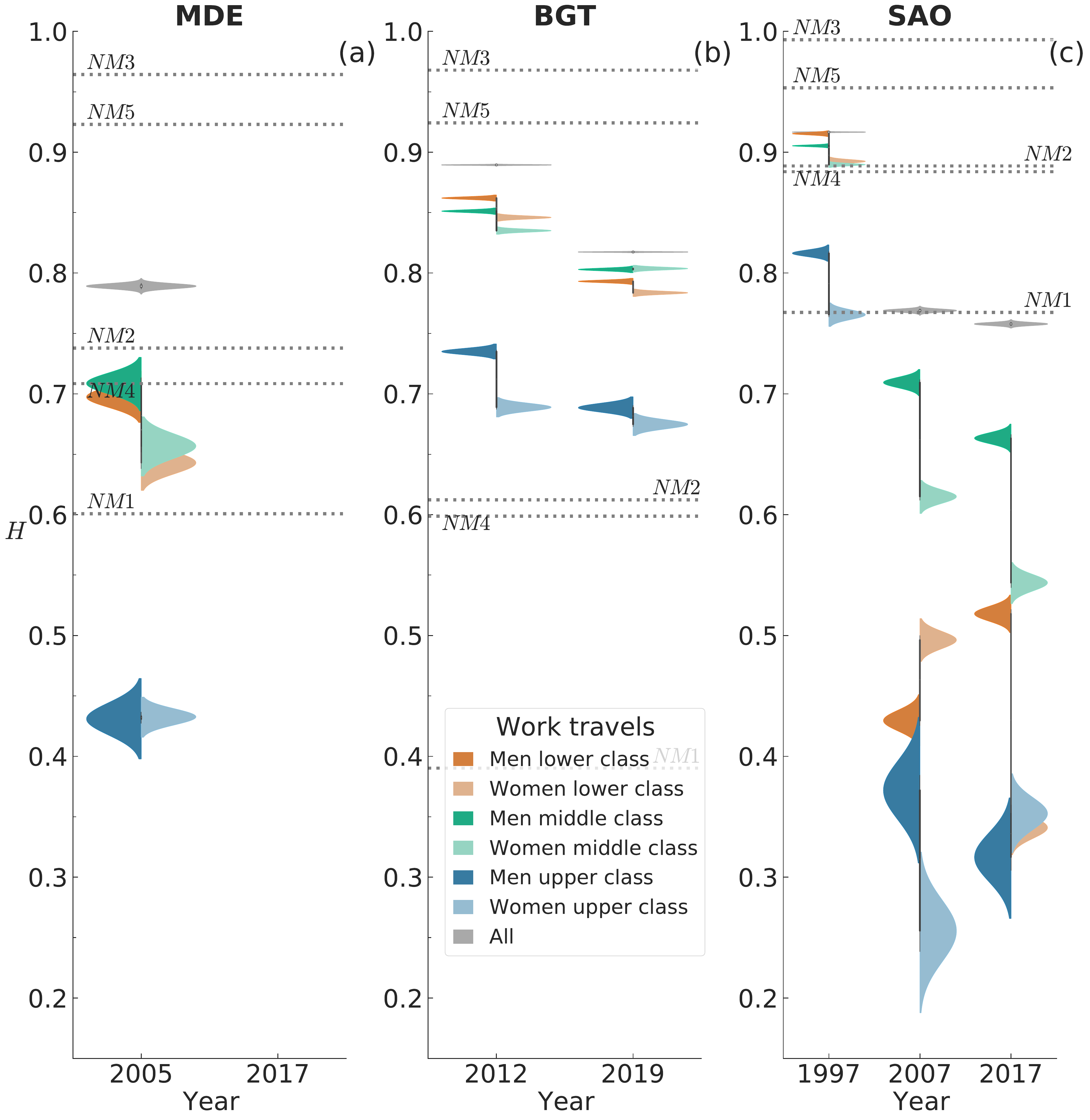}
		\caption{\textbf{Violin plots of the mobility diversity, $H$, of travels made for \work purposes that the origin and destination are different by travellers grouped according to their socioeconomic status and gender.} See the caption of \ref{fig:noselfloops_allgendersocioregions}~Fig for further details.}
		\label{fig:noselfloops_workgendersocioregions}
	\end{figure}

\begin{figure}[H]
		\centering
		\includegraphics[width=0.9\textwidth]{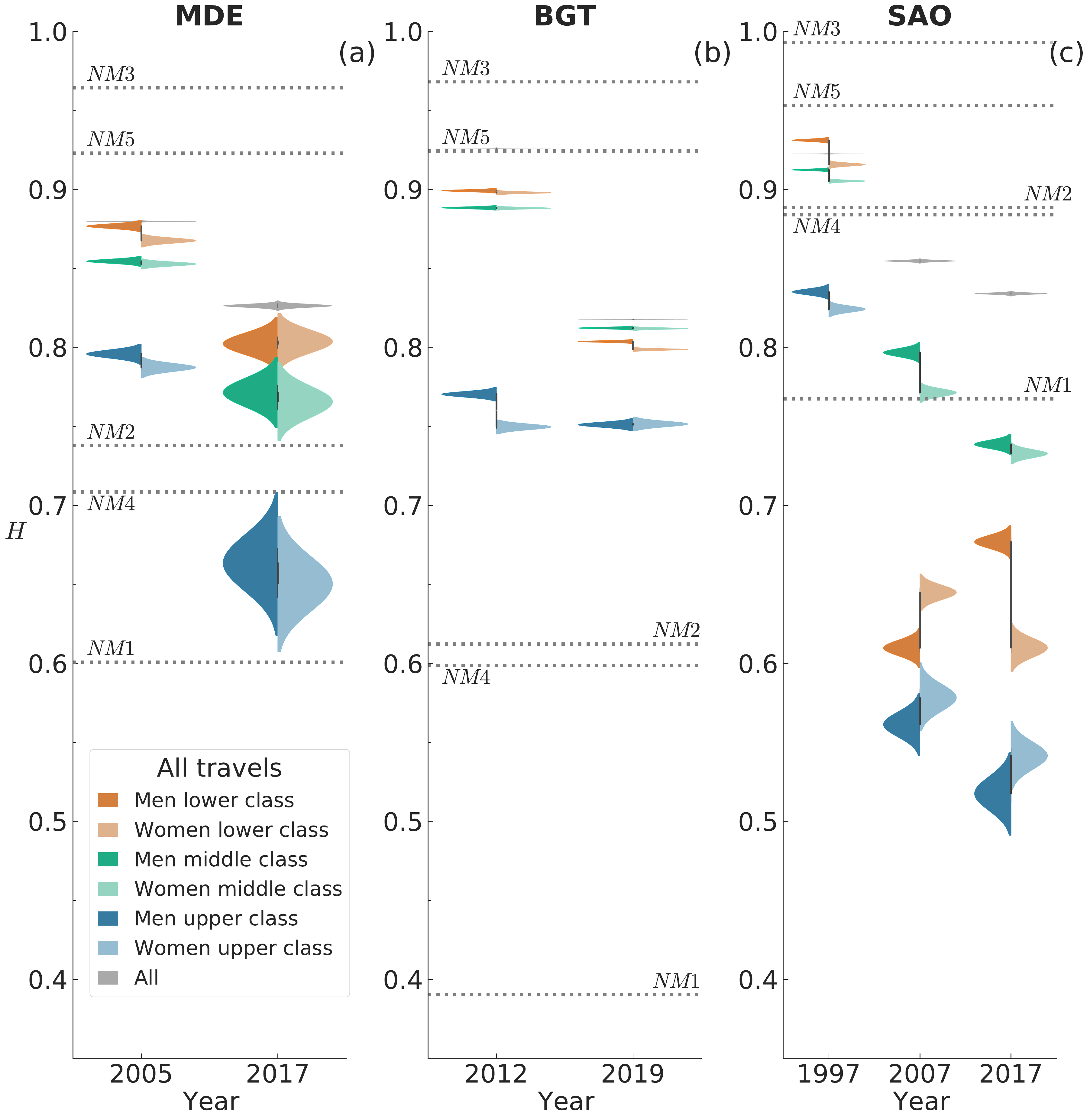}
		\caption{\textbf{Violin plots of the mobility diversity, $H$, of travels made for \all purposes whose destination zone is different from the traveller's home zone and made by travellers grouped according to their socioeconomic status and gender.} See the caption of \ref{fig:noselfloops_allgendersocioregions}~Fig for further details.}
		\label{fig:nohome_allgendersocioregions}
	\end{figure}

\begin{figure}[H]
		\centering
		\includegraphics[width=0.9\textwidth]{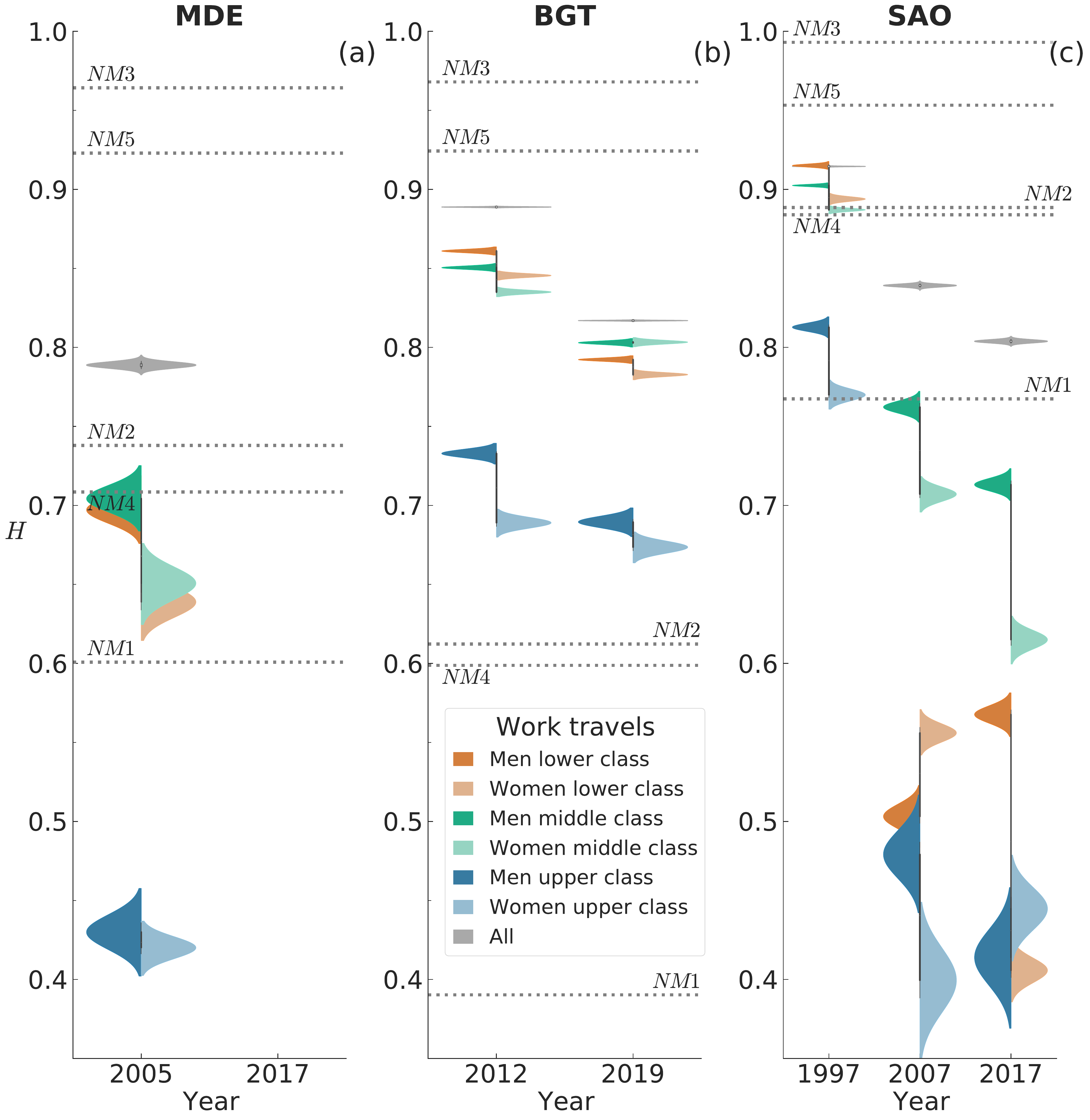}
		\caption{\textbf{Violin plots of the mobility diversity, $H$, of travels made for \work purposes whose destination zone is different from the traveller's home zone and made by travellers grouped according to their socioeconomic status and gender.} See the caption of \ref{fig:noselfloops_allgendersocioregions}~Fig for further details.}
		\label{fig:nohome_workgendersocioregions}
	\end{figure}

    \textcolor{black}{After looking at the effects of removing endogenous and residential mobility on travels grouped by gender only, we can repeat the same analysis on the travels grouped according to both gender and socioeconomic status. \ref{fig:noselfloops_allgendersocioregions} and \ref{fig:noselfloops_workgendersocioregions} Figs are the non-endogenous counterparts of Fig \ref{M-fig:allgendersocioregions} and \ref{fig:workgendersocioregions} Fig, whereas \ref{fig:nohome_allgendersocioregions} and \ref{fig:nohome_workgendersocioregions} Figs account for the non-residential case.}
    
    \textcolor{black}{Independently on the travel's purpose considered, pruning either endogenous or residential travels affects the average values of mobility diversity, $\avg{H}$. In particular, we observe a generalised decrease in $\avg{H}$ as well as, for a given socioeconomic status, an amplification of the gender based differences, $\Delta H = \left\lvert \avg{H^W} - \avg{H^M} \right\rvert$ (with indices $W$ and $M$ denoting \women{} and \men{}). A clear example is the case of \sao{} where both endogenous and residential mobility play a significant role on the values of $\avg{H}$ and the differences observed between women and men belonging to the same socioeconomic status. Still, the extent of both the decrease in $\avg{H}$ and increase of $\Delta H$ is not constant neither between regions nor across years or travel's purpose.}
    %
    %
	%
	%
    %
	%
	%
    %
	%
	%
    %

	\textcolor{black}{However, rather than limiting ourselves to a qualitative visual comparison of the violin plots, we adopted a more quantitative approach by measuring the peak-to-peak difference between pairs of $KDE(H)$, obtained for travels made by men ($M$) and women ($W$) belonging to socioeconomic status $S$, $\Delta g = median(H^M_{S}) - median(H^W_{S})$. Note that in the definition of $\Delta g$ the sign of the peak-to-peak difference determines the hierarchy existing between men and women.}
	
	\textcolor{black}{Tables \ref{tab:genderdifferences_noselfloop} and \ref{tab:genderdifferences_nohome} summarise the outcome of our analysis. In particular, we observe how in most cases the values of $\Delta g$ are remarkably bigger than the same quantity computed including also endogenous and residential travels, regardless of the travel's purpose. In addition, in some cases (\eg{} \sao) the values of $\Delta g$ are even bigger than the maximum difference computed considering all the travels, $\max(\Delta g^\star)$. Moreover, \work{} travels differences appear to be those more affected by the elimination of endogenous and residential mobility. Finally, we notice that sometimes the hierarchy between men and women gets inverted (denoted by highlighted cells).}
    
    \textcolor{black}{In conclusion, the analysis of the mobility diversity computed excluding endogenous and residential mobility seem to rule out the hypothesis that the gender differences observed in the mobility diversity are simply the byproduct of residential segregation and are due, instead, to intrinsic differences in the way women and men move.}
    
    %
    %
	\begin{table}[H]
	\caption{\textbf{Gender based differences of the Kernel Density Estimator $KDE(H)$ of the mobility diversity, $\Delta g$, for travels made for \all and \work purposes \textcolor{black}{having different origin and destination zones} made by travellers grouped according to their socioeconomic status, $S \in \{ \text{\lowerc, \middlec, \upperc} \}$, and gender.} The values of $\Delta g$ are multiplied by a factor of $10^{3}$. Highlighted cells represent the cases in which $\Delta g < 0$ (\ie{} $median(H^W_{S}) > median(H^M_{S})$).  \textcolor{black}{Column $G$ denotes the gender of the more diverse travellers (M for men and W for women). The symbol $>$ ($<$) in column $T$ denotes differences whose values are -- in absolute value -- bigger (smaller) than the one obtained taking into account all travels. The value of $\max(\Delta g^\star)$ is computed from \ref{tab:genderdifferences} Table.}}
	\label{tab:genderdifferences_noselfloop}
	\centering
	\resizebox{0.85\linewidth}{!}{
		\begin{tabular}{ccl|r|rrr|rrr|rrr|rrr}
		    \toprule
			\multirow{2}{*}{City} & \multirow{2}{*}{Year} & \multirow{2}{*}{Purpose} & \multirow{2}{*}{$\max(\Delta g^\star)$} & \multicolumn{3}{c|}{\all} & \multicolumn{3}{c|}{\lowerc} & \multicolumn{3}{c|}{\middlec} & \multicolumn{3}{c}{\upperc}\\
			 &  &  &  & $\Delta g$ & $G$ & $T$ & $\Delta g$ & $G$ & $T$ & $\Delta g$ & $G$ & $T$  & $\Delta g$ & $G$ & $T$ \\
			\midrule
				
			\multirow{4}{*}{\mde} 
			 & \multirow{2}{*}{2005} & \all & 12.10 & 4.21 & M & $<$ & 6.30 & M & $>$ & 1.45 & M & $<$ & 9.47 & M & $<$\\
			 & & \work & -8.07 & 32.15 & M & $>$ & 54.41 & M & $>$ & 51.83 & M & $<$ & \cellcolor{gray!25} 1.72 & \cellcolor{gray!25} W & \cellcolor{gray!25} $<$\\
				
			 & \multirow{2}{*}{2017} & \all & 16.15 &  3.11 & M &  $>$ &  \cellcolor{gray!25}1.55 & \cellcolor{gray!25} W & $\cellcolor{gray!25} >$ & 7.21 & M & $>$ & 15.33 & M & $<$\\
			 & & \work & 19.67 & 324.73 & M & $>$ & 201.37 & M & $>$ & 137.85 & M & $>$ & 50.04 & M & $>$\\[0.125cm]
			\hline
				
			\multirow{4}{*}{\bgt} 
			 & \multirow{2}{*}{2012} & \all & 15.39 & \cellcolor{gray!25} 2.97 & \cellcolor{gray!25} W & \cellcolor{gray!25} $<$ & 1.24 & M & $<$ & \cellcolor{gray!25} 0.06 & \cellcolor{gray!25} W & \cellcolor{gray!25} $<$ & 21.13 & M & $>$\\
			 & & \work & 33.89 & 10.36 & M & $>$& 16.05 & M & $>$ & 16.13 & M & $>$ & 46.09 & M & $>$\\
			 & \multirow{2}{*}{2019} & \all & 2.05 & 2.10 & M & $<$ & 5.28 & M & $>$ & 0.33 & M & $<$ & \cellcolor{gray!25} 2.22 & \cellcolor{gray!25} W & \cellcolor{gray!25} $<$\\
			 & & \work & 5.15 & 2.41 & M & $<$ & 9.46 & M & $>$ & \cellcolor{gray!25} 0.76 & \cellcolor{gray!25} W & \cellcolor{gray!25} $>$ & 13.76 & M & $>$\\[0.125cm]
			\hline
				
			\multirow{6}{*}{\sao} 
			 & \multirow{2}{*}{1997} & \all & 16.05 & 7.52 & M & $>$ & 14.55 & M & $>$ & 7.73 & M & $>$ & 17.20 & M & $>$ \\
			 & & \work & 30.75 & 13.18 & M & $>$ & 22.92 & M & $>$ & 15.57 & M & $>$ & 50.80 & M & $>$\\
			 & \multirow{2}{*}{2007} & \all & -5.97 & 39.92 & M & $>$ & \cellcolor{gray!25} 36.43 & \cellcolor{gray!25} W & \cellcolor{gray!25} $>$ & \cellcolor{gray!25} 62.20 &\cellcolor{gray!25} W & \cellcolor{gray!25} $>$ & 5.53 & M & $<$ \\
			 & & \work & 9.95 & 47.08 & M & $>$ & \cellcolor{gray!25} 66.70 & \cellcolor{gray!25} W & \cellcolor{gray!25} $>$ & 94.36 & M & $>$ & 116.40  & M & $>$\\
			 & \multirow{2}{*}{2017} & \all & 6.51 & 37.53 & M & $>$ & 62.91 & M & $>$ & 8.25 & M & $>$ & \cellcolor{gray!25} 19.63 & \cellcolor{gray!25} W & \cellcolor{gray!25} $>$ \\
			 & & \work & 15.34 & 110.94 & M & $>$ & 176.97 & M & $>$ & 119.60 & M & $>$ & \cellcolor{gray!25} 36.16 & \cellcolor{gray!25} W &  \cellcolor{gray!25} $>$ \\
			\bottomrule
        \end{tabular}
	}
	\end{table}
	%
	
	%
    %
	\begin{table}[H]
	\caption{\textbf{Gender based differences of the Kernel Density Estimator $KDE(H)$ of the mobility diversity, $\Delta g$, for travels made for \all and \work purposes \textcolor{black}{having not the residential zone of the traveller as destination zone} made by travellers grouped according to their socioeconomic status, $S \in \{ \text{\lowerc, \middlec, \upperc} \}$, and gender.} See the caption of  \ref{tab:genderdifferences_noselfloop}~Table for notations and definitions.}
	\label{tab:genderdifferences_nohome}
	\centering
	\resizebox{0.85\linewidth}{!}{
	    \begin{tabular}{ccl|r|rrr|rrr|rrr|rrr}
		\toprule
		\multirow{2}{*}{City} & \multirow{2}{*}{Year} & \multirow{2}{*}{Purpose} & \multirow{2}{*}{$\max(\Delta g^\star)$} & \multicolumn{3}{c|}{\all} & \multicolumn{3}{c|}{\lowerc} & \multicolumn{3}{c|}{\middlec} & \multicolumn{3}{c}{\upperc}\\
         &  & & & $\Delta g$ & $G$ & $T$ & $\Delta g$ & $G$ & $T$ & $\Delta g$ & $G$ & $T$ &
		 $\Delta g$ & $G$ & $T$ \\
		\midrule
				
		\multirow{4}{*}{\mde} 
		 & \multirow{2}{*}{2005}
		 & \all & 12.10 & 4.44 & M & $>$ & 9.11 & M & $>$ & 1.74 & M & $<$ & 8.51 & M & $<$\\
		 & & \work & -8.07 & 36.50 & M & $>$ & 58.29 & M & $>$ & 53.54 & M & $>$ & 9.84 & M & $>$\\
		 & \multirow{2}{*}{2017} 
		 & \all & 16.15 & 1.71 & M & $<$ & \cellcolor{gray!25} 1.49 & 				\cellcolor{gray!25} W & \cellcolor{gray!25} $>$ & 5.87 & M & $>$ & 13.23 & M & $<$\\
		 & & \work & 19.67 & 290.01 & M & $>$ & 157.93 & M & $>$ & 166.62 & M & $>$ & 52.31   & M & $>$\\[0.125cm]
	    \hline
				
		\multirow{4}{*}{\bgt} 
		 & \multirow{2}{*}{2012} & \all & 15.39 & \cellcolor{gray!25} 2.78  & \cellcolor{gray!25} W & \cellcolor{gray!25} $<$ & 1.25 & M & $<$ & 0.23 & M & $<$ & 20.61 & M & $>$ \\
		 & & \work & 33.89 & 9.76 & M & $>$ & 15.50 & M & $>$ & 15.43 & M & $>$ & 43.84 & M & $>$\\
		 & \multirow{2}{*}{2019} 
		 & \all & 2.05 & 2.16 & M & $>$ & 5.03 & M & $>$ & 0.32 & M & $<$ & \cellcolor{gray!25} 0.43 & \cellcolor{gray!25} W & \cellcolor{gray!25} $<$ \\
		 & & \work & 5.15 & 2.79 & M & $>$ & 9.48 & M & $>$ & \cellcolor{gray!25} 0.34 & \cellcolor{gray!25} W & \cellcolor{gray!25} $>$ & 15.93 & M & $>$\\[0.125cm]
		\hline
				
		\multirow{6}{*}{\sao} 
		 & \multirow{2}{*}{1997} & \all & 16.05 & 6.98 & M & $>$ & 14.55 & M & $>$ & 7.73 & M & $>$ & 17.20 & M & $>$ \\
		 & & \work & 30.75 & 12.15 & M & $>$ & 15.37 & M & $>$ & 7.09 & M & $>$ & 10.90 & M & -\\
		 & \multirow{2}{*}{2007} & \all & -5.97& 18.92 & M & $>$ & \cellcolor{gray!25} 35.16 & \cellcolor{gray!25} W & \cellcolor{gray!25} $>$ & 25.35 & M & $>$ & \cellcolor{gray!25} 16.93 & \cellcolor{gray!25} W & \cellcolor{gray!25} $>$ \\
		 & & \work & 9.95 & 34.27 & M & $>$ & \cellcolor{gray!25} 52.87 & \cellcolor{gray!25} W & \cellcolor{gray!25} $>$ & 55.11 & M & $>$ & 79.61 & M & $>$ \\
		 & \multirow{2}{*}{2017} & \all & 6.51 & 20.15 & M & $>$ & 67.04 & M & $>$ & 5.90 & M & $>$ & \cellcolor{gray!25} 24.00 & \cellcolor{gray!25} W & \cellcolor{gray!25} $>$ \\
		 & & \work & 15.34 & 93.56 & M & $>$ & 161.93 & M & $>$ & 98.13 & M & $>$ & \cellcolor{gray!25} 30.85 & \cellcolor{gray!25} W & \cellcolor{gray!25} $>$ \\
		\bottomrule
		\end{tabular}
		}
	\end{table}

	\newpage
	
	%
	%
	\bibliographystyle{plain} 
	\bibliography{references}
	
%
%

\makeatletter

\makeatother